\documentclass[12pt]{article}
\usepackage[utf8]{inputenc}
\usepackage[colorlinks=true,allcolors=blue]{hyperref}%
\usepackage[margin=1in]{geometry} 
\usepackage{graphicx} 
\usepackage{caption}
\usepackage{amsmath}
\usepackage[bottom]{footmisc}
\usepackage[authoryear]{natbib}
\usepackage{lscape} 
\usepackage{titling} 
\usepackage{setspace}
\usepackage{longtable}
\usepackage{mathptmx}
\usepackage{placeins}
\usepackage{chngcntr}
\usepackage{xcolor,colortbl}
\usepackage{dsfont}
\usepackage{setspace}
\usepackage{tabularx}
\newcolumntype{L}{>{\raggedright\arraybackslash}X}
\usepackage{multirow}
\usepackage{booktabs}
\usepackage{subcaption}
\usepackage{cleveref}
\DeclareMathAlphabet{\mathcal}{OMS}{cmsy}{m}{n}
\definecolor{darkgreen}{rgb}{0.20,0.43,0.09}
\definecolor{colhigh}{RGB}{50, 80, 200} 

\onehalfspace

\begin{document}

\begin{titlepage}
\title{\textsc{\textbf{Relatio}: Text Semantics Capture \\ Political and Economic Narratives}}
\author{Elliott Ash\thanks{ETH Zürich, \texttt{ashe@ethz.ch}},  Germain Gauthier\thanks{CREST -- Ecole Polytechnique,  \texttt{germain.gauthier@polytechnique.edu}}, Philine Widmer\thanks{University of St.Gallen, \texttt{philine.widmer@unisg.ch}}}
\date{\today}
\maketitle
\begin{abstract}
\noindent Social scientists have become increasingly interested in how \textit{narratives} -- the stories in fiction, politics, and life -- shape beliefs, behavior, and government policies. This paper provides an unsupervised method to quantify latent narrative structures in text documents. Our new software package \textsc{relatio} identifies coherent entity groups and maps explicit relations between them in the text. We provide an application to the \textit{United States Congressional Record} to analyze political and economic narratives in recent decades. Our analysis highlights the dynamics, sentiment, polarization, and interconnectedness of narratives in political discourse.\footnote{A special thanks to Andrei Plamada and ETH Scientific IT Services for pivotal contributions to the programmatic implementation \textsc{relatio} (\url{https://github.com/relatio-nlp/relatio}). We also thank Johannes Binswanger, Jordan Boyd-Graber, Paul Dutronc-Postel, Kfir Eliaz, Malka Guillot, Kai Gehring, Theresa Gessler,  Roland Hodler, Philip Resnick, Alessandro Riboni, Michael Roos, Ran Spiegler, and Ekaterina Zhuravskaya for insightful comments. We are grateful for feedback from conference participants at PolMeth U.S., PolMeth Europe, the Swiss Society of Economics and Statistics Conference, the Monash-Warwick-Zurich Text-as-Data Workshop, as well as seminar attendees at CREST, ETH, the Paris School of Economics, and the Universities of Zurich and St.Gallen. This work was supported by the Swiss National Science Foundation. We thank all (early) users of \textsc{relatio} for providing valuable feedback, in particular Andrea Sipka and Wenting Song.}

\bigskip


\end{abstract}
\end{titlepage}
\newpage

\setcounter{page}{2}
 
\section{Introduction}
 
If stories are the engines of political identity and collective action, then the best politicians will also be the best storytellers \citep{patterson1998narrative}. Motivated by such ideas, and building on the well-established work in psychology on how narratives shape human perceptions and constructions of social reality \citep[][]{white1980value,polkinghorne1988narrative,bruner1991narratives,armstrong2020stories}, an expanding social-science literature has begun to attend to how narratives drive political, social, and economic outcomes \citep[][]{BranchMcGoughZhu2017,BenabouFalkTirole2018,kuhle2020thought,eliaz2020model}. Across disciplines, there is a growing recognition that narratives play a fundamental role in both individual and collective decision-making \citep{akerlof2010animal,shiller2019narrative}. Hence, both science and policy would benefit from a better understanding of how narratives form, spread, and influence behavior. 

Nonetheless, quantitative analysis of narratives is still largely unexplored \citep{shiller2019narrative}. A principal impediment to the production of such evidence is the challenge of measuring narratives in written or spoken texts. In particular, such a measure must capture relationships between entities -- characters, concepts, organizations, or objects \citep[e.g.][]{sloman2005causal}. The existing text-as-data approaches in social science do not account for these relationships \citep{GrimmerStewart2013,GentzkowKellyTaddy2019JEL}. This paper describes a new method that satisfies this requirement -- by identifying \textit{who} does \textit{what} to \textit{whom}, and by mapping the relationships and interactions among entities in a corpus. The accompanying open-source package, \textsc{relatio}, allows researchers to measure interpretable narratives from plain text-documents. These narratives can be used as inputs in empirical social science analysis.

The starting point of the narrative mining method is semantic role labeling, a linguistic algorithm that takes in a plain-text sentence and identifies the action, the agent performing that action, and the patient being acted upon. The resulting feature space of agents, actions, and patients is much more informative about narratives than the feature space generated by standard text-as-data methods. Yet, that space is too high-dimensional to be useful for most narrative analyses. Hence, the next part of our narrative mining method is a set of dimensionality reduction procedures. Our entity clustering approach takes the multiple phrase variants referring to the same entity (e.g., ``taxes on income'' and ``income taxation'', or ``Former President Reagan'' and ``Ronald Reagan'') and resolves them into a single entity label.

We demonstrate the usefulness of the method in an application to speeches in the U.S. Congress (1994-2015). The set of extracted entities is interpretable and includes key actors in the U.S. political economy -- Republicans, Democrats, the budget, terrorists, or Medicare, to name just a few. For each entity, our approach highlights what actions (i.e., verbs) connect it to other entities. For instance, we capture that ``Republicans'' and ``Medicare'' are related such that ``Republicans end Medicare''. ``Medicare'', in turn, is also connected to other entities, such as ``Medicare provides healthcare''. Overall, the resulting narrative statements are intuitive and close to the original raw text. 
 
The method captures salient narratives around historical events, such as the September 11th attacks and the subsequent U.S. invasion of Iraq. We find that religious invocations such as ``God bless America'' and ``God bless the troops'' increased in response to these events. Ranking narratives by relative party usage reveals articulations of partisan values: Democrats are concerned about ``Americans losing unemployment benefits'', ``budget cuts to Medicare'', and ``oil companies making profits''. Republicans want ``Americans to keep their money'', decry ``government control of healthcare'', and affirm that ``Americans rely on oil''. 

Besides showing how narratives divide the parties, our approach also demonstrates how narratives connect up with each other to form a broader discourse. We propose a network-based approach to combine several narrative statements in a directed multigraph, linking up entities with their associated actions. The resulting narrative networks form a visual expression of political worldviews, establishing rich context for qualitative researchers. Further, applying node centrality and graph distance measures to these networks could help illuminate new dimensions of narrative discourse.

The paper concludes with a discussion of how narrative mining fits in with other text-as-data methods. The narrative statements produced by \textsc{relatio} are often more informative and interpretable than bag-of-words or n-gram representations of text documents. Finally, we discuss our method's limitations along with opportunities for improvement and extension. 

\section{Method: Mining Narratives from Plain Text Corpora} \label{sec:Methods}

According to the Oxford English Dictionary, a narrative is an ``account of a series of events, facts, etc., given in order and with the establishing of connections between them''. In a leading framework from social psychology, narratives are similarly defined as sets of relationships between entities that act on each other \citep{sloman2005causal}. In human language, such relationships are established through grammatical statements describing actors, actions, and the acted-upon.  

This distinctive relational aspect of narratives is missing from the standard text-as-data-tools in social science. Dictionary methods rely on matching particular words or phrases \citep{BakerBloomDavis2016QJoE,shiller2019narrative,enke2020moral}. Unsupervised learning methods such as topic models and document embeddings break sentences down into words or phrases and ignore grammatical information \citep{HansenMcMahonPrat2017TQJoE,bybee2020structure}. These previous methods recover information on mentioned entities and mentioned actions, but without connecting them. This section outlines our method for uncovering the recurring relationships established between latent entities and actions in a corpus. 

\subsection{Semantic roles as narrative building blocks} \label{sec:notation}

Semantic role labeling (SRL) is a computational-linguistics algorithm that answers basic questions about the meaning of written sentences -- in particular, \textit{who} is doing \textit{what} to \textit{whom}.\footnote{Linguistically, semantic roles are functional components of sentences that are defined based on their relationship with the main verb in a clause. For a detailed background, see \cite{jurafsky2020speech}, ch. 20 (and sources cited therein). For information on the particular schema we use, see \citet{bonial2012english}.} The \textbf{agent} (``who'') is the actor in an event -- e.g., the grammatical subject of the verb in an active clause. The \textbf{verb} (``what'') captures the action in the clause. The \textbf{patient} (``whom'') is the entity that is affected by the action -- i.e., the object or the target.\footnote{For simplicity, we use the word ``patient'' to refer to both linguistic patients and linguistic benefactives -- in most sentences, these refer to the direct object and indirect object, respectively.} For example, in the sentence ``Millions of Americans lost their unemployment benefits'', ``Millions of Americans'' is the agent, ``lost'' the verb, and ``their unemployment benefits'' the patient.\footnote{Any role can be empty (except for the verb). For instance, with ``Millions of Americans suffer'', the patient role is empty.} 

Not only do semantic roles differentiate actions and entities within a sentence, but they also map the relationships between them. For example, SRL would extract the same directional relation for ``Millions of Americans lost their unemployment benefits'' as the inverted sentence ``Unemployment benefits were lost by millions of Americans.'' In some cases, the directions of actions contain pivotal information to understand the narrative.\footnote{Consider for example: ``Terrorists threaten American troops.'' and ``American troops threaten terrorists''.} This robustness to word ordering is an important feature of SRL relative to other approaches in natural language processing, such as topic models, which ignore semantic relations in the sentence.  

\begin{figure}
\begin{center}
\caption{Examples of Semantic Role Labeling Annotations}
\includegraphics[width = 0.32\textwidth]{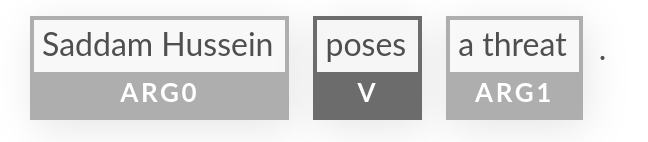}

        \includegraphics[width = 0.4\textwidth]{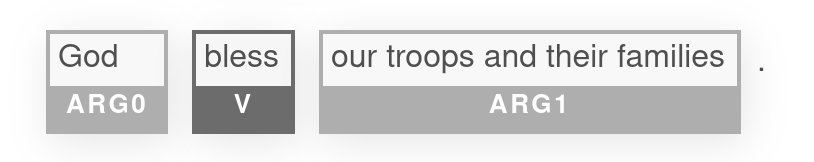}
        
        \includegraphics[width = 0.4\textwidth]{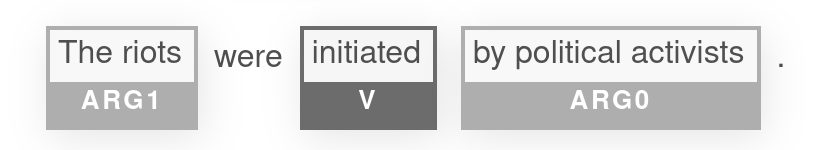}
        
        \includegraphics[width = 0.5\textwidth]{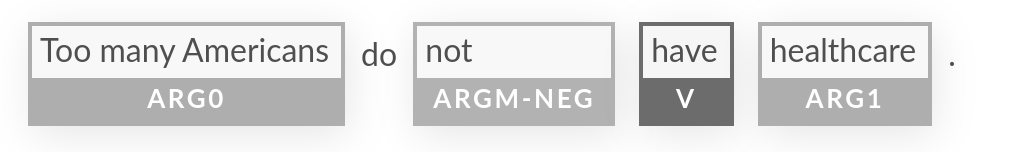}
        
        \includegraphics[width = 0.7\textwidth]{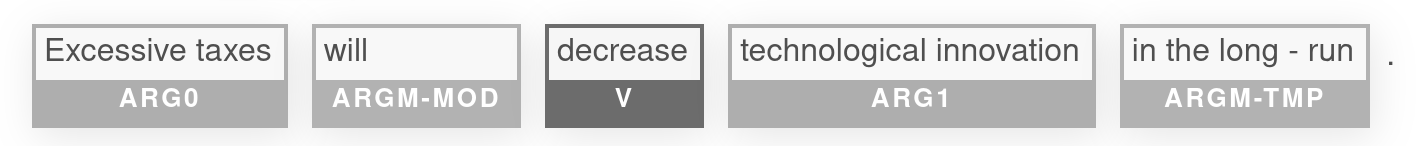}
\label{fig:SRLExample}
\end{center}
\scriptsize
\flushleft
\textbf{Note:} Examples of semantic role annotations based on \texttt{allennlp}'s programmatic implementation \citep{Gardner2017AllenNLP}. See \url{https://demo.allennlp.org/semantic-role-labeling} for additional examples. ARG0 refers to the agent, V to ther verb, and ARG1 to the patient. The last example shows additional semantic roles, modality (ARGM-MOD), negation (ARGM-NEG), and temporality (ARGM-TMP). Our implementation considers negations (ARGM-NEG). While not further discussed here, it also allows for modal indicators (but not yet for temporality).
\end{figure}

We show additional example sentences, with semantic role annotations, in Figure \ref{fig:SRLExample}. These examples come from the state-of-the-art model implemented by AllenNLP \citep{Gardner2017AllenNLP}, used in the empirical application below. The sentences illustrate the flexibility of SRL in capturing variation in sentence content and word order. For example, the agent and patient can both be at the beginning or end of the sentence. Further, while all complete sentences in English have a verb, some sentences lack an agent, patient, or both.\footnote{``Americans go'' has an agent, but not a patient; ``Americans change'' has a patient, but not an agent, and ``Go'' (the imperative) has a verb but neither an explicit agent or patient.} A number of additional semantic features, such as modality, negations, and temporal content, are labeled as well.

Our model of language is based on the information contained in semantic roles. Formally, define $\mathcal{A}_0$, $\mathcal{V}$, and  $\mathcal{A}_1$ as the sets of phrases respectively annotated as agents, verbs and patients. A sentence can be decomposed into a sequence of semantic roles:
\begin{equation} \label{eq:srl_role_sequence}
\textsc{Agent} \xrightarrow{\textsc{Verb}} \textsc{Patient} \in \mathcal{A}_0 \times \mathcal{V} \times \mathcal{A}_1
\end{equation}
For example, our previously discussed sentence would be represented as: 
\begin{equation*}
\text{Millions of Americans} \xrightarrow{\text{\, lost \,}} \text{their unemployment benefits}
\end{equation*}
Negation is encoded as a modification of the verb. For example, ``did not lose'' is replaced with the verb ``not-lost''. So, a similar statement with a negation would be represented as:
\begin{equation*}
\text{Americans} \xrightarrow{\text{\, not-lost \,}} \text{their unemployment benefits}
\end{equation*}

\subsection{From semantic roles to interpretable actions and entities \label{subsec:entity-extraction}}

Let $\mathcal{S}= \mathcal{A}_0 \times \mathcal{V} \times \mathcal{A}_1$ comprise the space of semantic roles observed in the corpus. These semantic roles capture the relationships between entities that are characteristic of narrative statements. In most real-world corpora, however, $\mathcal{S}$ is too high-dimensional for further analysis. Thus, the next step is to compress the set of actions and entities to a lower-dimensional yet sufficiently informative representation. 

For instance, most people would agree that ``Millions of Americans lost their unemployment benefits'' and ``Many Americans were losing their much-needed unemployment checks'' refer to the same underlying narrative. In particular, ``much-needed unemployment checks'' and ``their unemployment benefits'' both refer to ``unemployment benefits''. Similarly, ``Many Americans'' and ``Millions of Americans'' both refer to ``Americans''. These examples illustrate that observed agents and patients are drawn from a smaller set of latent entities, $E$, such that $|E| \leq |\mathcal{A}_0 \cup \mathcal{A}_1|$. Thus, our dimension reduction aims to infer these latent entities and their associated text realizations. 

We proceed in two steps for dimension reduction. First, we directly extract coherent entities using named entity recognition \citep[e.g.,][ch. 8]{jurafsky2020speech}. This algorithm automatically extracts references to specific people, organizations, events, and locations. In the second sentence from Figure \ref{fig:SRLExample}, for example, ``Saddam Hussein'' is identified as a named entity. In practice, we build a vocabulary of named entities as the $L$ most frequent named entities recognized in the corpus. If a semantic role refers to a named entity from the vocabulary, it is labeled as such. In applications, we have found that it is straight-forward to inspect the list of entities ranked by frequency and set the threshold $L$ to balance dimensionality and interpretability. Appendix \ref{app:sec:mined_entities} lists examples of frequent named entities.

The second step of dimension reduction, applied to the remaining agent and patient phrases that do not contain a named entity, consists of semantic clustering. These phrases usually refer to coherently separable entities or groups of entities that we would like to group together -- e.g. ``unemployment benefits'' and ``unemployment checks''. Our approach for clustering the phrases associated with such entities begins with a phrase encoder, applied to compress each plain-text entity snippet to a low-dimensional dense vector. As a computationally efficient default, we use a weighted average of the word vectors across each word in the agent or patient segment \citep{AroraLiangMa2016}. We then apply a clustering algorithm (e.g., K-Means) to the matrix of entity encodings to produce $K$ clusters. For interpretability, each cluster can be labeled by the most frequent term within the cluster.

Unlike the selection of $L$ for named entities, the selection of $K$ for clustered entities is not straightforward. Entities are not clearly demarcated and the preferred level of granularity will depend on the application. For example, should ``tax credit'' and ``tax rebate'' be clustered? What about ``Republicans'' and ``conservatives''? Hence, hyperparameter choices for clustering entity embeddings require more care. Automated cluster-quality metrics such as the silhouette score can work as a starting point. But for best results, the phrase clustering output should be produced and inspected for different $K$ and selected based on the goals of the application. 

So far, we have not discussed dimension reduction for verbs. In practice, we find that embedding-based clustering of verbs produces unreliable results, in particular because it frequently assigns antonyms (e.g., ``decrease'' and ``increase'') to the same cluster. In our application, the number of unique verbs is relatively small (60 times smaller than the number of unique agents and patients), so we decide not to dimension-reduce verbs. Hence, we do not make any changes to verbs, except for normalizing the verb tense and adding the ``not-'' prefix to negated verbs.

\subsection{Narrative statements and narrative structure} 
\label{sec:narrative-structure}

After dimension reduction, we obtain narratives of the form:
\begin{equation} \label{eq:narrative_space}
\textsc{agent entity} \xrightarrow{\textsc{(Negated) verb}} \textsc{patient entity} \in E \times V \times E
\end{equation}
The set $\mathcal{N} = E \times V \times E$ is the space of all potential narrative statements for a given corpus. Its dimensionality is determined by the number of unique entities and verbs. The set of entities contains named entities and clustered entities.

These narratives can then provide inputs to qualitative or quantitative analysis. In the Congressional Record, for example, we will produce counts by legislator and year for each item in $\mathcal{N}$. Such counts can be used for descriptive analysis, as variables in regressions, or as a feature set for machine learning algorithms.

Moving beyond such counts, a key feature of narratives is that they embed entities and relations in a broader, enmeshed structure. Consider for instance: ``Taxes fund hospitals and hospitals save lives''. Our method represents this sentence as two separate narrative statements:
\begin{align*}
& \text{taxes} \xrightarrow{\text{\, fund \,}} \text{hospitals} \\
& \text{hospitals} \xrightarrow{\text{\, save \,}} \text{lives}
\end{align*}
It is easy and intuitive to combine these two simple narrative statements to reveal a more complex narrative chain. This broader narrative has a network structure:
\begin{align*}
& \text{taxes} \xrightarrow{\text{\, fund \,}} \text{hospitals} \xrightarrow{\text{\, save \,}} \text{lives}
\end{align*}

More generally, a list of simple narrative statements (as captured by Equation \ref{eq:narrative_space}) can be represented as a directed multigraph, in which the edges are actions and the nodes are entities. Formally, let $n \subset \mathcal{N}$ be a subset of the narrative space. Let $e \subset E$ be the set of distinct entities in $n$. We define a narrative graph as a tuple ($e$, $n$), in which $e$ represents the vertices and $n$ represents the edges of the directed multigraph. 

\begin{figure}
\centering
\caption{Flowchart for \textsc{relatio}}
\includegraphics[width = 0.71\textwidth]{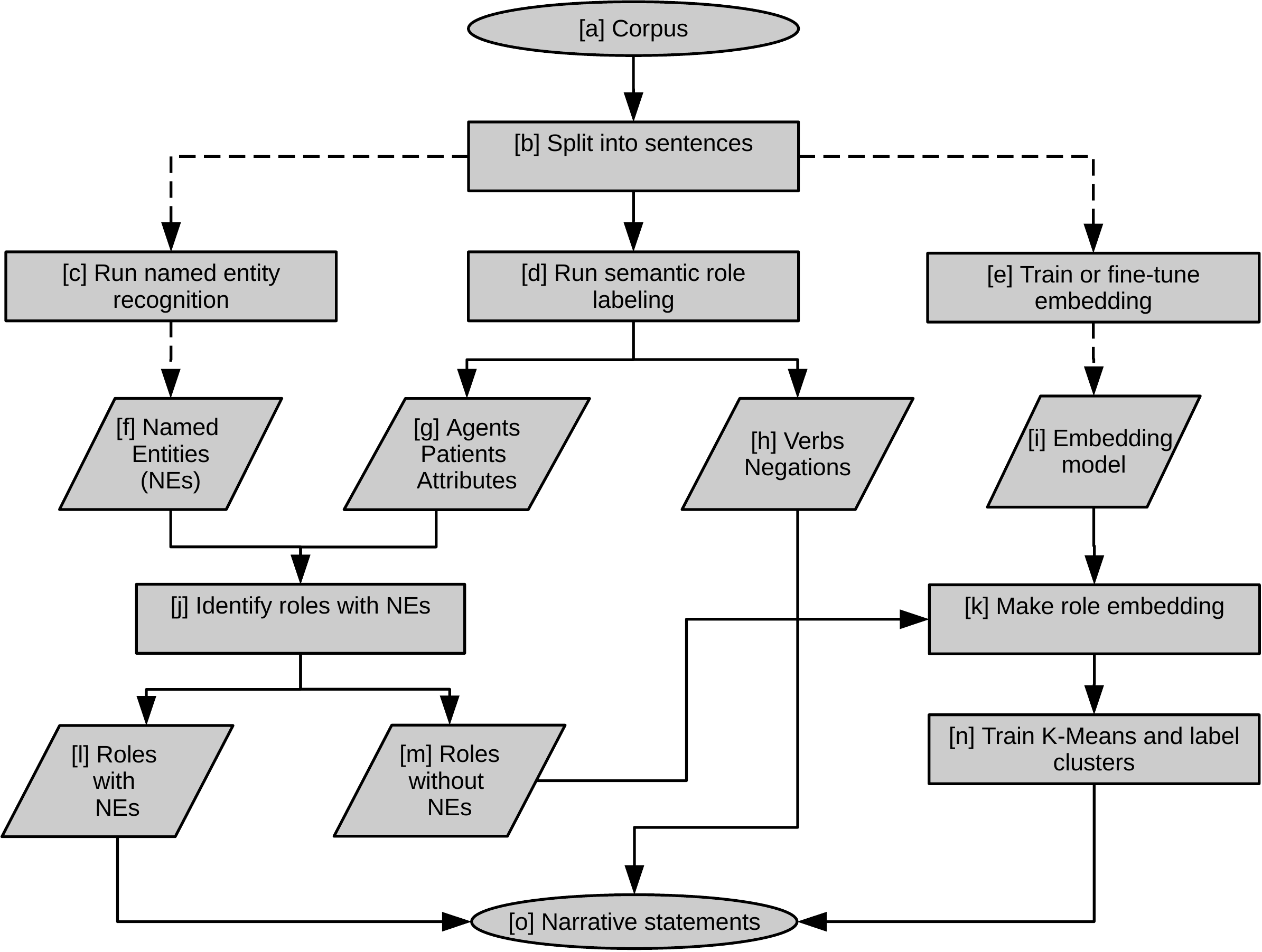}
\label{fig:flowchart}
\scriptsize
\flushleft
\textbf{Note:} Code flowchart for programmatic implementation, open-sourced as the Python \textsc{relatio} package (\url{github.com/relatio-nlp/relatio}). Circles represent the start and the end of the pipeline. Rectangles represent arithmetic operations and data manipulations. Parallelograms represent inputs and outputs.
\end{figure}

\subsection{Open-source implementation} 
\label{sec:open-source-implementation}

Figure \ref{fig:flowchart} provides an illustrated summary of our pipeline. The input is a plain-text corpus [a], which is segmented into sentences [b]. Those sentences are the inputs to named entity recognition [c], semantic role labeling [d], and training (or fine-tuning) a phrase embedding model [e]. The outputs of that first round of operations are a set of named entities [f], annotated semantic roles [g,h], and an embedding model for vectorizing phrases [i]. After tagging named entities [j], the roles containing named entities [l] are finalized for the output. The roles without named entities [m] are vectorized using the phrase embedding model [k], and then piped to the K-means clustering model to produce clustered entities [n]. The final narrative statements [o] are built from the named entities [j], verbs [h], and clustered entities [n]. 

This system is automated and streamlined as part of the accompanying Python package, \textsc{relatio}. A baseline narrative feature set can be produced with only a few lines of code. More advanced users can customize and adapt the inputs, settings, and outputs. For more details, see Appendix \ref{app:sec:implementation} or the package repository, \url{github.com/relatio-nlp/relatio}.

\section{Application to the U.S. Congressional Record}\label{results}

To demonstrate the workings of our narrative mining method, we apply it to a large corpus of floor speeches in U.S. Congress. The application is designed to illustrate how the method can be used for qualitative and quantitative analysis in social science and digital humanities. 

\subsection{Implementation}

\paragraph{Data.}

Our application uses the \textit{United States Congressional Record} for the years 1994-2015. The \textit{Record} is an exact transcript of speeches given on the floor of the House and the Senate, published in digital format since 1994 by the  U.S. Government Publishing Office (GPO). It has been widely used for text-as-data applications in the social sciences \citep[e.g.,][]{lauderdale2016measuring, ash2017elections, gentzkow2019measuring, gennaro2021emotion}. We link the Congressional speeches to the speakers' metadata (importantly, name and party affiliation). 

\paragraph{Choice of hyperparameters.} \label{sec:preprocessing}

We run our pipeline with the default settings on the entire Congressional Record. Appendix \ref{app:sec:implementation} provides some additional material on how the narrative mining system is adapted to the corpus. We limit our analysis to ``complete'' narratives containing an agent, a verb, and a patient. After some hyperparameter tuning, we select 1000 named entities and 1000 clustered entities (see Appendix \ref{app_human_validation} for details). Clustered entities are first labeled by the most frequent term, with some manual relabeling done after brief inspection.

As the distribution of narrative frequencies is heavily skewed to the left, and since we are interested in recurring narratives, we restrict our attention to narratives pronounced at least 50 times overall in the corpus (i.e., at least twice a year on average across all speakers).\footnote{This filtering comes at the end and does not influence the construction of our narrative model. Even legislators who speak infrequently show up in our data (see \citeauthor{proksch2012institutional}, \citeyear{proksch2012institutional}), unless the narratives pronounced by them are rare overall.}

\paragraph{Resulting narratives.}

Table \ref{tab:most_frequent} lists the ten most frequent narratives, after excluding sentences related to parliamentary procedure. For each narrative, we show two original sentences that the narrative represents. Appendix \ref{app:sec:most_freq_narr} shows more examples of sentences of these most frequent narratives in context. The narratives are easily interpretable and semantically close to the original raw text. Thus, the approach satisfies our objective of preserving the important information of \textit{who} does \textit{what} to \textit{whom}.

\begin{table}
\centering
\caption{Most Frequent Non-Procedural Narratives}\label{tab:most_frequent}
\small
\begin{tabularx}{\linewidth}{l|c|L}
\textbf{Narrative} & \textbf{Freq.} & \textbf{Spoken sentence(s)} \\
\hline \hline
people lose job & 606 & -- \textit{President, people are losing jobs every day.} \newline  -- \textit{People shouldn't have to lose their jobs to pay for the New York fund.}\\ \hline
american lose job & 448 & -- \textit{[...] Americans across this country have lost jobs, unemployment is at a high rate, people are having to make decisions.} \newline -- \textit{These funds will go a long way in supporting American workers who have lost their jobs [...].} \\ \hline
citizen abide law & 420 & \-- \textit{I felt kicked around and ignored by the very system the government has in place to protect law-abiding citizens.} \newline -- \textit{The vast majority of these private security officers are dedicated, hard-working, law-abiding citizens of this country [...].} \\ \hline
american have healthcare & 414 & -- \textit{[...] [W]e can contain costs and help enable every American to have access to health insurance coverage.} \newline -- \textit{[...] Americans already have universal health care because the emergency rooms cannot legally refuse to treat patients.} \\ \hline
government run healthcare & 359 & -- \textit{I don't understand what they are talking about: ``socialized medicine,'' ``Cuban-style, government-run health care.''} \newline -- \textit{Is that what you are talking about where you all of a sudden shift from people who figure out you can get the government to pay for everything, a government-run health care program?} \\ \hline
god bless american & 357 & -- \textit{And may God continue to bless America.} \newline -- \textit{God bless these heroes, their families and God bless America.} \\ \hline
worker lose job & 355 & -- \textit{Democrats want to help more workers who lose their jobs because of trade, especially workers providing services.} \newline -- \textit{If we are going to have a real trade package for this country, it has to benefit not just those who win from trade but those who lose from trade as well, including the workers who lose their job through no fault of their own.} \\ \hline
people need help & 338 & -- \textit{And that is what this welfare reform is all about -- to do something about people who are down on their luck and need help.} \newline -- \textit{We are supposed to represent the people who need help across this country.} \\ \hline
god bless troop & 329 & -- \textit{In conclusion, God bless our troops, and the President's actions should be based on remembering September the 11th in the global war on terrorism.} \newline -- \textit{In conclusion, may God bless our troops, and we will never forget September 11.} \\ \hline
small business create job & 304 & -- \textit{Small businesses create 80 percent of the jobs, so you would think a good piece of the relief would go to small business.} \newline -- \textit{Mr. President, small businesses represent more than 99 percent of all employers [...] and create about 75 percent of the new jobs in this country.} \\ \hline
\hline
\end{tabularx}
\end{table}

Appendix \ref{app:sumstats} provides summary tabulations on the statements, roles, entities, actions, and narratives extracted from the Congressional Record. The dimensionality reduction is substantial, with the 17.3 million plain-text sentences in the original corpus eventually reduced to 1,638 unique narratives. 
Besides clustering of entities (see Section \ref{subsec:entity-extraction}), this reduction is achieved by requiring ``complete'' narratives (with both an agent and a patient), by the filtering out infrequent narratives (see ``Choice of hyperparameters'' above), and by dropping narratives containing entities that are procedural or otherwise not related to politics or policy (Appendix \ref{app:procedural_and_noise} shows a complete list of these entities). 
The list of narratives, reported in full in Appendix \ref{sec:FinalNarratives} and readable in a few minutes, provides a concise summary of political discourse in the United States.

\subsection{Analysis} \label{sec:application:emp_analysis}

\begin{figure}
\caption{History, Sentiment, and Politics in Narrative Discourse}
\centering
\begin{subfigure}{\textwidth}
\caption{\small ``Terrorism and War'' Narratives}
\includegraphics[width=\textwidth]{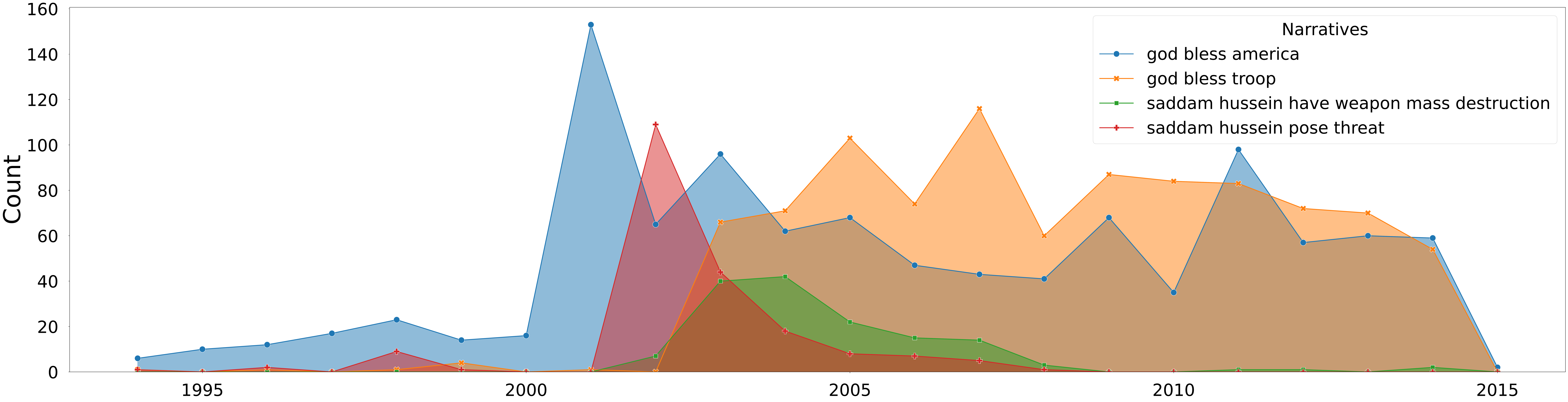}
\label{fig:TerrorNarratives}
\end{subfigure}
\begin{subfigure}{\textwidth}
\caption{\small Emotional Narratives}
\includegraphics[width=\textwidth]{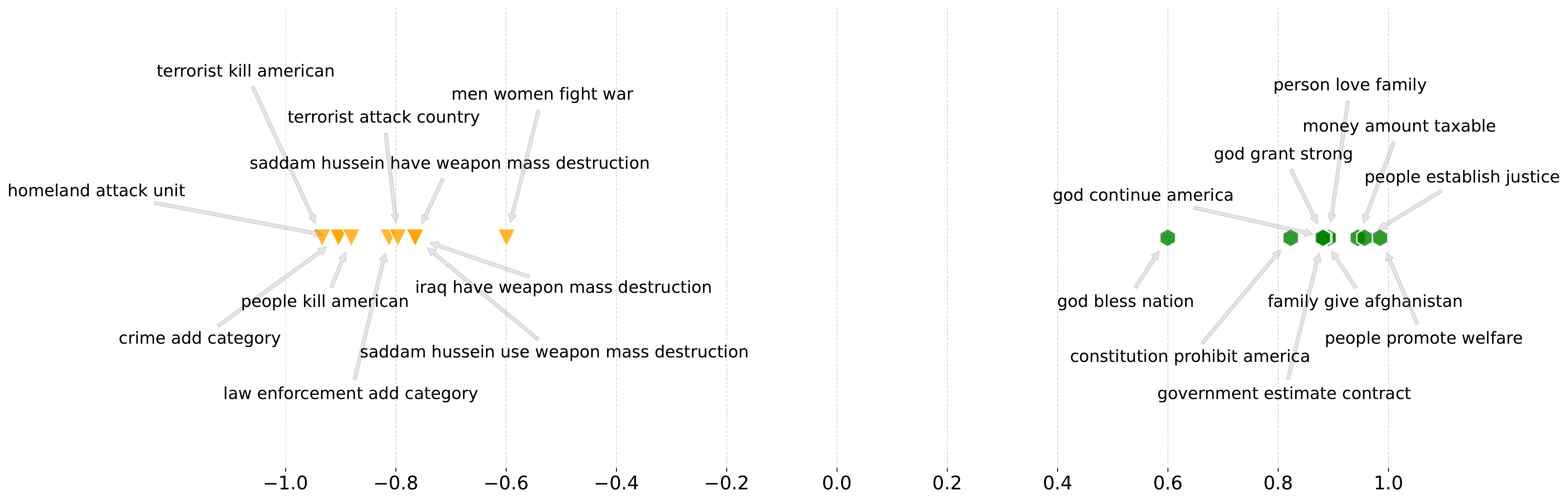}
\label{fig:SentimentNarratives}
\end{subfigure}
\begin{subfigure}{\textwidth}
\caption{\small Partisan Narratives}
\includegraphics[width=\textwidth]{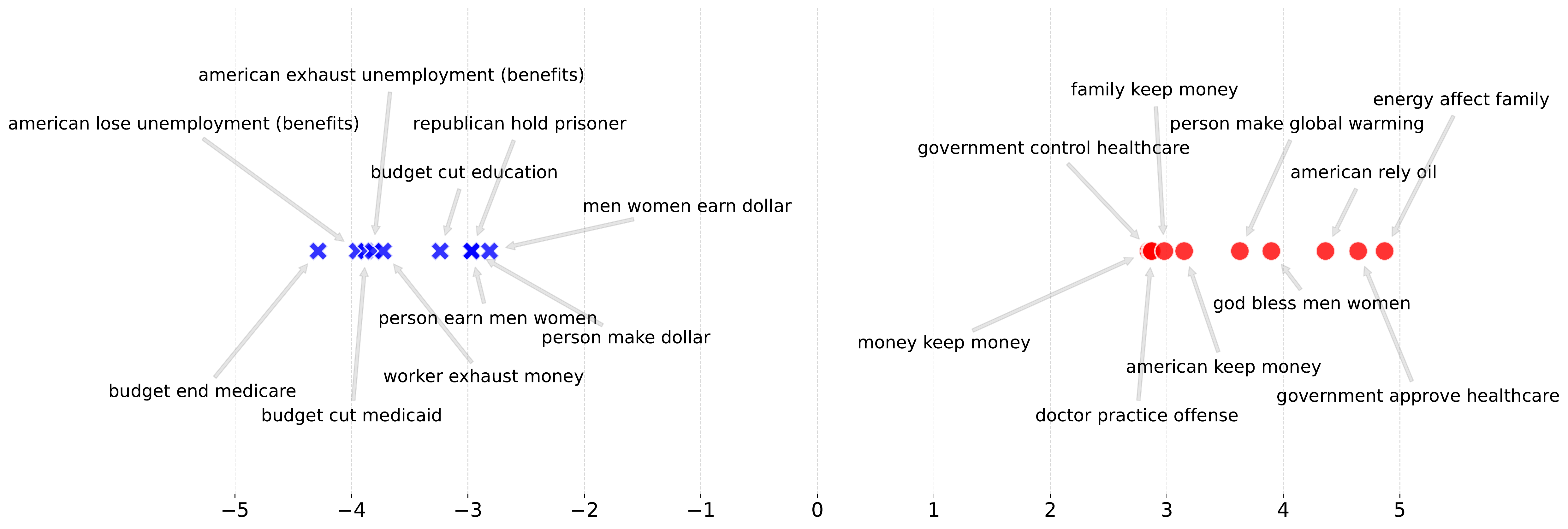}
\label{fig:PoliticalNarratives}
\end{subfigure}
\label{fig:AllNarratives}
\scriptsize
\flushleft
\textbf{Note:} This figure shows how narratives may help researchers make sense of speeches in the US Congress. To provide some historical perspective, Panel (a) presents time-series counts of a selection of prevalent narratives in the wake of the 9/11 attacks. The counted narratives are ``god bless america'' (blue), ``god bless troop'' (orange), ``saddam hussein have weapon mass destruction'' (green), and ``saddam hussein pose threat'' (red). Panels (b) and (c), respectively, plot the 20 most extreme narratives in the U.S. Congressional Record along the sentiment and partisanship dimensions. Panel (b) displays the most positive and negative narratives. The sentiment compound measure, computed using the NLTK VADER package, is averaged over all sentences in which the narrative appears. A high compound sentiment indicates positive sentiment (in green), whereas a low compound sentiment indicates negative sentiment (in orange). Panel (c) displays the most Republican and Democrat narratives. A high log-odds ratio reflects narratives pronounced more often by Republicans relative to Democrats (in red), and vice versa for a low log-odds ratio (in blue).
\end{figure}

Now, we use our mined narratives for a descriptive analysis of discourse in U.S. Congress. Starting with Figure \ref{fig:AllNarratives}, we first show that narratives capture salient historical events and how those events are framed. Second, we show that narratives contain emotional and partisan resonance. Third, we produce narrative graphs and analyze how political discourse relates entities to one another.

\paragraph{Narratives reflect key events in U.S. history.}

A first task for which our narrative features are useful is to describe changes in discourse over time. To illustrate how dynamic shifts in narratives reflect historical events, we explore the discourse around the September 11th attacks and the subsequent Global War on Terror.\footnote{Appendix \ref{app:additional-results} shows additional time-series of the most frequent narratives in the US Congress.}

Figure \ref{fig:TerrorNarratives} shows a time series for a selection of narrative statements pertaining to terrorism and war over the years 1994-2015. First, we see that the narrative on the threat posed by Saddam Hussein spikes in 2002, as the Bush Administration pushed its case for war and for congressional authorization. The supporting narrative of Hussein having or using weapons of mass destruction spikes at the same time \citep{kull2003misperceptions}, continuing over the subsequent years. Meanwhile, appeals to God blessing America surged in the wake of the September 11th attacks \citep{klocek2019war}. Further, starting in 2003 with the Iraq invasion, a second religious narrative of God blessing the troops gained hold and persisted for a decade. 

In Appendix \ref{app:narrative_cooccurrence}, we show that the statements in Figure \ref{fig:TerrorNarratives} are part of a broader political story on the American response to 9/11 and the intervention in Iraq. Narratives around Saddam Hussein and Iraq posing a threat are accompanied by a political case for intervention: on people needing help, the nation taking action, and the need to make sacrifices (Appendix Table \ref{tab:coocs}). The constellation of related narratives also includes those used by the anti-war opposition -- e.g., that the Bush administration was misleading Americans. 

\paragraph{Popular narratives have emotional resonance.} 

A commonly discussed feature of narratives is that they appeal to people's sentiments \citep{angeletos2013sentiments,eliaz2020model}. To analyze this dimension in the congressional speeches, we use a sentiment analyzer to score each narrative by the average sentiment of sentences where the narrative appears.\footnote{We work with Valence Aware Dictionary for Sentiment Reasoning, commonly known as VADER.} We then produce a ranking of narratives by sentiment, both in the positive and negative directions.

The highest-ranked narratives by positive and negative tone are presented in Figure \ref{fig:SentimentNarratives}. The most positive narratives include those related to the Constitution and Founding, of the benefits of healthcare, and of small businesses providing jobs. The negative set includes narratives about providing help in times of need and of the threats posed by terrorists.\footnote{``Crime add category'' shows up as negatively valent because it includes many sentences related to hate crime. Specifically, ``crime'' includes ``hate crime legislation'', while ``category'' includes ``new categories [to current hate crimes law]''. This narrative was reiterated many times by Senator Gordon Smith, in sentences such as ``\textit{Each congress, Senator Kennedy and I introduce hate crimes legislation that would add new categories to current hate crimes law, sending a signal that violence of any kind is unacceptable in our society.}''} Again, the results provide a qualitative window into the priorities and values held by U.S. Congressmen.

\paragraph{Partisan narratives map ideological disagreements.} 

Scholars have long studied how competing ideologies are reflected in speech, using alternatively n-grams \citep{ash2017elections,gentzkow2019measuring} or topic models \citep{QuinnMonroeColaresiCrespinRadev2010AJoPS}. These previous approaches identify those entities and concepts that, when mentioned, tend to signal party affiliation. Our narrative features provide another angle by identifying the \textit{connections} between entities and concepts that, when expressed, signal partisanship. 

To explore this dimension, we produce an odds ratio for each narrative as its relative usage by Republicans or Democrats \citep{monroe2008fightin}. Figure \ref{fig:PoliticalNarratives} displays the narratives that are most partisan, with a negative coefficient indicating Democrat-slanted and a positive coefficient indicating Republican-slanted. For example, Democrats are concerned about budget cuts for public programs and Americans losing unemployment benefits, while Republicans care about government interference with healthcare and Christian values.\footnote{The Republican narrative ``government approve healthcare'' reflects sentences such as ``\textit{The Virginia court held that the individual mandate requiring every American to purchase government-approved health insurance was unconstitutional.}'' or `\textit{`This bill takes away that freedom, requiring every American to purchase a government-approved health plan, pay a tax, or even go to jail.}''} These narratives highlight the mirroring of partisan policy priorities during this recent time period.

To see better how the expressed \textit{connections} between entities signal partisanship, we compute a narrative divisiveness score for each entity, as the average (absolute value) log-odds ratio of the narratives where the entity appears, minus the log-odds ratio of the entity itself. The highest-ranked entities on this score are those agents and patients for which the political parties most differ in their articulated connections to other entities, after adjusting for the partisanship of the entity itself.

\begin{table}[]
\caption{Entities Associated with the Most Divisive Narratives} \label{tab:divisive-entities-main}
 \begin{center}
 \renewcommand{\arraystretch}{1.1}
\resizebox{0.75\linewidth}{!}{%
\begin{tabular}{lll}
\hline \\[-1em]
\textbf{Entity} & \textbf{Democrat Narratives} & \textbf{Republican Narratives} \tabularnewline [1ex]
\hline \\[-1em]
\multicolumn{3}{l}{\textbf{A. Policy-Related Entities}} \tabularnewline [1ex]
\hline
 \multirow{2}{*}{oil} & oil pay fee & oil create job\tabularnewline
  & oil make profit & american rely oil \tabularnewline
\cline{2-3} \cline{3-3} 
 \multirow{2}{*}{budget} & budget end medicare & budget raise tax\tabularnewline
  & budget cut medicaid & budget increase tax \tabularnewline
\cline{2-3} \cline{3-3} 
 \multirow{2}{*}{healthcare} & child lose healthcare & government control healthcare\tabularnewline
  & family afford healthcare & government approve healthcare\tabularnewline
\cline{2-3} \cline{3-3} 
 \multirow{2}{*}{interest/rate} & reserve raise interest/rate & individual bearing interest/rate\tabularnewline
  & fed raise interest/rate & capital gain interest/rate\tabularnewline
\cline{2-3} \cline{3-3} 
 \multirow{2}{*}{worker} & worker exhaust money & employee hire worker\tabularnewline
  & worker have exhaust & worker support retiree\tabularnewline
\cline{2-3} \cline{3-3} 
 \multirow{2}{*}{loan} & small business receive loan & people get loan\tabularnewline
  & student take loan & company hold loan\tabularnewline
\cline{2-3} \cline{3-3} 
 \multirow{2}{*}{insurance} & insurance deny coverage & american buy insurance\tabularnewline
  & insurance drop coverage & government run insurance\tabularnewline
\cline{2-3} \cline{3-3} 
 \multirow{2}{*}{medicare} & budget end medicare & medicare run money\tabularnewline
  & republican end medicare & doctor see medicare\tabularnewline
\cline{2-3} \cline{3-3} 
 \multirow{2}{*}{energy} & energy create job & american produce energy\tabularnewline
  & nation need energy & energy affect family\tabularnewline
\cline{2-3} \cline{3-3} 
 \multirow{2}{*}{job} & company move job & american not-do job\tabularnewline
  & company ship job & tax kill job\tabularnewline
\hline \\[-1em]
\multicolumn{3}{l}{\textbf{B. Identity and Symbolic Entities}} \tabularnewline [1ex]
\hline 
 men/women & person earn men/women & men/women defend nation\tabularnewline
  & men/women earn dollar & god bless men women\tabularnewline
\cline{2-3} \cline{3-3} 
 democrat & republican join democrat & democrat do nothing\tabularnewline
  & democrat balance budget & democrat raise tax\tabularnewline
\cline{2-3} \cline{3-3} 
 constitution & constitution prohibit america & constitution give authority\tabularnewline
  & men women write constitution & constitution give power\tabularnewline
\cline{2-3} \cline{3-3} 
 american & american lose unemployment (benefits) & american keep money\tabularnewline
  & american exhaust unemployment (benefits) & american rely oil\tabularnewline
\cline{2-3} \cline{3-3} 
 community & community need help & community perform abortion\tabularnewline
  & program provide community & community promote abortion\tabularnewline
\cline{2-3} \cline{3-3} 
 republican & republican hold prisoner & republican gain power\tabularnewline
  & republican refuse action & democrat join republican\tabularnewline
\cline{2-3} \cline{3-3} 
 nation & nation sign treaty & men women defend nation\tabularnewline
  & nation make progress & nation face fight\tabularnewline
\cline{2-3} \cline{3-3} 
 people & people lose unemployment (benefits) & people call tax\tabularnewline
  & people face mortgage & people keep money\tabularnewline
\cline{2-3} \cline{3-3} 
 family & family afford healthcare & family keep money\tabularnewline
  & family lose healthcare & energy affect family\tabularnewline
\cline{2-3} \cline{3-3}  
 person & person earn men/women & person love family\tabularnewline
  & person make dollar & person make global warming\tabularnewline
\hline 
\end{tabular}
}
    \end{center}
\scriptsize
\flushleft
\textbf{Note:} This table shows the set of entities that appear in narratives with a high average log-odds ratio by partisan mentions, adjusted for the log-odds ratio of the entity itself. Panel A includes policy-related entities, while Panel B includes entities are related to groups or ideas. Democrat Narratives and Republican Narratives show the associated narratives with the lowest and highest log odds ratios, respectively.
\end{table}

The entities with the highest narrative divisiveness are listed in Table \ref{tab:divisive-entities-main}.\footnote{We filtered out some infrequent entities that ranked highly in the list due to noise. To be included in this list, the entity must appear in at least six unique narratives, and have at least three Democrat-slanted narratives and three Republican-slanted narratives. Appendix Table \ref{tab:more-divisive-entities} shows a longer, unfiltered list of divisive entities, with the associated scores. For comparison, a list of the least divisive entities are shown in Appendix Table \ref{tab:least-divisive-entities}.} Panel A shows the top-ten policy-related entities, while Panel B shows the top-ten entities related to identity groups or symbols. The second and third columns, respectively, show the most Democrat-slanted and most Republican-slanted narratives associated with the entity, again as measured by the log-odds ratio.

The list of divisive entities illustrate how the same entities can be used in very different narratives by different parties. On the policy side (Panel A), Democrats lament the profits earned by \textit{oil} companies, while Republicans retort that Americans rely on \textit{oil}. Democrats worry about \textit{budget} cuts to Medicare and Medicaid, while Republicans are concerned that the \textit{budget} will increase taxes. Meanwhile, Democrats complain that companies are shipping \textit{jobs} overseas, while Republicans assert that taxes kill \textit{jobs}. On the identity side (Panel B), Democrats attend to \textit{men and women's} earnings, while Republicans celebrate their defense of the nation and invoke God's blessings.  For \textit{Americans}: While Democrats emphasize that ``Americans lose unemployment benefits'', Republicans stress that ``Americans rely on oil''.  Overall, the narratives associated with divisive entities illuminate the key divergences in political priorities between Democrats and Republicans. Identifying such partisan connections would be infeasible with standard text-as-data approaches, such as n-grams. 

A lingering question is whether mentions of narratives -- that is, connections between entities -- are overall more polarized than mentions of singular entities. We test this possibility formally following  the approach in \cite{peterson2018classification}. Specifically, we train machine classifiers to predict a speaker's partisanship (Republican or Democrat) in held-out data, using as features either the entities or the narratives pronounced in speeches.\footnote{Our machine learning model is L2-penalized logistic regression. We select the regularization strength using five-fold cross-validated grid search in a 75\% training set and evaluate performance in a 25\% test set.} The narrative features predict partisanship more accurately: we obtain an out-of-sample accuracy of 81\% with narratives, but only 74\% for entities.\footnote{To obtain the confidence intervals for these accuracy measures, we employ a five-fold cross-validation in the test set. Averaged across folds, the accuracy is 79\% and 73\% for narratives and entities, respectively. The corresponding 95\%-confidence intervals are [74\%, 83\%] and [67\%, 78\%]. The test set contains only 244 speakers; this small sample size across folds likely contributes to the wide confidence intervals.} The higher test-set accuracy for narratives suggest that they are more informative about partisanship  than mentions of topics or specific entities. The \textit{connections} between entities framed by narratives help politicians tell stories in line with their partisan values.

\paragraph{Narratives reveal the connected structure of political debates.} 

So far, we have analyzed narratives in isolation, as two entities connected by one verb. Yet, a defining feature of narratives is that simple statements link up to form more complex stories.\footnote{For example, conspiracy theory narratives have shown to become extremely complex in their proposed connections between disparate entities \citep[e.g.,][]{wood2015online,tangherlini2020automated}.} As previewed in Section \ref{sec:narrative-structure}, we map the narrative space of U.S. congressional speeches as a directed multigraph, with entities composing nodes and verbs composing link between nodes.

The resulting graph of linked entities can be used for a variety of network-based analyses. In particular, centrality measures can be used to determine which entities are pivotal to political narrative structure \citep[e.g.][]{hanneman2005introduction}. Entities with high closeness centrality, for example, are the most connected to the broader narrative network.\footnote{In our corpus, the top-five entities by closeness centrality are job, family, healthcare, funding, program.} Out-degree centrality captures influential agents who tend to act on many other entities,\footnote{The top-five entities by out-degree centrality are people, Americans, program, government, and person.} while in-degree centrality captures the most receptive patients -- for example, policies to be enacted and the beneficiaries of those policies.\footnote{The top-five entities by in-degree centrality are job, healthcare, money, family, and services.} Such analyses can be further enriched with metadata, notably political party of the speakers.

\begin{figure}
\centering
\caption{Top 100 Most Frequent Narratives in the U.S. Congress}
\includegraphics[width=\linewidth]{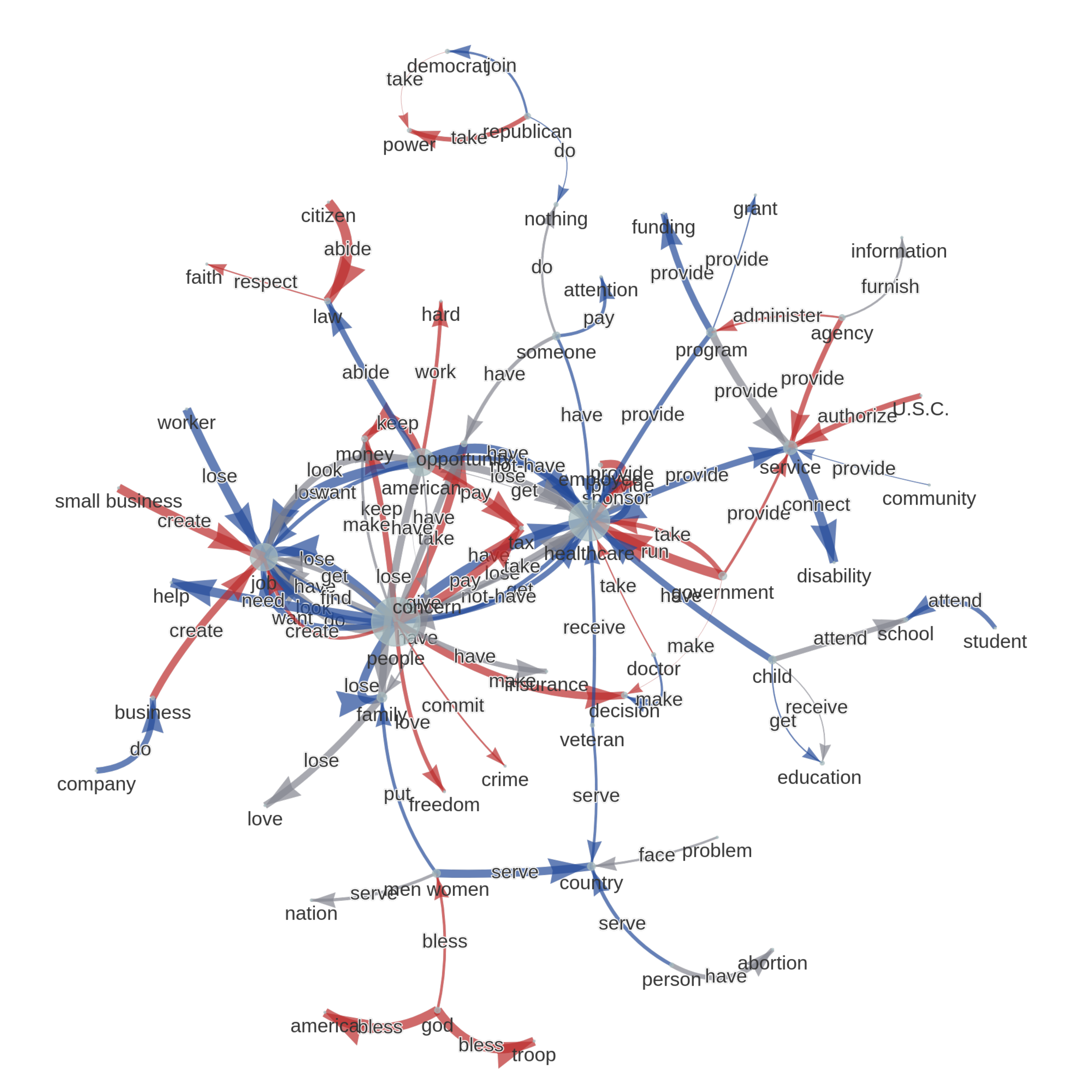}
\label{fig:StructureNarratives}
\scriptsize
\flushleft
\textbf{Note:} This figure displays the 100 most frequent narratives in the U.S. Congressional Record. We represent our narrative tuples in a directed multigraph, in which the nodes are entities and the edges are verbs. Node and edge sizes are respectively scaled by node degree and narrative frequency. The resulting figure is obtained via the Barnes Hut force-directed layout algorithm. The direction of edges reflects the direction of the actions undertaken. The color of edges indicates partisan narratives -- statistically significant log-odds ratios (95\% level) are colored in red for Republicans and in blue for Democrats, with nonpartisan narratives in gray.
\end{figure}

The narrative graphs are useful to support qualitative descriptive analysis of political corpora. Using an interactive web browser applet, even the full set of 1,638 congressional narratives -- 353 entities interwoven by 140 verbs -- can be explored efficiently and informatively. Since the entire network is too large to depict as a static figure, we visualize the 100 most frequent narratives here. Figure \ref{fig:StructureNarratives} shows this subset of the narrative graph, with thicker arrows indicating higher frequency and colors indicating partisanship -- blue for Democrat, and red for Republican. The network summarizes in a single figure many of the stylized facts we have previously discussed. For example, while Democrats lament the problem of ``people losing jobs'', Republicans applaud the success of ``small businesses creating jobs''. While Democrats want that ``Americans have healthcare'', Republicans worry about ``government-run healthcare''. Overall, the graph provides a distilled view on the \textit{worldviews} expressed in the U.S. Congress. 

Figure \ref{fig:StructureNarratives} is just one example of many potentially informative visualizations of the narrative networks. Appendix \ref{app:sec:more_on_partisan} provides party-specific network graphs (see Figures \ref{fig:PartisanWorldviews_Dem} and \ref{fig:PartisanWorldviews_Rep}). Moreover, our GitHub repository provides interactive versions of such network graphs.\footnote{See \url{https://github.com/relatio-nlp/relatio} and the links therein.}

\section{Discussion} \label{sec:discussion}


To recap, we have designed and implemented a new text-as-data method that provides an intuitive mapping from a corpus of plain-text sentences to a sequence of low-dimensional narrative statements. In this section, we discuss how narrative mining fits in with other text-as-data methods, identify some of its limitations, and propose potential extensions.

\subsection{When should researchers use \textsc{relatio}?}

In the application sections above, we focus on results that would be difficult to produce with other text-as-data methods. Mainly, this means looking closely at the narratives themselves. In Figure \ref{fig:TerrorNarratives}, the depicted narratives capture the dynamics of War on Terror discourse better than plots of specific phrases like ``Saddam Hussein'' or ``God'' by themselves. Further, the most emotive and partisan narratives from Table \ref{tab:divisive-entities-main} and Figure \ref{fig:AllNarratives} are more informative than standard representations (like bag-of-words) because they map explicit relationships, requiring less contextual knowledge to interpret them. Finally, the narrative graphs exemplified by Figure \ref{fig:StructureNarratives} have no analogue in other commonly used text methods. 

The promise of this mode of quantitative description using narratives is not limited to our particular setting. \textsc{relatio} can be used to track the evolution of language in a range of corpora, such as social media and newspapers. In particular, the narrative graphs offer a novel opportunity for data-driven study of worldviews. Appendix \ref{app:trump_tweet_archive} provides an additional application of our method to President Donald Trump's tweets from 2011 through 2020. Furthermore, since we open-sourced our Python package in August 2021, two working papers have come out using the code for descriptive analysis in different settings, including \cite{sipka2021comparing} on social media posts about Q-Anon conspiracy theories and \cite{ottonello2022financial} on newspaper coverage of banks.

To take a broader view, we now discuss how narrative mining complements existing text-as-data methods. First, the narrative features output by \textsc{relatio} can be examined and analyzed the same way that specific word or phrase patterns are used in common dictionary methods. For example, \citet{BakerBloomDavis2016QJoE} count articles mentioning both an economy-related word and an uncertainty-related word to measure ``economic policy uncertainty.'' A \textsc{relatio}-based alternative could count narratives containing economy-related entities linked with uncertainty-related attributes or entities. The advantage of the narrative approach over dictionaries is that narratives can more easily specify semantically subtle links, such as distinguishing uncertainty caused by the economy, versus uncertainty caused by policy.

Second, the outputs of \textsc{relatio} are similar to approaches using syntactic dependency parsing to identify agents, actions, and objects. In principle, semantic role labeling is designed to out-perform syntax-based methods for this purpose; so, for example, \textsc{relatio} could be used to produce measures of relative worker-firm authority in labor union contracts, as an alternative to the dependency-based approach from \cite{ash2020unsupervised}. A caveat is that SRL is a more complex and error-prone linguistics task than dependency parsing. The costs of those parsing errors might outweigh the benefits from using SRL, but the methods have not been systematically compared.

Third, topic models such as LDA \citep{BleiNgJordan2003JoMLR} are, like \textsc{relatio}, designed to perform informative dimension reduction of documents. Narrative statements could provide an alternative to topics for research questions concerning the prevalence of specific concepts or issues. For example, \cite{catalinac2016pork} applies a topic model to parliamentary speeches, manually inspects and labels topics as being related to local pork or national policy, and then examines changes in the pork/policy categories in response to an institutional reform. An alternative approach using narrative mining would be to extract narratives (rather than topics) from those speeches and then label the narratives as being related to pork or policy. Whether topics or narratives are preferred will depend on the use case. Broadly speaking, topic models will identify the prevalence of relatively generic topics in documents, while \textsc{relatio} will detect the presence of distinctive arguments or claims. 

Besides being useful by themselves, the narrative statements produced by \textsc{relatio} can be used as an informative and interpretable feature set in downstream text-as-data applications. For example, document-level counts or frequencies over narrative statements provide an alternative to N-gram frequencies, topic shares, or document embeddings as inputs to text classifiers or text regressions. Similarly, document distance can be computed using cosine similarity between narrative frequency vectors the same way it can be done between N-gram, topic, or embedding vectors. Finally, topic models like LDA can be applied on top of the narrative representation of documents, allowing narrative statements to be topically bundled for further descriptive or empirical analysis.\footnote{Along those lines, Appendix \ref{app:AlternativeFeatures} presents a topic model on the Congress corpus and compares the words associated with each topic to the narratives associated with each topic. The top narratives tend to provide additional information about topic content which top words fail to capture. Consider the topic we labeled as ``economy''. The top (stemmed) words for this topic are ``busi'', ``small'', ``compani'', ``capit'', ``invest'', ``contract'', ``administr'', and ``job''. The associated narratives describe explicit relationships such as ``small business employ workforce'', ``individual receive funding'', or ``founder start business''.} 

A systematic comparison of \textsc{relatio} features to these existing alternatives for such tasks is a high priority for future work. Our analysis above showing that narrative features are more predictive of partisanship than entity features provides a promising indication in this direction. Holding the quality of the algorithmic outputs constant, narratives might be preferred to other feature sets due to their interpretability and tractability. \textsc{relatio} extracts distinctive entity-action tuples which summarize the core claims made in a corpus. This type of interpretable dimension reduction serves the same goals as other text-as-data methods, such as TF-IDF weighting \citep{GrimmerStewart2013}, supervised feature selection \citep{monroe2008fightin}, or filtering on parts of speech to extract noun phrases \citep{handler2016bag}. For either supervised or unsupervised learning algorithms, the dimension-reduced inputs will be computationally tractable, while the learned outputs will be informative and interpretable.


\subsection{Limitations and potential extensions}

While our narrative mining system shows some promising preliminary results, there are still a number of limitations and opportunities for improvement.

\paragraph{SRL quality.}

The quality of narrative outputs depends on the quality of the semantic role labeling (SRL) tags. When SRL fails, it produces nonsense data that will mostly be dropped in the pipeline's filtering steps. The quality of SRL tags correspondingly depends on the quality of text inputs. Hence, messy text with digitization errors, for example due to optical character recognition (OCR), may not produce usable results. A further source of SRL errors is grammatical complexity; SRL performs best on simpler language, such as that in social media posts. The \textit{Congressional Record} -- where long, potentially rambling spoken sentences are the norm -- is actually a difficult test case for the method. 

From the perspective of empirical research design, a relationship between speech complexity and SRL quality is concerning because the performance of our method may differ across subgroups. In our setting, for example, we have evidence that Democrats use somewhat more sophisticated language than Republicans.\footnote{The number of words per narrative is higher for Democrats, meaning that a narrative entails a marginally higher abstraction. The difference is not statistically significant, however.} If SRL errors are higher for these more sophisticated sentences, then they will tend to be selectively dropped from the sample, potentially biasing downstream results. 

Our hope is that these issues will diminish as more robust automated SRL models are introduced. For more difficult corpora, such as historical speeches digitized by OCR, it may be fruitful to adapt the approach to work with syntactic dependencies rather than semantic roles.

\paragraph{Entity extraction.}

A persistent practical difficulty is in the extraction of clustered entities. With more clusters, the entities become more coherent and specific, at the expense of higher dimensionality. With fewer clusters, semantically related yet contradictory phrases -- e.g., ``tax hike'' and ``tax cut'' -- will often be combined into the same entity. The standard cluster quality metrics, such as silhouette scores, provide limited guidance.

As with SRL, this is partly due to limitations with the associated algorithms. It could be that improved or domain-specific named entity recognition, alternative phrase encoding approaches \citep[e.g.,][]{reimers2019sentence}, or alternative clustering approaches \citep[e.g.,][]{stammbach2021docscan}, could mitigate these problems. The software package will be continuously updated following such developments.

Even with improved algorithms, however, entity extraction may produce imperfect outputs. For example, say one wants to resolve ``american'' and ``people'' to one entity -- e.g., for ``american lose job'' and ``people lose job''. This is currently not straightforward: by default, ``american'' is a named entity while ``people'' comes from a clustered embedding. A further wrinkle is that the optimal clustering may be context-dependent: for instance, clustering ``people'' and ``worker'' together might make sense for the ``[...] lose job'' narrative, but it may erase useful information in others -- e.g., ``government support worker'' has a different ideological valence from ``government support people''. Finally, our entity extraction algorithm cannot directly identify implicit references to entities, such as using the phrase ``White House'' to refer to the president. In sum, obtaining the entities most relevant to an application with \textsc{relatio} is art as much as science.

\paragraph{Further extensions and limitations.}

There are a number of additional extensions to consider. In the SRL step, additional semantic roles could be included, such as temporality or adverbial clause modifiers. The named entity recognizer could be improved to resolve co-references to the same entity. Further experimentation could be done with dimension-reducing verbs, for example by combining an embedding-based and dictionary-based approach (to prevent clustering of antonyms).

Another notable limitation is that the current version only works in the English language. In principle, however, the approach should be applicable in any language where a pre-trained SRL model is available. For small corpora, a pre-trained phrase encoder might also be required. Given the growing plethora of NLP tools across languages, we suspect that these requirements will not pose a significant impediment to diverse applications.

\newpage

\clearpage
\bibliographystyle{style/aea}
\bibliography{bib}

\clearpage

\newpage

\appendix 
\counterwithin{figure}{section}

\onecolumn
\setcounter{page}{1}
\setcounter{table}{0}
\setcounter{figure}{0}
\begin{center}
\vspace{3cm}
\vspace{3cm}
\end{center}
\begin{center}
\Large{Text Semantics Capture Political and Economic Narratives \\ \textbf{Online Appendix} }
\end{center}

\renewcommand{\thetable}{\thesection.\arabic{table}}
\renewcommand{\thefigure}{\thesection.\arabic{figure}}
\renewcommand{\thesection}{\Alph{section}}

\section{Python Implementation with \textsc{relatio}} \label{app:sec:implementation}
\setcounter{table}{0}
We detail the programmatic implementation of the procedure described in this paper, which is freely distributed as the Python package \textsc{relatio}. Our pipeline takes a collection of unstructured text documents as input and returns narrative statements (see Equation \ref{eq:narrative_space}). Figure \ref{fig:flowchart} presents a graphical outline of the different steps. 

First, we split the raw text into sentences using \texttt{spaCy} \citep{spacy}. Second, we annotate the semantic roles (SRL). For this task, we rely on a state-of-the-art pre-trained model implemented by \cite{Gardner2017AllenNLP}.\footnote{See \url{https://demo.allennlp.org/semantic-role-labeling}} SRL decomposes every sentence into one or multiple role sequences as defined in Equation \ref{eq:srl_role_sequence}. The current implementation covers agents (\texttt{ARG0}), verbs (\texttt{V}), patients (\texttt{ARG1}), attributes (\texttt{ARG2}), modals (\texttt{ARGM-MOD}) and negations (\texttt{ARGM-NEG}). An example of semantic role labeling annotations is provided in Figure \ref{fig:SRLExample}.

There are hundreds of thousands of unique semantic roles. We seek to regroup them into a smaller set of interpretable groups. An easy step in this direction is to ``clean'' all roles. We thus lowercase and lemmatize each word. We also drop punctuation, numbers and words from a list of pre-defined stopwords (for the complete list, see \ref{sec:stopwords}). After these cleaning steps, we drop roles of more than four tokens as they are often uninterpretable.

We do not process the verbs (\texttt{V}) any further, but connect them to their negation (if any). For agents (\texttt{ARG0}), patients (\texttt{ARG1}) and attributes (\texttt{ARG2}), we rely on two methods: Named Entity Recognition (NER) and phrase embedding clustering. We first run \texttt{spaCy}'s Named Entity Recognizer (NER) to identify all named entities mentioned in the corpus. After careful inspection of this list, we keep the 1000 most frequent entities and assign to them their own cluster. Beyond this threshold, named entities are pronounced less than 40 times and are often idiosyncratic. If a role contains one of these named entities, then it is designated as that entity and not processed any further.

\begin{figure}
    \centering
    \caption{Clustered Embeddings Recover Coherent Entities}
    \includegraphics[width = 0.8\textwidth]{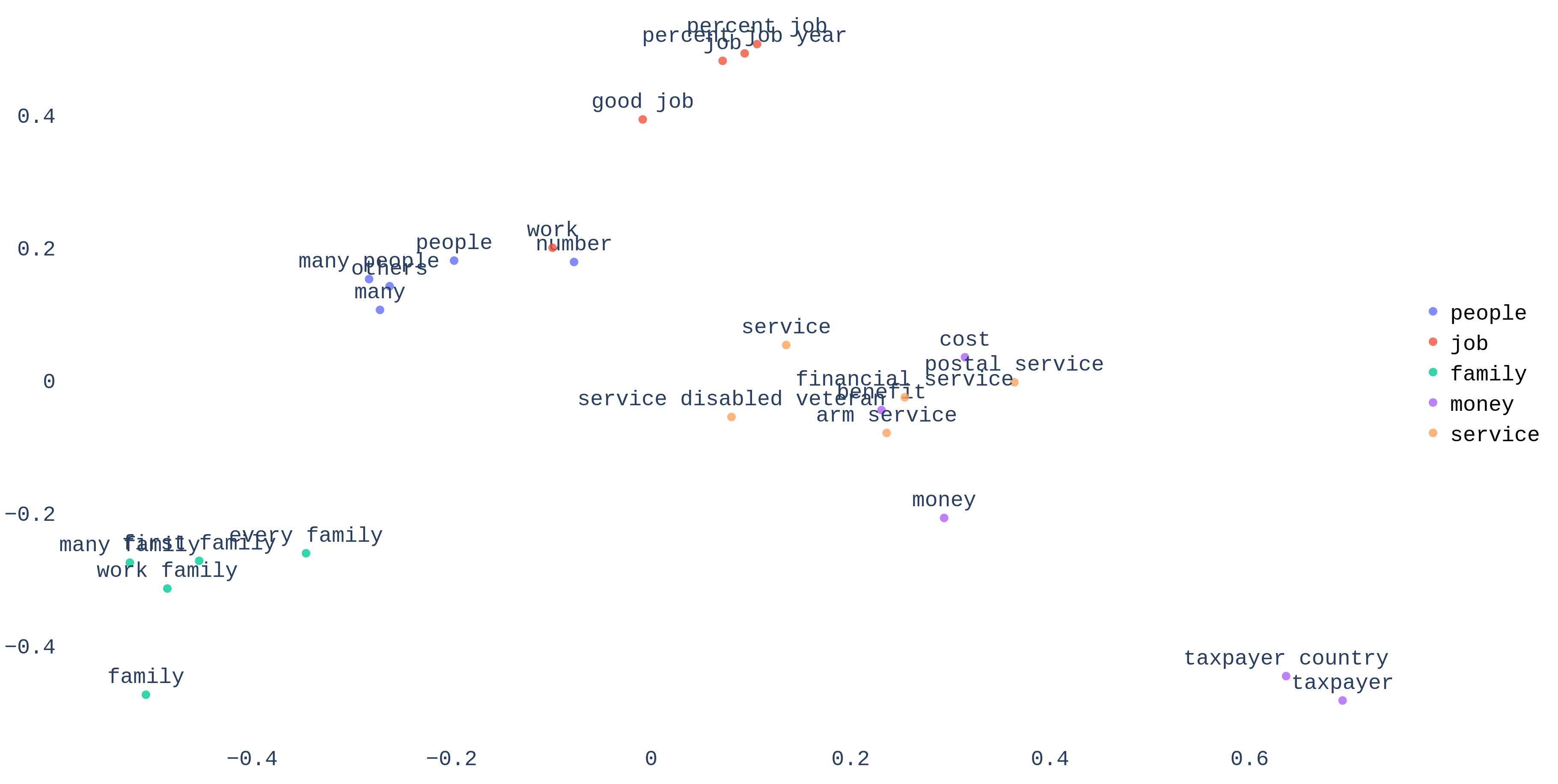}
    \label{fig:clustering}
    \footnotesize
    \flushleft
    \textbf{Note:} This figure illustrates the embedding-cluster process. We take the five most frequent entity clusters (produced using $K$-means applied to the phrase embeddings): people, job, family, money, and service. The embedding locations of the most frequent phrases in each cluster are illustrated here is a two-dimensional projection. The embedding space both co-locates and clusters similar phrases, providing dimension reduction while preserving semantically coherent entities. Embeddings are projected to two dimensions using PCA. 
\end{figure}

For all of the remaining roles that do not contain a frequent named entity, we produce text-embedding clusters as follows. First, we apply a phrase encoder to compress the plain text snippet to a low-dimensional dense vector. There are many encoders fit for this purpose (e.g. Sentence-BERT \citep{reimers2019sentence} and its variants). As a computationally efficient default, we use \citeauthor{AroraLiangMa2016}'s ``simple but tough-to-beat'' sentence embeddings, which consist of the weighted average of the GloVe word vectors \citep{PenningtonSocherManning2014} across each word in the role. The weights are chosen to be inversely proportional to the frequency of the word in the corpus. This means that relatively rare, more distinctive words are up-weighted in the phrase embedding. We use \texttt{gensim}'s pre-trained GloVe embeddings wih 300 dimensions. The result of this procedure is a vector in the embedding space for each agent, patient and attribute, where linguistically similar phrases (e.g. ``higher taxes'' or ``high tax rates'') have geometrically proximate vector representations. A graphical depiction is presented in Figure \ref{fig:clustering}.

Exploiting this geometry, we further reduce the dimensionality of our role sets by unsupervised clustering applied to the associated matrix of encoded vectors. As with the phrase encoding step, there are a number of possible approaches for clustering (e.g., DBSCAN, agglomerative clustering). We obtain effective performance with the conventional K-means clustering algorithm, which is also computationally lightweight. For K-means, the only hyperparameter is the number of clusters. Each remaining agent, patient and attribute in the corpus is assigned to a cluster. For interpretability, each cluster is labeled by the most frequent term within the cluster. Given the size of the corpus, we randomly select 50,000 speeches (roughly 5\% of the entire corpus) to train the K-Means model, and predict out-of-sample for the remaining speeches. We cluster agents, patients and attributes together. In turn, the same entity can appear as an agent, a patient or an attribute. This joint clustering ensures that roles can be connected easily in a directed multigraph.

Finally, we focus on all narratives that have either an Agent-Verb-Patient or an Agent-Verb-Attribute structure. As the distribution of narrative frequencies is heavily skewed to the left, and since we are interested in recurring narratives, we restrict our attention to narratives pronounced at least 50 times in the corpus (i.e., at least twice a year on average).

\clearpage

\section{Human Validation} \label{app_human_validation}
\setcounter{table}{0}

\subsection*{Tuning the number of clusters}

The number of mined entities is the core hyperparameter in our analysis. It depends on the number of named entities recognized, as well as the number of clusters we specify for K-Means. To pick the correct number of named entities in our analysis, we manually inspect the most frequent named entities in the corpus. We fix a threshold at 1000 named entities, above which the named entities are pronounced fewer than 40 times and tend to become idiosyncratic. We keep this threshold fixed throughout the rest of the analysis.

Contrary to the number of named entities, the number of clusters to specify for K-Means is not trivial for at least two reasons. First, it is unclear how many unnamed entities are being referred to in the corpus. Second, there is a trade-off between the informativeness of clusters and the magnitude of the dimension reduction. If a small number of clusters is specified, then clusters are likely to become abstract and to encompass a broad variety of phrases. If a large number of clusters is specified, then clusters will be more coherent (and thus potentially more interpretable), but they can also become redundant, with multiple clusters referring to the same underlying entity. This trade-off suggests that we would like to maximize cluster coherence while minimizing cluster redundancy.

We investigate the optimal number of clusters with a simple human validation task. The task is performed by freelancers on Upwork who were based in the U.S. and whose primary language is English. First, we run four clustering scenarios and specify respectively 100, 500, 1000, and 2000 clusters. We then provide annotators with a random sample of semantic roles and their associated cluster labels (from different cluster numbers). Additionally, we add placebo observations, where we show again a random sample of semantic roles, but instead of the predicted cluster label, we combine every role with a random cluster label (drawn from the entire pool of available labels). Annotators are then asked to rank the similarity of the semantic role and the cluster label on a scale of 1 (not similar at all) to 10 (extremely similar/the same). Our random validation sample consists of 4000 unique role and label combinations (500 true and 500 placebo combinations for every clustering scenario). Every unique combination is, on average, annotated by 2.5 freelancers, resulting a total sample of 10000 annotated combinations. Annotators do no see the number of clusters that were used to generate the cluster label, nor whether the semantic role is combined with a predicted cluster label or a placebo label. This procedure jointly evaluates cluster coherence and redundancy (coherence is reflected in how similar the semantic roles are to their predicted label; redundancy is captured by similarities between random combinations of roles and clusters).

\subsection*{Annotator recruitment}

In June 2021, we hired three different freelancers on Upwork to annotate 4000 combinations (applies to two freelancers) or 2000 combinations (applies to one freelancer) of roles and labels.
The instructions as communicated in the job post are shown in Figure \ref{fig:job_post}.

\begin{figure}[ht!]
\begin{center}
\caption{Wording of the validation job post on Upwork} \label{fig:job_post}

``\textit{We need freelancers (U.S.-based, fluent in English) to check the similarity between two phrases. In the attached file, please compare the phrase in column 1 and 2.
The question is: for every row, how similar are the entities described in column 1 and 2?
Please indicate from 0 (not similar at all) to 10 (very similar/the same). Note that you can select values in between 0 and 10 (e.g., 5).
This task is about evaluating a computer-based language model from a human perspective. It is about your ``hunch'' or intuition of how similar the phrases are. Do they refer to a similar entity?}''
\end{center}
\end{figure}

\begin{figure}[ht!]
\caption{Screenshot of Human Annotation File} \label{fig:annotation_file}
\begin{center}
\begin{tabular}{cc}
\includegraphics[width =0.6\textwidth]{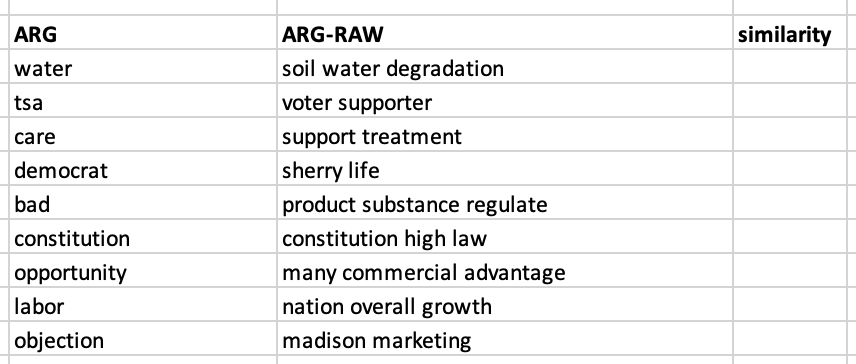}
\end{tabular}

\end{center}
\end{figure}

Upon posting the job, we sent a sample of 20 combinations to the first five freelancers who expressed their interest in the job. Out of these five, we hired the first three freelancers who submitted the annotated sample (all of these first three had annotated the sample correctly). In Figure \ref{fig:annotation_file}, we feature the form presented to the freelancers for annotation.

\subsection*{Human validation Results}

Table \ref{tab:sumstats_human_validation} summarizes the key figures of the human validation across numbers of clusters. The average similarity of clustered roles is 4.49; for the placebos, this figure amounts to 2.57.

\begin{table}[ht!]
\centering
\caption{Summary Statistics on the Human Annotations} \label{tab:sumstats_human_validation}
\begin{tabular}{l|cr}
\toprule
Total combinations annotated & 10000 & \\
Unique combinations annotated & 4000 & \\
Number of annotators & 3 & \\
Average number of annotators per combination & 2.5 & \\
Average number of annotations per annotator & 3334 \\
Average similarity (true combinations) & 4.49 \\
Average similarity (placebo combinations)  & 2.57 \\
Average inter-annotator similarity standard deviation & 2.42 \\ 
\bottomrule
\end{tabular}
\bigskip 

\end{table}

\begin{figure}[ht!]
\begin{center}
\caption{Similarity of semantic roles and cluster labels as reported by humans} \label{fig:human_validation}
\begin{tabular}{cc}
\includegraphics[width = 0.5\textwidth]{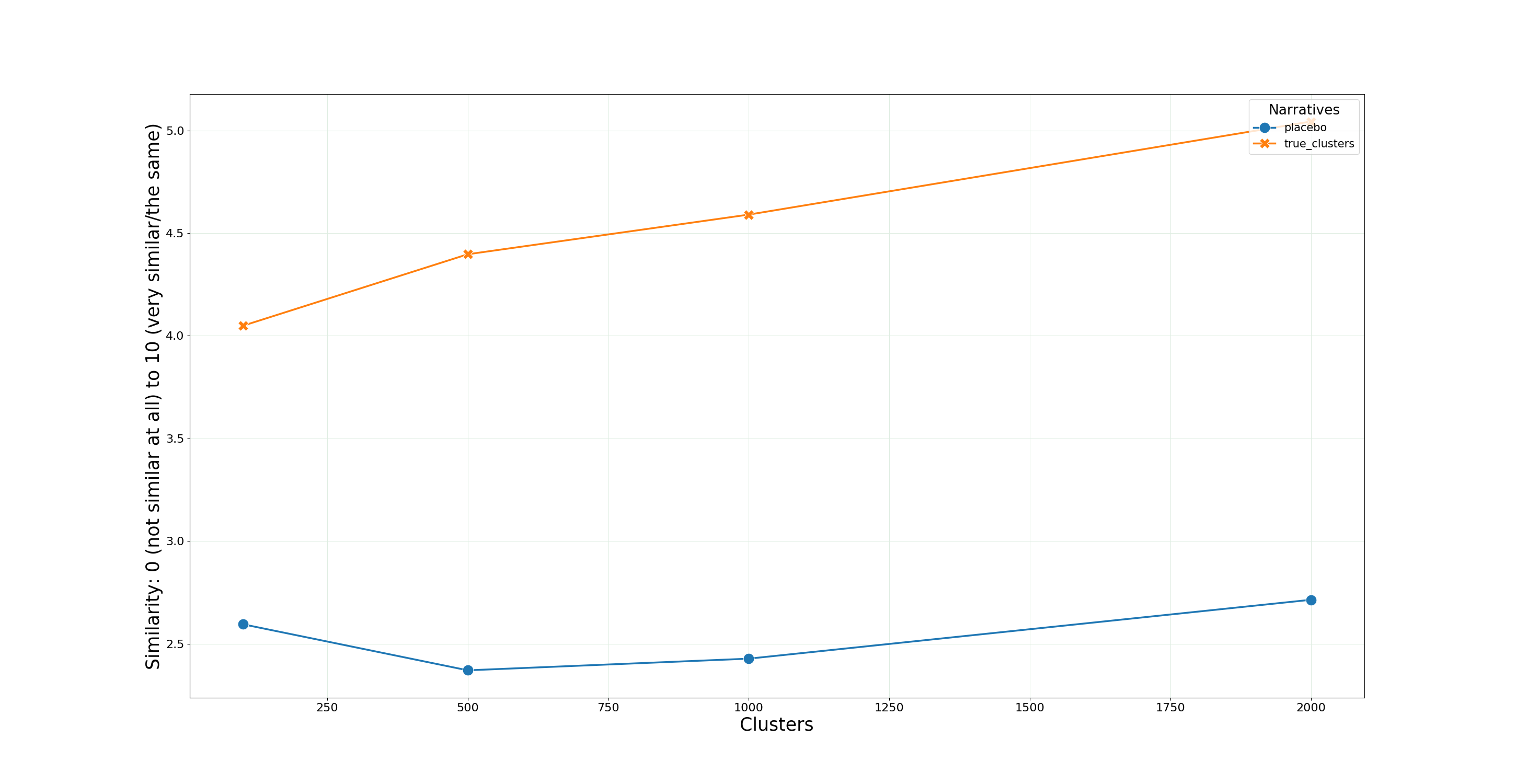} & \includegraphics[width = 0.5\textwidth]{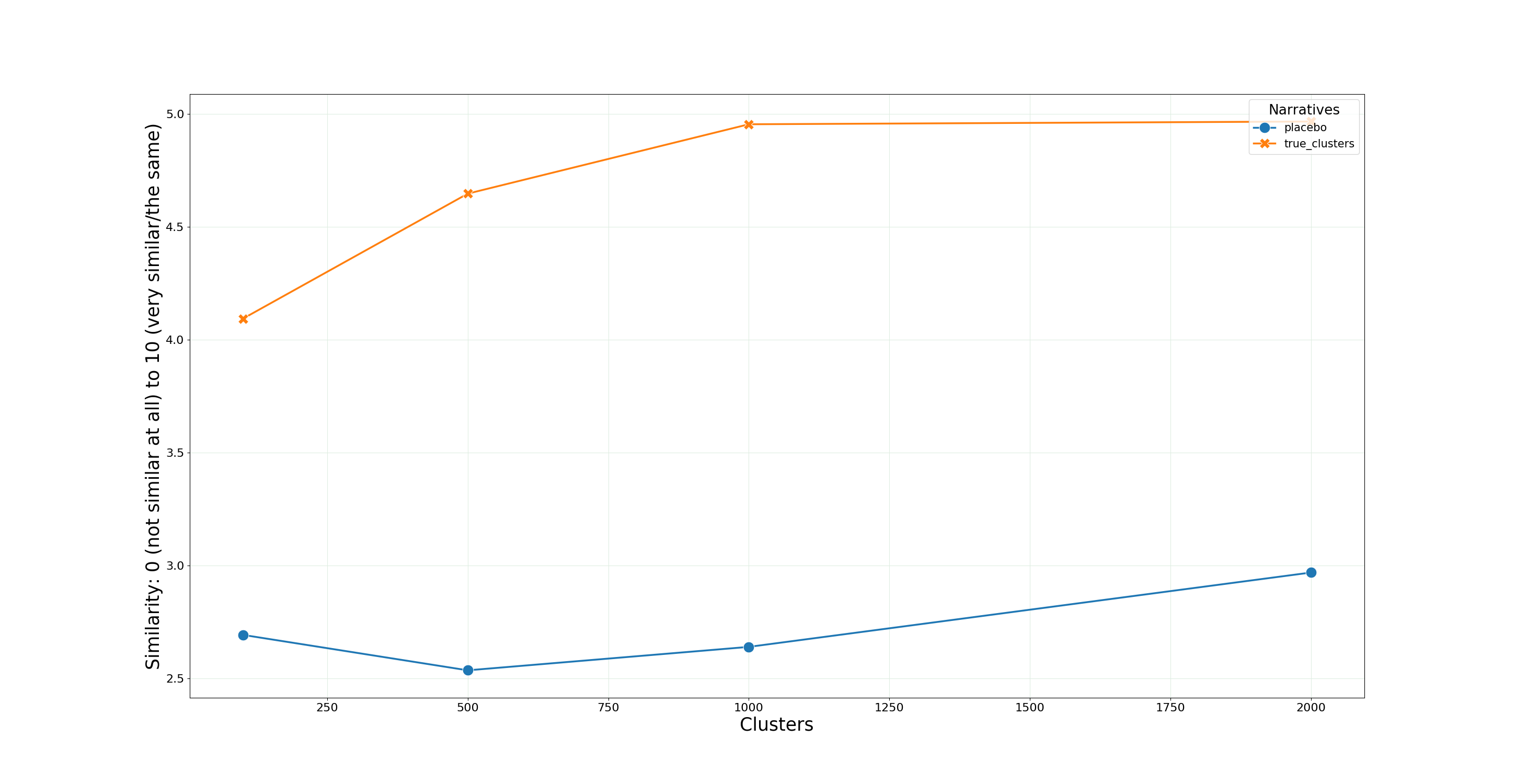} \\
All arguments  & Top 25\% arguments (by frequency)  \\
(1) & (2)  \\[6pt] \\
\end{tabular}
\end{center}
\footnotesize
\vspace{0.2cm}
\textbf{Note:} This figure shows the similarity between a random semantic role token and its cluster label for actual (role, cluster) tuples (yellow) and placebo tuples (blue; where roles are combined with a random cluster label), as annotated by human freelancers. Subfigure (1) shows the results for all arguments (that appear in narratives above our usual frequency threshold of 50). Here, the similarity increases for the clustering range studied here. However, after 1000 cluster, the placebo similarity also increases. Subfigure (2) shows the subset of validated arguments that belong to the top 25\% of entities (for the respective cluster numbers). Here, increasing the number of clusters seems to yield lower similarity gains after 1000 clusters.
\end{figure}

Figure \ref{fig:human_validation} decomposes the results by number of clusters: it shows the similarity between a random semantic role token and its cluster label for actual (role, cluster) tuples (yellow) and placebo tuples (blue; where roles are combined with a random cluster label), as annotated by human freelancers. Subfigure (1) shows the results for all arguments (that appear in narratives above our usual frequency threshold of 50, see Section \ref{sec:preprocessing}). Here, the similarity increases for the clustering specifications studied here (100, 500, 1000, 2000). However, after 1000 clusters, the placebo similarity also increases.

Next, we seek to understand whether more clusters lead to precision gains for frequent or infrequent entities. Thus, in Subfigure (2), we show the subset of validated arguments that belong to the top 25\% of entities (for the respective cluster numbers). Here, increasing the number of clusters seems to yield lower similarity gains after 1000 clusters. These results suggest that higher cluster numbers may be beneficial for less frequent entities. Again, however, the placebo similarity increases more after 1000 clusters. In our application, we focus on 1000 clusters.

\clearpage

\section{Inspecting the Quality of Mined Entities} \label{app:sec:mined_entities}
\setcounter{table}{0}
\begin{table}[!ht]
\begin{center}
\caption{Most Frequent Clustered Entities (i)} 
\label{tab:clustered_entities1}
\footnotesize
\renewcommand{\arraystretch}{1.5}
\begin{tabularx}{\textwidth}{lXc}
\toprule
      \textbf{Label} &                                                                                                                                                                                                                                                  \textbf{Most frequent phrases} &  \textbf{Frequency} \\
\midrule
          people &                                                                                          people - 182086 | many - 28671 | others - 17168 | many people - 7433 | number - 6196 | have - 5907 | several - 2299 | many others - 1413 | include - 883 | many country - 684 &     266530 \\
         someone &                                                                                          someone - 17689 | anyone - 13401 | everyone - 12217 | everybody - 10819 | somebody - 9924 | anybody - 8159 | nobody - 7671 | whoever - 1030 | someone else - 833 | else - 739 &      96520 \\
      healthcare & health care - 7917 | health insurance - 7407 | health human service - 1906 | health - 1817 | health care reform - 1508 | health care system - 1489 | health plan - 1410 | health care provider - 1080 | health care coverage - 1007 | health insurance coverage - 1005 &      94753 \\
           money &                                                                                                    money - 30048 | taxpayer - 15388 | benefit - 9898 | cost - 9534 | pay - 2496 | expense - 1298 | amount money - 831 | cash - 744 | taxpayer money - 657 | earn - 529 &      87652 \\
          family &                                                                        family - 37326 | life - 25338 | work family - 2310 | great - 1981 | many family - 1444 | real problem - 585 | family business - 537 | well life - 524 | great country - 438 | real people - 430 &      83902 \\
          person &                                                                            person - 30540 | woman - 22444 | man - 10881 | every person - 755 | many woman - 582 | woman child - 545 | first woman - 454 | another person - 404 | man woman - 336 | million woman - 266 &      74314 \\
             job &                                                                                              job - 40808 | work - 16726 | good job - 1357 | job do - 648 | create job - 584 | great job - 548 | work people - 510 | well job - 405 | good pay job - 356 | do job - 353 &      69466 \\
         country &                                                 country - 53529 | people country - 2428 | around - 725 | along - 626 | people across country - 388 | across country - 354 | across - 350 | people around world - 311 | around world - 255 | country around world - 229 &      67619 \\
           child &                                                                         child - 40835 | million child - 1278 | every child - 1224 | peer - 779 | many child - 714 | child family - 593 | child care - 508 | child support - 440 | care need - 439 | family child - 289 &      52698 \\
             law &                                                      law - 41489 | public law - 1375 | tax law - 488 | legislative branch - 384 | legal system - 361 | civil action - 327 | legal status - 307 | judicial review - 246 | judicial system - 223 | judicial branch - 172 &      51540 \\
\bottomrule
\end{tabularx}
\end{center}
\end{table}
\bigskip

\begin{table}[h!]
\begin{center}
\caption{Most Frequent Clustered Entities (ii)} 
\label{tab:clustered_entities2}
\footnotesize
\renewcommand{\arraystretch}{1.5}
\begin{tabularx}{\textwidth}{lXc}
\toprule
      \textbf{Label} &                                                                                                                                                                                                                                                  \textbf{Most frequent phrases} &  \textbf{Frequency} \\
\midrule
            plan &                                                               plan - 19094 | proposal - 13111 | initiative - 3198 | propose - 2008 | agenda - 1479 | president plan - 573 | president proposal - 422 | legislative proposal - 312 | be propose - 247 | blueprint - 213 &      50304 \\
        business &                                                       business - 20863 | industry - 8018 | many company - 719 | management - 719 | enterprise - 691 | foreign company - 628 | financial - 433 | business community - 397 | many business - 386 | management plan - 297 &      49657 \\
        employee &                                        employer - 20943 | employee - 16335 | contractor - 3404 | small employer - 475 | employer employee - 416 | many employer - 393 | private contractor - 334 | percent employer - 232 | many employee - 182 | large employer - 168 &      49211 \\
      individual &           individual - 33657 | account - 3647 | individual family - 643 | many individual - 463 | personal information - 283 | personal responsibility - 235 | individual business - 207 | every individual - 207 | million individual - 179 | family individual - 172 &      45713 \\
       community &                                         community - 13902 | organization - 8763 | local community - 2305 | international community - 2100 | youth - 1386 | membership - 564 | behalf - 487 | many community - 487 | nonprofit organization - 462 | participation - 405 &      43074 \\
             tax &                                                                                     tax - 15994 | tax cut - 5609 | spending - 4017 | cut - 3774 | tax increase - 1480 | raise - 648 | raise tax - 399 | spending cut - 318 | billion tax cut - 229 | percent cut - 221 &      41583 \\
         company &                                                                                             company - 34769 | consortium - 288 | subsidiary - 222 | affiliate - 196 | ag - 178 | venture - 147 | ge - 129 | foreign subsidiary - 63 | startup - 63 | conglomerate - 62 &      39520 \\
attorney (judge) &                                                              judge - 12108 | attorney - 11995 | lawyer - 4360 | prosecutor - 1935 | solicitor - 229 | hearing officer - 210 | superior - 184 | magistrate judge - 132 | local prosecutor - 132 | president lawyer - 92 &      37601 \\
          school &                                                          school - 16925 | college - 2924 | local school - 1676 | high school - 1305 | public school - 1178 | college student - 519 | academy - 460 | school system - 417 | many school - 403 | community college - 335 &      34830 \\
            fund &                                                           fund - 17564 | investment - 3647 | investor - 2817 | foreign investor - 293 | invest - 228 | fund provide - 208 | hedge fund - 192 | foreign investment - 182 | private investment - 175 | mutual fund - 158 &      33931 \\
\bottomrule
\end{tabularx}

\end{center}
    \footnotesize
    \flushleft
	\textbf{Note:} This table presents the 20 most frequent clustered entities identified in the U.S. Congress. The column ``Label'' refers to the final label assigned to the entity after manual inspection. The column ``Most frequent phrases'' lists the ten most frequent phrases which were assigned to these clustered entities, as well as their frequency of occurrence (i.e., phrase - frequency). The column ``Frequency'' is the total number of mentions of this clustered entity in the corpus.
\end{table}

\begin{table}[h!]
\begin{center}
\caption{Most Frequent Named Entities (i)} 
\label{tab:named_entities1}
\footnotesize
\renewcommand{\arraystretch}{1.5}
\begin{tabularx}{\textwidth}{lXc}
\toprule
         \textbf{Label} &                                                                                                                                                                                                                                                                                                         \textbf{Most frequent phrases} &  \textbf{Frequency} \\
\midrule
      american &                                                                 american - 49293 | american people - 35168 | million american - 8209 | american public - 3212 | american taxpayer - 3187 | many american - 3118 | american family - 3074 | every american - 3031 | american worker - 2682 | american citizen - 1956 &     196166 \\
       program &                                                                         program - 44125 | pilot program - 1125 | federal program - 958 | important program - 617 | government program - 580 | medicaid program - 448 | cop program - 403 | education program - 314 | many program - 293 | entitlement program - 280 &     132721 \\
    government &                                                  government - 62923 | local government - 5866 | foreign government - 1085 | government official - 553 | government shutdown - 437 | russian government - 383 | tribal government - 357 | government spending - 344 | big government - 303 | branch government - 285 &     114088 \\
    republican &                                              republican - 51157 | republican leadership - 6332 | republican majority - 3678 | republican friend - 1246 | republican plan - 1169 | republican side - 614 | republican president - 486 | many republican - 478 | republican proposal - 471 | republican control - 410 &      82419 \\
       service &                                                                          service - 14224 | arm service - 3166 | forest service - 2589 | postal service - 2282 | internal revenue service - 896 | wildlife service - 862 | good service - 812 | custom service - 690 | public service - 637 | military service - 607 &      79560 \\
administration &  administration - 56284 | president administration - 606 | administration official - 484 | previous administration - 476 | administration policy - 268 | administration proposal - 217 | administration budget - 200 | republican administration - 149 | administration president - 125 | last administration - 121 &      73593 \\
        nation &                                                                                                    nation - 34474 | great nation - 918 | nation world - 445 | people nation - 351 | many nation - 280 | entire nation - 277 | nation economy - 272 | rogue nation - 239 | nation child - 231 | foreign nation - 226 &      71985 \\
        budget &                                                                             budget - 28857 | republican budget - 2091 | president budget - 1780 | budget deficit - 663 | democratic budget - 446 | budget plan - 427 | budget agreement - 392 | budget proposal - 345 | budget process - 319 | budget surplus - 298 &      63469 \\
        agency &                                                agency - 23605 | federal agency - 6048 | local educational agency - 1605 | government agency - 942 | educational agency - 859 | law enforcement agency - 723 | head agency - 592 | intelligence agency - 530 | public housing agency - 514 | department agency - 389 &      62556 \\
       america &                                                                                      america - 24636 | people america - 971 | rural america - 481 | corporate america - 306 | america senior - 258 | job america - 246 | america family - 209 | america child - 202 | bless america - 197 | everybody america - 191 &      58065 \\
\bottomrule
\end{tabularx}

\end{center}
\end{table}

\bigskip

\begin{table}[h!]
\begin{center}
\caption{Most Frequent Named Entities (ii)} 
\label{tab:named_entities2}
\footnotesize
\renewcommand{\arraystretch}{1.5}
\begin{tabularx}{\textwidth}{lXc}
\toprule
         \textbf{Label} &                                                                                                                                                                                                                                                                                                         \textbf{Most frequent phrases} &  \textbf{Frequency} \\
\midrule
      national &              national security - 2623 | national - 1426 | national guard - 1159 | national debt - 1153 | national academy science - 672 | national interest - 563 | national labor relation - 471 | national intelligence director - 422 | director national intelligence - 407 | national science foundation - 358 &      53177 \\
      democrat &                                      democrat - 25073 | democrat republican - 2377 | republican democrat - 2183 | democrat majority - 459 | democrat leadership - 418 | president democrat - 406 | republican democrat alike - 297 | many democrat - 280 | democrat republican alike - 260 | liberal democrat - 220 &      40348 \\
        energy &                                                                             energy - 5592 | department energy - 2259 | energy policy - 1119 | energy commerce - 652 | high energy - 327 | energy company - 312 | renewable energy - 305 | energy efficiency - 294 | energy independence - 258 | energy crisis - 246 &      35776 \\
         group &                                                                             group - 8480 | bipartisan group - 464 | special interest group - 458 | work group - 413 | terrorist group - 389 | group people - 256 | environmental group - 252 | interest group - 250 | group health plan - 246 | outside group - 231 &      34463 \\
       project &                                                       project - 9652 | demonstration project - 610 | pilot project - 310 | construction project - 241 | program project - 104 | important project - 100 | research project - 100 | military construction project - 98 | project authorize - 93 | water project - 92 &      30883 \\
           war &                                                                                                                                   war - 6299 | war iraq - 831 | world war ii - 819 | war terrorism - 482 | civil war - 476 | cold war - 408 | war terror - 338 | war power - 217 | war drug - 201 | world war - 196 &      28666 \\
     authority &                                         authority - 7485 | local authority - 527 | palestinian authority - 508 | housing authority - 310 | administer authority - 299 | federal authority - 256 | public housing authority - 236 | budget authority - 212 | trade promotion authority - 170 | exist authority - 158 &      28472 \\
small business & small business - 14213 | small business owner - 1592 | small business administration - 538 | small business concern - 517 | many small business - 439 | small business people - 183 | percent small business - 156 | small business community - 131 | individual small business - 128 | family small business - 124 &      27522 \\
      medicare &                                                              medicare - 10868 | medicare beneficiary - 1220 | medicare patient - 808 | medicare system - 309 | medicare trust fund - 215 | medicare benefit - 199 | medicare cut - 150 | medicare recipient - 149 | cut medicare - 138 | traditional medicare - 127 &      27256 \\
       veteran &                                                                                veteran - 11324 | veteran affair - 1688 | many veteran - 412 | disabled veteran - 377 | nation veteran - 328 | veteran family - 189 | homeless veteran - 188 | million veteran - 181 | veteran benefit - 156 | america veteran - 156 &      27199 \\
\bottomrule
\end{tabularx}

\end{center}
    \footnotesize
    \flushleft
	\textbf{Note:} This table presents the 20 most frequent named entities identified in the U.S. Congress. The column ``Label'' refers to the final label assigned to the entity after manual inspection. The column ``Most frequent phrases'' lists the ten most frequent phrases which were assigned to these named entities, as well as their frequency of occurrence (i.e., phrase - frequency). The column ``Frequency'' is the total number of mentions of this named entity in the corpus.
\end{table}

\clearpage

\section{Summary Statistics}
\label{app:sumstats}
\setcounter{table}{0}

\begin{table}[h!]
\caption{Summary Statistics on the Congressional Record Corpus}
    \label{tab:SummaryStatistics}
    \begin{center}
    \resizebox{0.7\textwidth}{!}{
    \def\arraystretch{1.25}
    \begin{tabular}{lrrrr}
\toprule
                  Variable &  All Narratives &  Complete Narratives &  Frequent Narratives &  Relevant Narratives \\
\midrule
                  Speeches &         1190969 &               550561 &               187379 &               105514 \\
                 Sentences &        17312358 &              4196393 &               279044 &               160827 \\
                Statements &        76395962 &              5774295 &               346685 &               180964 \\
        Narratives, unique &         5963192 &              3570119 &                 2838 &                 1638 \\
                Agents raw &        76395962 &              5774288 &               346685 &               180964 \\
        Agents raw, unique &         1604754 &               945412 &                21041 &                15035 \\
  Agents clustered, unique &            1884 &                 1879 &                  415 &                  245 \\
              Patients raw &        76395962 &              5774293 &               346685 &               180964 \\
      Patients raw, unique &         8322792 &              1794942 &                33774 &                21411 \\
Patients clustered, unique &            1884 &                 1877 &                  499 &                  353 \\
                 Verbs raw &        76395962 &              5774295 &               346685 &               180964 \\
         Verbs raw, unique &           44889 &                18642 &                  391 &                  276 \\
     Verbs cleaned, unique &           34780 &                11807 &                  183 &                  140 \\
\bottomrule
\end{tabular}}
\end{center}
\footnotesize
\flushleft
\textbf{Note:} This table shows descriptive statistics on the U.S. Congressional Record at different steps of the pipeline. ``All narratives'' are those with at least an agent or a patient. ``Complete narratives'' are those with both an agent and a patient. Among those, ``frequent narratives'' were pronounced at least 50 times. Finally, we focus on politically or economically relevant entities (``relevant narratives'') -- as opposed to procedural/uninterpretable entities.
\end{table}

\begin{figure}[ht!]
\begin{center}
\caption{Share of Times An Entity appears as an Agent (as opposed to a Patient)}
\label{fig:arg0_arg1_shares}
\includegraphics[width = 0.6\textwidth]{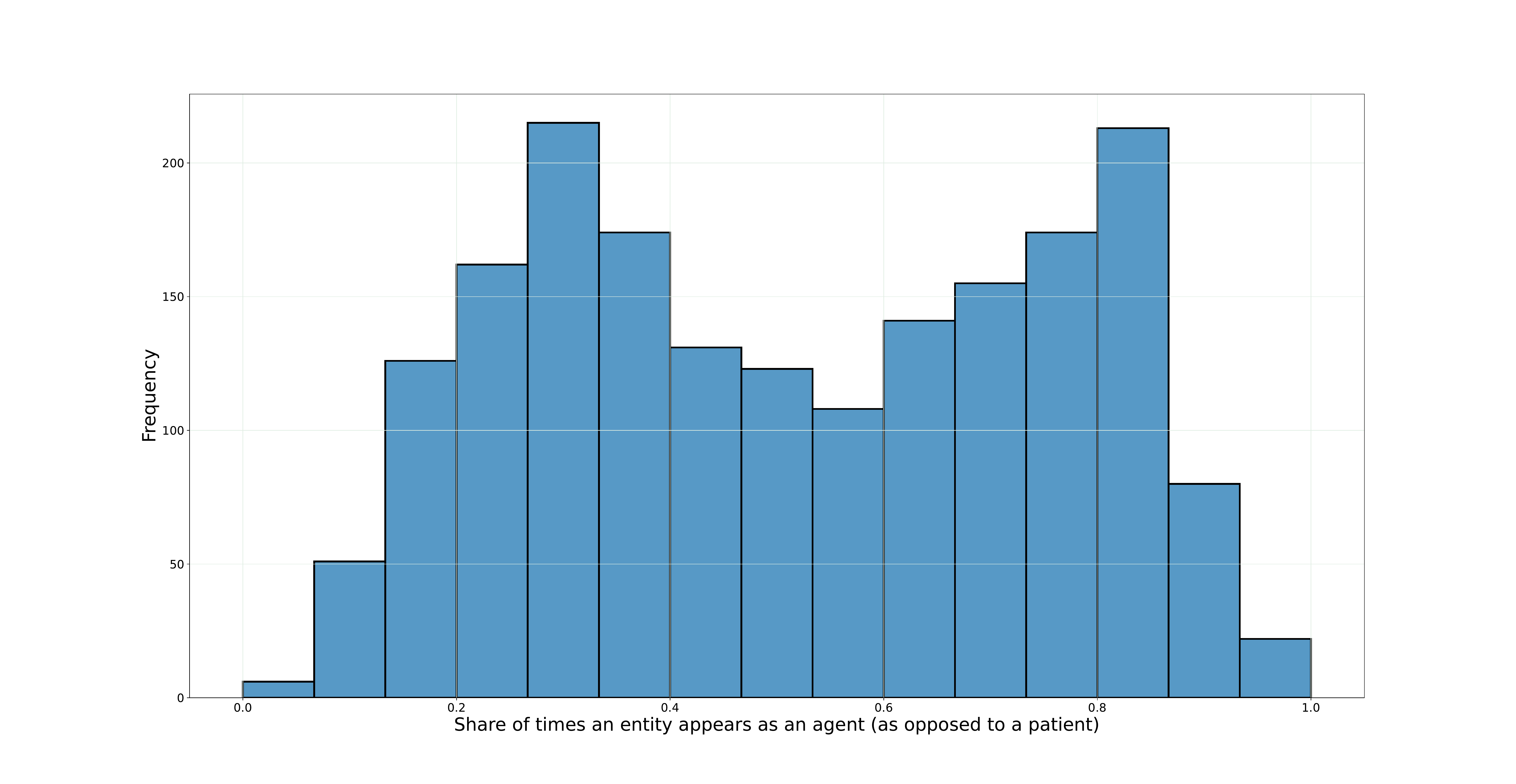}
\end{center}
\footnotesize
\flushleft
\textbf{Note:} This figure presents, for every entity, the share of times it appears as an agent (as opposed to a patient). For example, if an entity occurred as an agent 8 times and as a patient 2 times, the corresponding share would be 0.8. Most entities appear both as an agent and a patient. However, entities also tend to be ``specialized'': Most of them appear as either an agent or a patient more frequently.
\end{figure}

\clearpage
\section{Examples of Narratives and Sentences}
\label{app:examples}
\setcounter{table}{0}

{\setlength{\parindent}{0cm}

This appendix shows examples of random sentences and their associated narratives as well as the most frequent narratives and random examples of sentences they appear in. In both cases, we focus on narratives that contain (at least) an agent, a verb, and a patient (defined as ``complete'' narratives, see Section \ref{sec:application:emp_analysis} in the main part).

\subsection*{Sampled sentences and their associated narratives} \label{sec:random_sent_narr}

The following list shows a random sample of sentences, along with the complete narrative(s) we mine for each sentence.

\dotfill

\textit{The Surgeon General serves as ``America's Doctor.''}

\textbf{physician serve america}

\dotfill

\textit{Also, a county can opt simply to have the money sent back to the U.S. Treasury without pursuing projects.}

\textbf{county pursue project}

\dotfill

\textit{I have heard from several Montanans who say they want the security to view an individual paper version of the ballot before it is cast and counted.}

\textbf{military view ballot}

\dotfill

\textit{What on Earth could they be talking about, setting up trade policies with other countries that undercut our producers and undercut our workers?}

\textbf{country undercut production}

\textbf{country undercut worker}

\dotfill

\textit{This Nation has come a long way from the terrible day of September 11 and the impact that had both on the infrastructure in New York and in Washington, but also on the psyche of the American public, the confidence.}

\textbf{effect have infrastructure}

\dotfill

\textit{In my view, the issue we need to debate today is how we're going to help that individual who can't meet the challenges and wants to stay in their home.}

\textbf{individual not-meet challenge}

\dotfill

\textit{It is our model for how a safety net should work.}

\textbf{profit should job}

\dotfill

\textit{We know that our children's future depends on their education.}

\textbf{future depends education }

\dotfill

\textit{Since the Supreme Court has delineated it as a tax, it's clear that obligations or commitments to not raise taxes on the middle class have gone by the wayside.}

\textbf{supreme have promulgate regulation}

\dotfill

\textit{How did the capitalists threaten capitalism?}

\textbf{ideology threaten ideology}

\dotfill

\textit{Another business person in my State of West Virginia just sent me his tax collection for next year for the ObamaCare health plan.}

\textbf{person sent collection}

\dotfill

\textit{With respect to commercial supply needs, taxpayer subsidies in the form of loan guarantees should be restricted to those areas not undertaken by the private sector.}

\textbf{public not-undertaken area}

\dotfill

\textit{the United States Government, nor any other government, never made a dime; they took it from you.}

\textbf{government not-make dime}

\dotfill

\textit{Now the contrast to that is the Republican substitute which will be offered has a permanent, government-wide Hyde amendment, meaning unequivocally, no Federal funds can be used for abortion anywhere in any bill that passes.}

\textbf{republican have government}

\dotfill

\textit{Many college students--including 37 percent of Hispanic students-- receive Pell Grants each year, and this bill will now allow students to receive these vital grants year round.}

\textbf{student receive grant}

\dotfill

\textit{The bill before us does nothing to address the sequestration cuts that the FAA will have to make in Fiscal Year 2014 and beyond.}

\textbf{F.A.A. make sequestration}

\dotfill

\textit{Equally important, we cannot permit any international tribunal to interpret and to apply the trade laws of the United States.}

\textbf{prosecution apply trade}

\textbf{prosecution interpret trade}

\dotfill

\textit{I think Americans care about Social Security too, but instead of talking constructively about fixing Social Security, we demonize the President, we demonize the plan.}

\textbf{american care social security}

\dotfill

\textit{All of our Nation's children should be able to take advantage of technology and a ride on the information superhighway.}

\textbf{nation take opportunity}

\dotfill

\textit{People come to the floor talking about how to save taxpayer dollars, but we have not undertaken to make reforms that would protect taxpayers in the first place.}

\textbf{reform protect money}

\dotfill

\subsection*{Most frequent narratives and examples of sentences they appear in} \label{app:sec:most_freq_narr}

The following list shows the most frequent narratives in our corpus (ordered by frequency), along with a random sample of sentences in which they appear.

\noindent

\dotfill

\textbf{people lose job}

\vspace{0.2cm}

\textit{Mr. Speaker, I urge my colleagues to defeat the previous question so that we can immediately bring up H.R. 4415, which would restore unemployment benefits to 2.8 million Americans, people who have lost their job and are simply trying to find their next job and want to prevent their families from losing everything they have worked for in that period.}

\textit{Our colleague from Ohio just said this is an opportunity for people who have lost their jobs.}

\textit{Providing health care benefits--again, none of us subscribes to the notion that people who are unemployed or lose their jobs are anywhere near as much a victim as those victims on September 11, at the World Trade Center, or the Pentagon, or aboard that airplane in Pennsylvania.}

\textit{It appears, we've been hearing over and over from the Democratic leadership, and even from the President, people are losing jobs every day.}

\textit{People shouldn't have to lose their jobs to pay for the New York fund.}

\dotfill

\textbf{citizen abide law}

\vspace{0.2cm}

\textit{Conceivably, the agencies could develop more narrow regulations that focus on the money laundering issue, in an effort to curb criminal activity, that would not carry with it the heavy burdens of regulation on the banks and the potential intrusion into the financial privacy of ordinary, law-abiding citizens, which none of us wants to do.}

\textit{The vast majority of these private security officers are dedicated, hard-working, law-abiding citizens of this country, and are properly screened before hiring and trained before deployment.}

\textit{Parents have a sacred duty to rear their children in love and righteousness, to provide for their physical and spiritual needs, to teach them to love and serve one another to observe the commandments of God and to be law-abiding citizens wherever they live.}

\textit{I felt kicked around and ignored by the very system the government has in place to protect law-abiding citizens.}

\textit{The American people appropriately look to their Government to maintain a peaceable society but do not want law enforcement to stray into the private lives of law-abiding citizens.}

\dotfill

\textbf{american lose job}

\vspace{0.2cm}

\textit{These funds will go a long way in supporting American workers who have lost their jobs due to the economic slowdown and last year's terrorist attacks.}

\textit{As a direct result of the hasty legislation, experts have estimated that over 1,000 Americans will lose their jobs unless Congress takes immediate action to correct and clarify the Affordable Care Act's impact on expatriate health care plans.}

\textit{To raise taxes while these Americans are losing their jobs is irresponsible.}

\textit{We have all kinds of issues, and Americans across this country have lost jobs, unemployment is at a high rate, people are having to make decisions.}

\textit{Mr. Speaker, I hope that my colleagues fully understand that since the current administration took office, an average of 157,000 Americans are losing their jobs every month.}

\dotfill

\textbf{government run healthcare}

\vspace{0.2cm}

\textit{While Republicans offered positive solutions to tackle the spending, Democrats created a massive new government-run health care plan that hurts the economy, interferes with patient choices, and does nothing to bring down the cost of health care.}

\textit{Is that what you are talking about where you all of a sudden shift from people who figure out you can get the government to pay for everything, a government-run health care program?}

\textit{I don't understand what they are talking about: ``socialized medicine,'' ``Cuban-style, government-run health care.''}

\textit{Well, unfortunately, there are people who love people but think that by the government running health care--which will inevitably lead to rationing of health care--that somehow that's a better thing.}

\textit{114 million--that's the number of people who could lose their current health care coverage--coverage, of course, that they like--under the proposed government-run health plan in H.R. 3962.}

\dotfill

\textbf{american have healthcare}

\vspace{0.2cm}

\textit{We were told by Senator Kennedy and by President Clinton we have to give up this freedom because we have 30 million American families who have no health insurance.}

\textit{The fact is that we can contain costs and help enable every American to have access to health insurance coverage.}

\textit{They deserve to know why all this money is being spent on a war of choice, when one on eight Americans lives in poverty, and when 46 million Americans have no health insurance, including 13 million children.}

\textit{One of the Presidents of the United States once stated that Americans already have universal health care because the emergency rooms cannot legally refuse to treat patients.}

\textit{I remind my colleagues one more time that we are talking about trying to assist 44 million Americans who have no health insurance.}

\dotfill

\textbf{god bless america}

\vspace{0.2cm}

\textit{May God bless their souls and may God bless America.}

\textit{And may God continue to bless America.}

\textit{God bless these heroes, their families and God bless America.}

\textit{God bless America, and God bless Texas.}

\textit{May God bless Ronald Reagan and Mrs. Reagan and may God bless America.}

\dotfill

\textbf{people need help}

\vspace{0.2cm}

\textit{But I think we ought to help people who need help because it is the right thing to do.}

\textit{And that is what this welfare reform is all about--to do something about people who are down on their luck and need help.}

\textit{It was so typical of Cassandra; she was always thinking of others who might need help.}

\textit{We are supposed to represent the people who need help across this country.}

\textit{But bit by bit, we will be taking all of these changes that are in this bill and showing you how it is not going to help the people who really need the help, who are not getting that help.}

\dotfill

\textbf{god bless troop}

\vspace{0.2cm}

\textit{In conclusion, God bless our troops, and we will never forget September 11th in the global war on terrorism.}

\textit{In conclusion, God bless our troops, and we'll never forget September the 11th.}

\textit{In conclusion, God bless our troops, and we will always remember September 11.}

\textit{In conclusion, God bless our troops, and the President's actions should be based on remembering September the 11th in the global war on terrorism.}

\textit{In conclusion, may God bless our troops, and we will never forget September 11.}

\dotfill

\textbf{worker lose job}

\vspace{0.2cm}

\textit{Democrats want to help more workers who lose their jobs because of trade, especially workers providing services.}

\textit{There are going to be workers who lose their jobs because of this rule.}

\textit{In the cities, workers lose their jobs too.}

\textit{If we are going to have a real trade package for this country, it has to benefit not just those who win from trade but those who lose from trade as well, including the workers who lose their job through no fault of their own.}

\textit{It does not provide unemployment benefits to workers who have lost their jobs, or extend health care coverage to those employees, nor does it prohibit the airlines from abrogating their contracts with workers; and it mandates no job protections, or a system for rehiring when our airline industry recovers.}

\dotfill

\textbf{small business create job}

\vspace{0.2cm}

\textit{The time and money required to keep up with government paperwork prevents small businesses from growing and creating new jobs.}

\textit{President, our Nation's small businesses have created 64 percent of all new jobs in the last 15 years, yet in the last year nearly 85 percent of the jobs lost have come from small businesses.}

\textit{Small businesses create 80 percent of the jobs, so you would think a good piece of the relief would go to small business.}

\textit{Mr. President, small businesses represent more than 99 percent of all employers, employ 53 percent of the private work force, and create about 75 percent of the new jobs in this country.}

\textit{Stop blocking job-training programs and initiatives by the President, because everyone is not going to college, community colleges, where we need to train people for changing jobs in technology opportunities that we are missing and helping small business, not hurting small business to create jobs so we can have people working in the future.}

\dotfill

\textbf{service connect disability}

\vspace{0.2cm}

\textit{Documentation described in this paragraph is a rating decision or a letter from the Department of Veterans Affairs that confirms that the veteran is totally disabled because of one or more service-connected injuries or service-connected disability conditions.}

\textit{That is why section 301 of this bill would allow veterans who lack that access, who do not have a service-connected disability, and who do not have affordable health insurance, to enroll in the VA's health care system.}

\textit{House Bill Section 302 of H.R. 1716, as amended, would extend eligibility for specially adapted housing grants to veterans with permanent and total service-connected disabilities due to the loss, or loss of use, of both arms at or above both elbows.}

\textit{Finally, I find it disturbing that during a time of war an anonymous member of Congress is willing to halt legislation that would help Persian Gulf War veterans with service-connected disabilities and}

\textit{The bill specifically aims at helping vulnerable veterans--known in veterans parlance as ``Category A'' veterans--who have either low income or a service-connected disability.}

\dotfill

\textbf{constitutional (thing) balance budget}

\vspace{0.2cm}

\textit{and I think there are probably a half dozen or dozen other Members who would similarly vote for it and we would have 70 or 75 votes for a constitutional amendment to balance the budget.}

\textit{When and if we pass a constitutional amendment to balance the budget by the year 2002, and if that is ratified by 75 percent of the States, that is not going to cure all of our problems.}

\textit{I simply say in tackling this proposition this Senator, and I expect two-thirds of the Senate, are strongly in support of and will pass a constitutional amendment to balance the budget.}

\textit{It is a provision that requires that the Senate and House of Representatives, before December 31 of this year vote on a constitutional amendment to balance the budget.}

\textit{We fell just one vote short of getting a constitutional amendment to balance the budget.}

\dotfill

\textbf{people have healthcare}

\vspace{0.2cm}

\textit{So here we are in a nation with a health care crisis, in a nation where each day fewer people have health insurance, a nation where each day the cost of health care is going up, a nation where businesses are struggling to keep health insurance on the owners of the business and their employees, where labor unions are at their wit's end about how to provide the basic benefit package and still increase take-home pay, here we are in a certifiable American crisis when it comes to health care.}

\textit{They basically agree because they know that with much more coverage, with many more people having health insurance, they could spread out their business.}

\textit{When Senator Kennedy, Senator Chafee, Senator Rockefeller and I worked on the original legislation in 1997, our goal was to cover the several million children who had no health insurance.}

\textit{I do not think you will find 17 people in the gallery who do not want to have health insurance coverage.}

\textit{people who have no health insurance are saying, Yes, we don't want to go throughout life worrying about whether we are going to go bankrupt or whether we are going to be able to go to a doctor, and they are trying to get more information about the Affordable Care Act, and they are signing up in huge numbers--higher than people had anticipated.}

\dotfill

\textbf{public held debt}

\vspace{0.2cm}

\textit{During the 1- year period between December 2004 and December 2005, foreigners purchased 96 percent of the new debt held by the public.}

\textit{According to the CBO, assuming the continuation of many current policies, debt held by the public as a share of our GDP is projected to reach an implausibly high 947 percent of GDP by 2084.}

\textit{In 1998, we paid down over \$50 billion in debt; 1999 paid down over \$80 billion in debt; and in fact, with this Republican budget that is here on the floor today, we will reduce the debt held by the public \$450 billion over just 4 years.}

\textit{The debt held by the public now exceeds \$9 trillion.}

\textit{What we projected was that the debt that this country owes, much of which is owed to the public, these people out there that are buying all these Treasury notes, Treasury bills, and Treasury bonds every time the Treasury has an auction, we projected a year ago that there would be no debt held by the public after 10 years.}

\dotfill

\textbf{people pay tax}

\vspace{0.2cm}

\textit{The only way we can do that, with the Tax Code is to create a similar deadline, and that is to say to the people of this country that we are going to do away with the existing code and that we are going to start over, with a new Tax Code that makes sense to the people who have to pay the taxes in this country.}

\textit{Well before the IRS started getting involved in your health care and sharing your information and forcing people to pay even more taxes, let's first look at the job the IRS is already doing.}

\textit{We are all winners when people get a better chance to be more effective, more productive, pay more taxes, and have America to maintain its leadership in the world.}

\textit{The more people you have working, the less taxes you have to pay from everybody, and the less taxes you have to put on business, and the more people they can hire, and the more people that can pay taxes.}

\textit{As our economy grows and new jobs are added, people pay more in taxes.}

\dotfill

\textbf{people lose family}

\vspace{0.2cm}

\textit{I am going to come down here week after week and tell the simple stories of the dozens of people who lose their lives every day due to gun violence.}

\textit{I've also stood at the site of the grim tragedy of the Mesaba Airlines commuter crash, only 6 miles from my home in Chisholm, Minnesota; the flight path toward the Chisholm Hibbing Airport in December 1993, where 19 people lost their lives because that aircraft didn't have a ground proximity warning system.}

\textit{All of us attended funerals of people who lost their lives on September 11, and the pain is still there.}

\textit{In memory of these 17 people and more--I assume, since we do not reflect communities of 12,000 or more who lost their lives, that almost that many will lose their lives today somewhere in this country--it is our fervent hope that we will do a better job in reducing this level of violence in our country.}

\textit{How many more will lose their lives if the local Coast Guard stations must devote the majority of their time to homeland security alone?}

\dotfill

\textbf{american pay tax}

\vspace{0.2cm}

\textit{This always depends, again, on Americans paying taxes.}

\textit{Americans work for more than 4 months just to pay their taxes.}

\textit{It costs Americans \$160 billion a year to comply with the code, let alone the taxes Americans pay.}

\textit{This legislation is for all Americans who have never broken the law and pay taxes out of their hard-earned money.}

\textit{According to the nonpartisan Tax Foundation, Americans will have to work 108 days this year just to be able to pay their taxes--108 days.}

\dotfill

\textbf{employee provide healthcare}

\vspace{0.2cm}

\textit{An employer must provide health insurance for their employees if they have more than 50 employees or 50 full-time equivalents.}

\textit{Under the Affordable Care Act, if they have 50 or more employees and they work 30 hours a week, then the employers have to provide health insurance or pay a fine.}

\textit{What does ObamaCare mean for a working family that has been receiving employer-provided health insurance from their small business?}

\textit{Then tell me if an employer in carrying out their fiduciary duties in providing health care for their employees-- including plan determinations, reporting, enrolling people, choosing plans, maybe an optional plan, and so on--tell me they do not do more than what the exemptions are here.}

\textit{if they are lucky enough to work for an employer that provides health care, the government will subsidize it with a deduction to that business; but if, by pure happenstance, they are either unemployed or they are employed by an employer who cannot offer them or does not offer them health insurance coverage, we are going to punish them and we are going to say they ought to go out and buy insurance}

\dotfill

\textbf{people do job}

\vspace{0.2cm}

\textit{There is nothing like the view of people who have actually done this work.}

\textit{They are the people doing the work it takes to put the food on our kitchen tables, not just those on the farm but those who manufacture, sell farm equipment, the people who ship the crops from one place to another, the people who have the farmers markets and local food hubs, the people who work in food processing and crop protection and crop fertility, not to mention the researchers and the scientists who worked hard every day to fight pests and diseases that threaten our food supply.}

\textit{If there are not enough people to do the job, hire them.}

\textit{This country is strong and able and creative, but it is not going to stay that way if we do not leave the power of this country in the hands of the people who do the work, because the working men and women of this country are the ones who make it work and prosper and make it creative.}

\textit{There is something that, based on the work that people did in 2014 and the predecessor years and all the incredible progress that has been made, that there is some day in the future.}

\dotfill

\textbf{men women serve country}

\vspace{0.2cm}

\textit{We must ensure that those men and women who served this country during the Gulf War, on active duty, in the reserves, and those who have since left the military services, receive proper medical attention.}

\textit{This bill ensures that the men and women of the military who go into harm's way and bravely serve our country will have their vote counted.}

\textit{We are talking about men and women who have served our country.}

\textit{Let us stand here today, this is an opportunity, let us stand here today and send a bipartisan relationship message to all of the men and women who are bravely serving our country today and tell them, as we have told them in a bipartisan fashion in the past, that we do indeed care about them and that we do indeed care about their safety.}

\textit{In the 2 months they have been in Iraq, these men and women have been serving under the leadership of LTC Mark Warnecke, having truly served their country in the true tradition of the National Guard.}

\dotfill

}

\clearpage

\section{Comparison to Alternative Text Features} \label{app:AlternativeFeatures}
\setcounter{table}{0}

To identify latent topics in political discourse, we run Latent Dirichlet Allocation (LDA) on the U.S. Congressional Record. We then compare our method to topics and n-grams by presenting the ten most representative words as well as the ten most representative narratives associated to each topic share in Tables \ref{tab:topics1}, \ref{tab:topics2}, and \ref{tab:topics3}. We restrict ourselves to narratives that were at least pronounced 50 times in the corpus.

\bigskip

\newcolumntype{b}{X}
\newcolumntype{s}{>{\hsize=.5\hsize}X}
\newcolumntype{t}{>{\hsize=.3\hsize}X}

\begin{table}[h!]
\caption{Topics, Words and Narratives (i)} 
\label{tab:topics1}
\centering 
\scriptsize
\renewcommand{\arraystretch}{1.5}
\begin{tabularx}{\textwidth}{tsb}
\toprule
                       \textbf{Topic} &                                                                                 \textbf{Top Words} &                                                                                                                                                                                                                                                                                                                 \textbf{Top Narratives} \\
\midrule
                     banking &             bank, financi, compani, credit, consum, drug, card, institut, account, market &                                                                                                         bank issue card, company hold business, money fund outlay, measure provide authority, company hold loan, asset back military, american want reform, mortgage back military, action have effect, consumer make instruct \\
              communications &        page, line, internet, servic, amount, communic, comput, increas, access, broadcast &                                                           program serve child, provider offer service, tax benefit american, reconciliation carry recommendation, budget receive recommendation, facility provide service, company provide service, network provide service, provider provide service, citizen have healthcare \\
                       crime &                   crime, law, enforc, victim, crimin, violenc, justic, state, prison, act &                                               law enforcement add category, crime add category, injustice motivate crime, juvenile commit crime, violent crime serve (legal) sentence, program meet regulation , juvenile commit violent crime, offender serve (legal) sentence, child witness violence, offender commit crime \\
                     culture &                art, smith, museum, delawar, idaho, cultur, music, endow, recognit, artist &                                                                          american have service, university conduct study, american do job, parent care child, someone seek recognition, program help community, federally fund project, program have (positive/negative) effect, applicant receive grant, people leave country \\
                     economy &              busi, small, compani, capit, invest, contract, administr, job, creat, econom &                                      person control small business, eligible participate program, program provide fund, small business employ workforce, agency give priority, agency require information , veteran control small business, service control small business, individual receive funding, founder start business \\
                   education &                educ, school, student, teacher, program, colleg, year, high, state, higher &                                                                         student use technology, student leave discipline, school reduce class size, school hire student, agency serve school, school use fund, student receive degree (education), agency receives grant, agency require information , student have education  \\
        employment and labor &                         worker, employ, employe, work, labor, job, wage, union, pay, hour &                                                                                                  employee pay overtime, employee hire permanent, person earn dollar, men women earn dollar, worker join union, worker form union, person make dollar, market take healthcare, person make workforce, commonwealth perform duty \\
                 environment &             environment, wast, water, regul, epa, facil, state, clean, protect, administr &                            individual take opportunity, child have disability, toxic (thing) cause cancer, national preserve social security, program need reform, environmental protection agency do job, supreme declare unconstitutional, republican want repeal (something), money justify money, republican want medicare \\
                      family &                  children, famili, child, parent, welfar, care, young, work, support, kid &                                                                                          girl have baby (boomer), action have effect, family adopt child, good start provide service, parent care child, good start serve child, family leave welfare, child witness violence, child have child, community participate program \\
                     farming &                food, product, farm, agricultur, farmer, tobacco, produc, price, fda, crop &                                                                     kid graduate school, F.D.A. regulate tobacco, tobacco cause cancer, child start smoking, food drug administration regulate tobacco, agency require information , nothing limit authority, law provide remedy, F.D.A. have authority, uniform serve country \\
\bottomrule
\end{tabularx}
\bigskip

\end{table}

\begin{table}[h!]
\centering 
\caption{Topics, Words and Narratives (ii)} 
\label{tab:topics2}
\scriptsize
\renewcommand{\arraystretch}{1.5}
\begin{tabularx}{\textwidth}{tsb}
\toprule
                       \textbf{Topic} &                                                                                 \textbf{Top Words} &                                                                                                                                                                                                                                                                                                                 \textbf{Top Narratives} \\
\midrule
fiscal policy and taxation &                     tax, percent, american, incom, famili, pay, job, increas, year, would &                                                                measure breach budget, measure would budget, measure provide authority, program require appropriation, budget receive recommendation, reconciliation carry recommendation, measure not-increase deficit, public held debt, fact impose tax, budget add trillion \\
              foreign policy &                  state, unit, nuclear, treati, weapon, nation, presid, peac, intern, iran &                                                                  wealthy pay fair, iran obtain nuclear (weapon), iran acquire nuclear (weapon), iran develop nuclear (weapon), iran have nuclear (weapon), iran pose threat, nuclear (weapon) pose threat, bank do business, constitution prohibit america, people help people \\
                gun violence &                       china, gun, chines, law, weapon, firearm, taiwan, kill, year, state &                                                                                                                   weapon pierce tank, national provide service, institute conduct study, customer sell gun, veteran need help, soldier pierce tank, gun abide law, crime get gun, witness give witness, american scratch staff \\
                  healthcare &              hospit, rural, nurs, johnson, home, physician, mitchel, area, facil, lungren &                                                                person perform abortion, pregnancy feel pain, healthcare deny coverage, person have corrective action, physician deliver baby (boomer), insurance drop coverage, insurance deny coverage, physician perform abortion, doctor acquire disease, patient get child \\
                     housing &               loan, hous, home, mortgag, bankruptci, rate, debt, famili, credit, interest &                                                                                         american make fund, homeowner face mortgage, soldier serve country, family face mortgage, law provide protection, federally assist housing, trade create job, people face mortgage, program provide housing, authority provide funding \\
                human rights &                 unit, govern, state, wherea, right, peopl, human, freedom, intern, religi &                                                                      program help community, message transmit concurrent, attorney (judge) represent client, people save family, american travel cuba, money amount taxable, burmese contain export, small business provide service, government make payment, business pay tax \\
                 immigration &                  secur, immigr, border, homeland, state, law, illeg, unit, alien, countri &                                                               bible fulfil self-employed, american hold market, country participate program, terrorist enter country, service administer program, employee hire illegal immigrant, people cross border, people enter country, employee hire immigrant, immigrant enter country \\
                   judiciary &                 fee, claim, patent, liabil, properti, damag, file, compani, lawsuit, case &                                                     party seek remedy, supreme have (legal) statute, association give grade, bush nominate circuit (courts), attorney (judge) follow law, attorney (judge) apply law, bush nominate attorney (judge), law confer power, agreement enters force, attorney (judge) interpret law \\
                 legislation &                amend, motion, bill, ask, resolut, consent, unanim, read, question, presid & (sick/paid/unpaid) leave be columbia, (sick/paid/unpaid) leave be guam, (sick/paid/unpaid) leave be virgin , money amount taxable, message transmit concurrent, burmese contain export, corporation  make charity, natural resource recommend substitute, service recommend substitute, individual make legislative (election) \\
                    military &          veteran, servic, disabl, militari, care, member, benefit, mental, health, provid &                                                                 air force request application, force base defense, service connect condition, veteran suffer illness, veteran earn money, veteran suffer diabetes, veteran receive compensation, defense submit service, service connect veteran, defense submit congressional \\
                       names &                 vote, smith, johnson, lewi, brown, taylor, davi, miller, weldon, peterson &                                     application decrease appropriation, law authorize study, brave men women give family, government make payment, corporation  make charity, veteran control small business, veteran administer law, nothing limit authority, person commit (legal) charge, congressional budget define money \\
              national pride &                        american, nation, year, serv, great, honor, mani, today, life, one &                                                                                                                 people remember courage, god wipe shot, god grant strong, american hold market, word do justice, commonwealth perform duty, african make contribution, church offer prayer, hero give family, god bless family \\
            native americans &                    indian, tribe, nativ, alaska, tribal, land, mine, trust, reserv, state &                                                                                             decision have effect, budget do job, someone get money, program carry program, employee perform service, healthcare save money, good start provide service, institute conduct study, someone know nothing, raw material  meet road \\
                     
\bottomrule
\end{tabularx}
\bigskip

\end{table}

\begin{table}[h!]
\centering 
\caption{Topics, Words and Narratives (iii)} 
\label{tab:topics3}
\scriptsize
\renewcommand{\arraystretch}{1.5}
\begin{tabularx}{\textwidth}{tsb}
\toprule
                       \textbf{Topic} &                                                                                 \textbf{Top Words} &                                                                                                                                                                                                                                                                                                                 \textbf{Top Narratives} \\
\midrule                     
natural resources / energy &                        energi, oil, gas, price, fuel, electr, product, use, produc, natur &                                                                                            someone buy gun, school use fund, ethanol blend gasoline, nation need energy, production shut supply and demand, production use american, oil account america, american rely oil, O.P.E.C. increase production, america need energy \\
                      nature &                     land, nation, area, park, forest, manag, conserv, resourc, state, acr &                                                                                           county mean county, city  mean city , authority provide funding, worker earn minimum wage, grant receive grant, agency provide technical assistance, family leave welfare, someone buy gun, nothing affect authority, service do job \\
                    no label &                  presid, veto, congress, aid, reagan, item, bush, earmark, clinton, power &                                                                                                                        company bring job, story detail fight, energy affect family, people not-want job, american make facility, people send kid, people buy car, country lose job, small business hire people, people try job \\
            political debate &             speaker, time, gentleman, chairman, member, hous, would, vote, bill, colleagu &                                (sick/paid/unpaid) leave be guam, (sick/paid/unpaid) leave be columbia, (sick/paid/unpaid) leave be virgin , someone seek recognition, period offer appropriate, republican refuse action, people have commitment, democratic offer alternative, law authorize appointment, friend make comment \\
                    politics &                      senat, clerk, follow, pro, read, tempor, amend, presid, legisl, rule &                                                                 commonwealth perform duty, judiciary conduct promulgate regulation, air force indicate appointment, self-employed executes budget, commitment have witness, news sign text, party raise soft money, witness limit medicine, group run ad, witness give witness \\
             public services &             shall, secretari, fund, provid, program, act, section, state, appropri, avail &                                                         law authorize uniform, U.S.C. authorize service, application decrease appropriation, fund subsidize deficit, U.S.C. authorize uniform, fund make funding, congressional budget define money, fund carry funding, person perform abortion, government estimate contract \\
                     science & nation, develop, research, technolog, communiti, program, support, scienc, center, improv &                                                            commonwealth perform duty, student pursue degree (education), director award grant, agency participate program, (sick/paid/unpaid) leave be columbia, federally fund study, fund pay profit, company participate program, program award grant, program achieve goal \\
              social welfare &         drug, medicar, senior, prescript, benefit, cost, plan, program, beneficiari, year &                                                           oil account america, production shut supply and demand, medicare operate prescription drug, american rely oil, production use american, senior join H.M.O., senior have coverage, medicare have prescription drug, senior need prescription drug, senior pay premium \\
                      sports &                    univers, team, year, colleg, game, state, nation, wherea, award, first &                                                             people have exhaust, parent take child, fund provide service, small business employ workforce, individual perform service, defense prescribed regulation , people sign petition, budget cut education , american obtain healthcare, individual participate program \\
                      states &               water, project, river, flood, corp, construct, lake, feder, cost, louisiana &                                                                water resource development authorize river, person commit offense, university conduct study, law authorize project, project provide water, F.D.A. approve medicine, program have effect, person take healthcare, worker earn minimum wage, agency carry program \\
                       trade &            trade, u.s, countri, agreement, unit, state, foreign, export, market, american &                                                                                         administration negotiate trade, china join wto, export create job, trade create job, china manipulate dollar, american make life quality, agreement enters force, country have deficit, constitution give authority, bank provide loan \\
              transportation &            transport, highway, safeti, airport, system, state, passeng, vehicl, air, rail &                                                              grant fund program, amtrak provide service, good start provide service, project meet regulation , railroad provide service, railroad provide transportation, project receive funding, association give grade, carrier provide service, government receive funding \\
                         war &            iraq, war, troop, terrorist, american, attack, presid, militari, terror, iraqi &               iraqi  take responsibility, iraqi  take power, iraq pose jeopardy, iraq have weapon mass destruction, disability receive education , saddam hussein pose threat, iraq pose threat, saddam hussein use weapon mass destruction, saddam hussein use chemical (weapon), saddam hussein have weapon mass destruction \\
\bottomrule
\end{tabularx}
\bigskip

\end{table}

\clearpage

\section{Additional Results}  \label{app:additional-results}
\setcounter{table}{0}

\begin{figure}[ht!]
\begin{center}
\caption{Most Frequent Narratives Over Time} \label{fig:most_freq_narr_over_time}
\includegraphics[width = \textwidth]{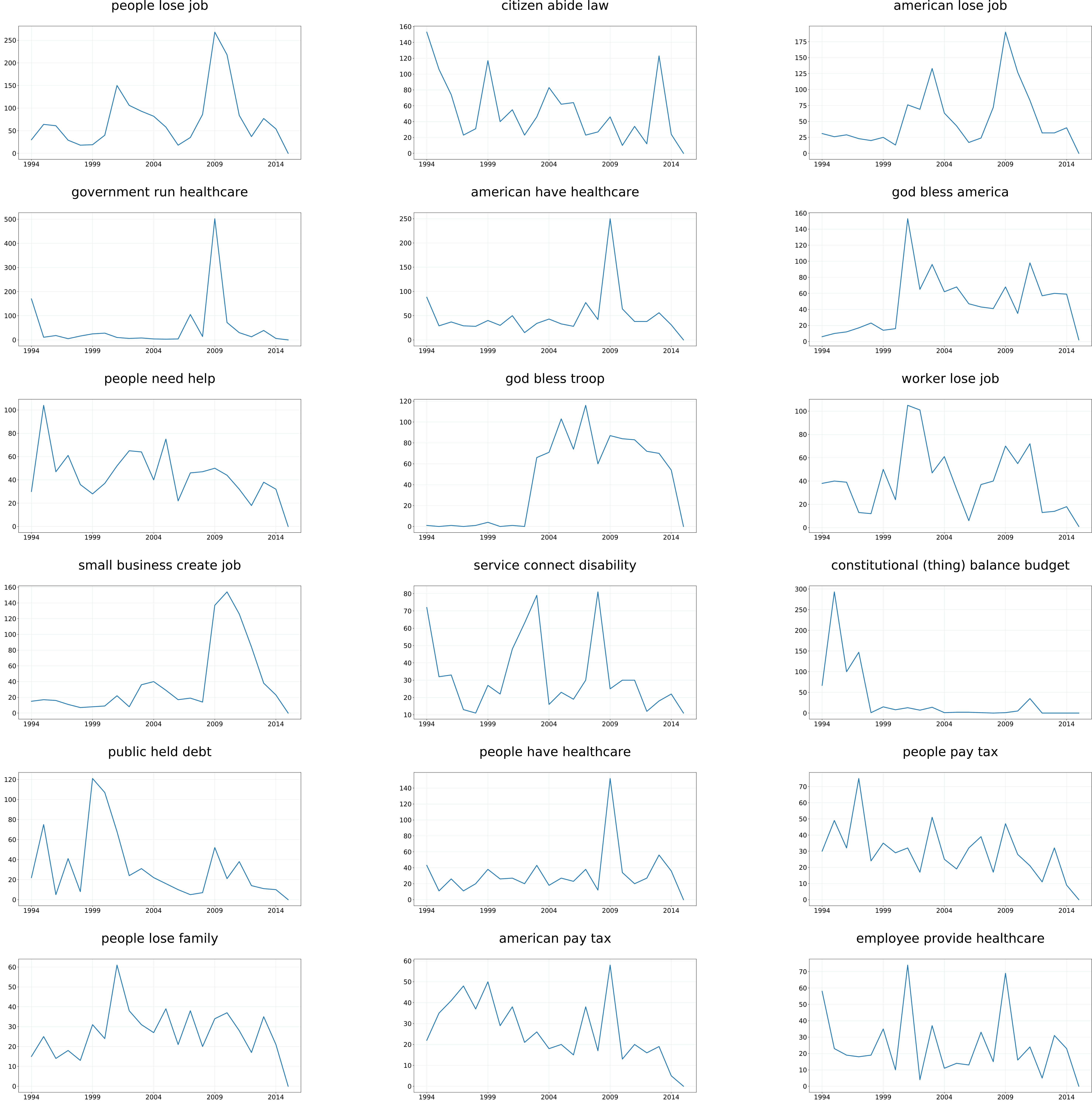}
\end{center}

\footnotesize
\flushleft
\textbf{Note:} In this figure, we plot the occurrences of the 18 most frequent narratives in the U.S. Congressional Record over the period 1994 to 2015. We focus on narratives pronounced at least 50 times and remove entities labeled as procedural or noise.
\end{figure}
\clearpage

\begin{table} 
\begin{center}
\caption{Most Frequent Narratives Per Entity} 
\label{tab:frequent_agent_narratives} 
\tiny
\renewcommand{\arraystretch}{1.1}
\begin{tabularx}{0.24\textwidth}{l} 
\hline 
family\\ 
\hline 
people lose family\\ 
american lose family\\ 
family lose love\\ 
men women put family\\ 
american give family\\ 
men women risk family\\ 
people give family\\ 
men women give family\\ 
god bless family\\ 
healthcare save family\\ 
\hline 
\end{tabularx} 
\begin{tabularx}{0.24\textwidth}{l} 
\hline 
money\\ 
\hline 
american keep money\\ 
people keep money\\ 
people make money\\ 
people get money\\ 
government take money\\ 
people have money\\ 
family keep money\\ 
people earn money\\ 
people put money\\ 
hard earn money\\ 
\hline 
\end{tabularx} 
\begin{tabularx}{0.24\textwidth}{l} 
\hline 
someone\\ 
\hline 
someone have opportunity\\ 
someone pay attention\\ 
someone have healthcare\\ 
someone do nothing\\ 
someone have idea\\ 
someone take healthcare\\ 
someone pay tax\\ 
someone do job\\ 
someone lose job\\ 
someone make money\\ 
\hline 
\end{tabularx} 
\begin{tabularx}{0.24\textwidth}{l} 
\hline 
agency\\ 
\hline 
agency provide service\\ 
agency administer program\\ 
agency furnish information \\ 
agency do job\\ 
agency receive grant\\ 
agency take action\\ 
agency provide funding\\ 
agency submit application\\ 
agency receive fund\\ 
agency participate program\\ 
\hline 
\end{tabularx} 

\bigskip 

\begin{tabularx}{0.24\textwidth}{l} 
\hline 
program\\ 
\hline 
program provide service\\ 
program provide funding\\ 
program provide healthcare\\ 
agency administer program\\ 
program provide grant\\ 
program make difference\\ 
program create job\\ 
student participate program\\ 
federally fund program\\ 
government run program\\ 
\hline 
\end{tabularx} 
\begin{tabularx}{0.24\textwidth}{l} 
\hline 
american\\ 
\hline 
american lose job\\ 
american have healthcare\\ 
american pay tax\\ 
american lose healthcare\\ 
american keep money\\ 
american lose family\\ 
american look job\\ 
american abide law\\ 
american not-have healthcare\\ 
american want job\\ 
\hline 
\end{tabularx} 
\begin{tabularx}{0.24\textwidth}{l} 
\hline 
child\\ 
\hline 
child have healthcare\\ 
child attend school\\ 
child get education \\ 
child receive education \\ 
child have education \\ 
child receive healthcare\\ 
child have opportunity\\ 
person have child\\ 
child need help\\ 
child not-have healthcare\\ 
\hline 
\end{tabularx} 
\begin{tabularx}{0.24\textwidth}{l} 
\hline 
service\\ 
\hline 
service connect disability\\ 
program provide service\\ 
healthcare provide service\\ 
U.S.C. authorize service\\ 
agency provide service\\ 
government provide service\\ 
community provide service\\ 
service provide service\\ 
company provide service\\ 
provider provide service\\ 
\hline 
\end{tabularx} 

\bigskip 

\begin{tabularx}{0.24\textwidth}{l} 
\hline 
person\\ 
\hline 
person have abortion\\ 
person serve country\\ 
person have child\\ 
person make difference\\ 
person do job\\ 
person earn dollar\\ 
person make decision\\ 
person have healthcare\\ 
person provide service\\ 
person make disaster\\ 
\hline 
\end{tabularx} 
\begin{tabularx}{0.24\textwidth}{l} 
\hline 
budget\\ 
\hline 
constitutional (thing) balance budget\\ 
budget balance budget\\ 
republican balance budget\\ 
budget raise tax\\ 
budget increase tax\\ 
budget do nothing\\ 
plan balance budget\\ 
budget cut program\\ 
government balance budget\\ 
budget reduce deficit\\ 
\hline 
\end{tabularx} 
\begin{tabularx}{0.24\textwidth}{l} 
\hline 
law\\ 
\hline 
citizen abide law\\ 
american abide law\\ 
law respect faith \\ 
people violate law\\ 
bush sign law\\ 
gun abide law\\ 
people break law\\ 
law prescribed witness\\ 
veteran administer law\\ 
law make appropriation\\ 
\hline 
\end{tabularx} 
\begin{tabularx}{0.24\textwidth}{l} 
\hline 
republican\\ 
\hline 
republican take power\\ 
republican join democrat\\ 
republican do nothing\\ 
republican balance budget\\ 
republican shut government\\ 
democrat join republican\\ 
republican cut medicare\\ 
republican put together\\ 
republican cut program\\ 
republican give tax break\\ 
\hline 
\end{tabularx} 

\bigskip 

\begin{tabularx}{0.24\textwidth}{l} 
\hline 
employee\\ 
\hline 
employee provide healthcare\\ 
employee sponsor healthcare\\ 
employee offer healthcare\\ 
employee drop coverage\\ 
employee provide coverage\\ 
employee take (paid/unpaid) leave\\ 
employee sponsor plan\\ 
employee sponsor coverage\\ 
employee provide service\\ 
employee hire worker\\ 
\hline 
\end{tabularx} 
\begin{tabularx}{0.24\textwidth}{l} 
\hline 
healthcare\\ 
\hline 
government run healthcare\\ 
american have healthcare\\ 
people have healthcare\\ 
employee provide healthcare\\ 
american lose healthcare\\ 
healthcare provide healthcare\\ 
healthcare provide service\\ 
child have healthcare\\ 
people lose healthcare\\ 
people get healthcare\\ 
\hline 
\end{tabularx} 
\begin{tabularx}{0.24\textwidth}{l} 
\hline 
business\\ 
\hline 
business create job\\ 
company do business\\ 
government sponsor business\\ 
company hold business\\ 
american do business\\ 
corporation  do business\\ 
government do business\\ 
people do business\\ 
business do business\\ 
business hire worker\\ 
\hline 
\end{tabularx} 
\begin{tabularx}{0.24\textwidth}{l} 
\hline 
people\\ 
\hline 
people lose job\\ 
people need help\\ 
people have healthcare\\ 
people pay tax\\ 
people lose family\\ 
people do job\\ 
people take opportunity\\ 
people make decision\\ 
people look job\\ 
people have concern\\ 
\hline 
\end{tabularx} 

\bigskip 

\begin{tabularx}{0.24\textwidth}{l} 
\hline 
government\\ 
\hline 
government run healthcare\\ 
government take healthcare\\ 
government provide service\\ 
government make decision\\ 
government take money\\ 
republican shut government\\ 
government sponsor business\\ 
government do nothing\\ 
government take property\\ 
government run program\\ 
\hline 
\end{tabularx} 
\begin{tabularx}{0.24\textwidth}{l} 
\hline 
job\\ 
\hline 
people lose job\\ 
american lose job\\ 
worker lose job\\ 
small business create job\\ 
people do job\\ 
people look job\\ 
business create job\\ 
american look job\\ 
people have job\\ 
people want job\\ 
\hline 
\end{tabularx} 
\begin{tabularx}{0.24\textwidth}{l} 
\hline 
tax\\ 
\hline 
people pay tax\\ 
american pay tax\\ 
budget raise tax\\ 
capital  gain tax\\ 
democrat raise tax\\ 
budget increase tax\\ 
tax kill job\\ 
family pay tax\\ 
someone pay tax\\ 
people not-pay tax\\ 
\hline 
\end{tabularx} 
\begin{tabularx}{0.24\textwidth}{l} 
\hline 
country\\ 
\hline 
men women serve country\\ 
person serve country\\ 
veteran serve country\\ 
problem face country\\ 
people serve country\\ 
challenge face country\\ 
people enter country\\ 
country face challenge\\ 
american serve country\\ 
country produce oil\\ 
\hline 
\end{tabularx} 
\end{center}

\bigskip

\footnotesize
\flushleft
\textbf{Note:} In this table, we plot the ten most frequent narratives for the 20 most frequent entities in the U.S. Congressional Record (1994-2015).

\end{table}

\clearpage

\section{Additional Material on Divisive Entities} \label{app:sec:more_on_partisan}
\setcounter{table}{0}

\subsection*{Partisan and Neutral Narratives}

\begin{figure}[ht!]
\begin{center}
\caption{Partisan Narratives in U.S. Congress}
\label{fig:PartisanNarratives}
\centering
\includegraphics[width = 0.5\textwidth]{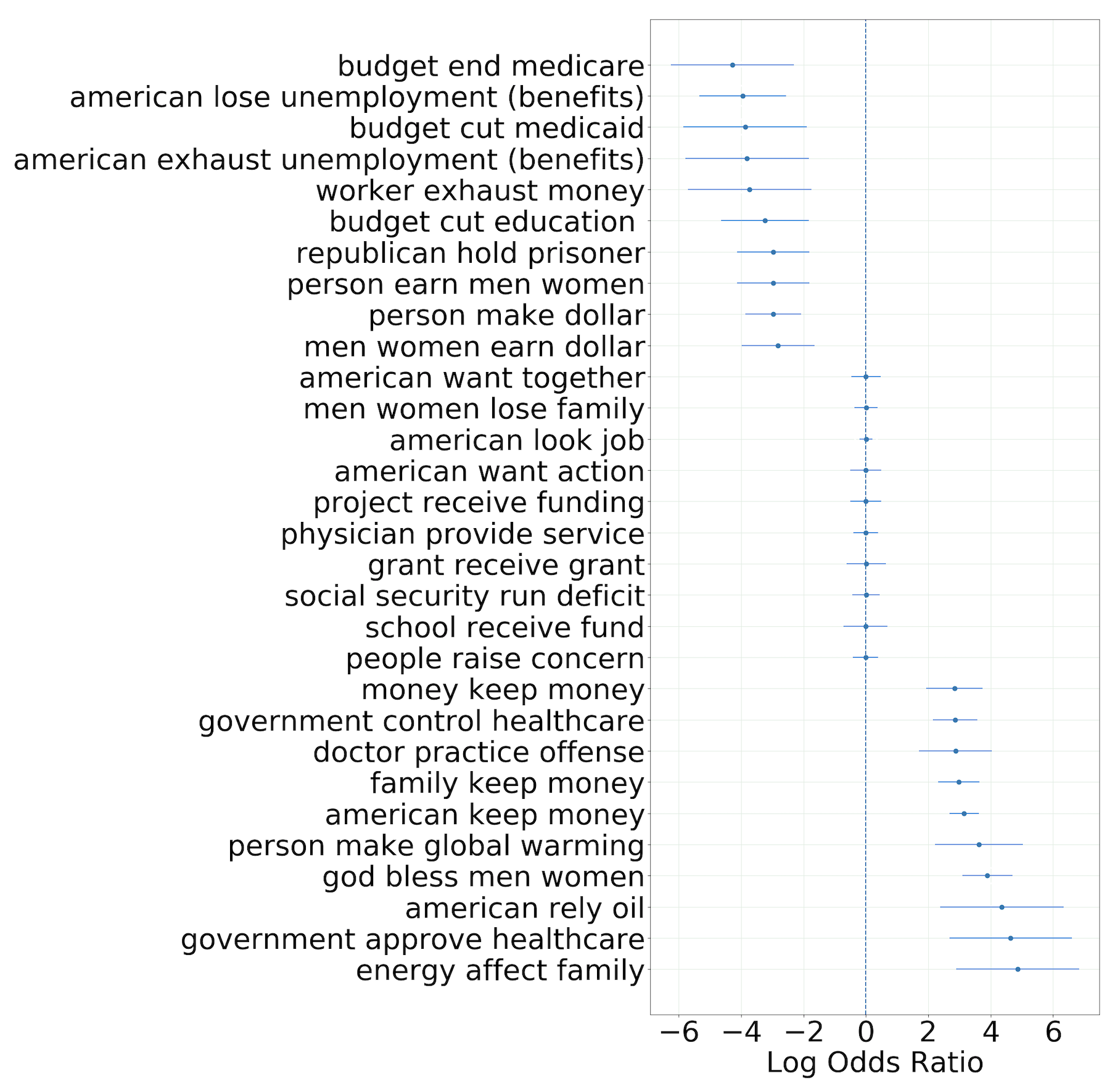}
\end{center}
\footnotesize
\vspace{0.2cm}
\textbf{Note:} This figure presents the most partisan narratives in the U.S. Congressional Record, as well as the least partisan narratives (i.e., with log-odds ratio closest to 0). For each narrative, we present the associated log-odds ratio and 95\% confidence interval. A high log-odds ratio reflects narratives pronounced more by Republicans (relative to Democrats), and vice versa for a low log-odds ratio. One of the least partisan narratives, ``men women lose family'', was pronounced most often in 2001 -- suggesting that his narrative captures a reaction to the 09/11 terrorist attacks.
\end{figure}

Figure \ref{fig:PartisanNarratives} shows again the most Republican and Democrat narratives, now along with the most neutral narratives.

\clearpage

\begin{table}[ht!]
    \caption{Longer, Unfiltered List of Divisive Entities}
    \label{tab:more-divisive-entities}
\begin{center}

\resizebox{.65\linewidth}{!}{%
\begin{tabular}{llcccccc}
\hline 
\multirow{2}{*}{Rank} & \multirow{2}{*}{Entity} & \multirow{2}{*}{Score} & \multicolumn{2}{c}{\underline{Log-Odds Ratio}} & \multicolumn{3}{c}{\underline{\# Narratives}}\tabularnewline
 &  &  & Narratives & Entity & Total & Dem & Repub \tabularnewline
\hline 
1 & filibuster & 1.93 & 2.10 & 0.16 & 2 & 1 & 1\tabularnewline
\textbf{2} & \textbf{oil} & \textbf{1.12} & \textbf{1.39} & \textbf{0.27} & \textbf{8} & \textbf{4} & \textbf{4}\tabularnewline
3 & H.M.O. & 0.96 & 1.05 & 0.10 & 2 & 1 & 1\tabularnewline
4 & remedy & 0.91 & 0.94 & 0.02 & 2 & 1 & 1\tabularnewline
5 & contract & 0.87 & 1.18 & 0.31 & 4 & 2 & 2\tabularnewline
6 & mortgage & 0.79 & 1.03 & 0.24 & 5 & 3 & 2\tabularnewline
\textbf{7} & \textbf{budget} & \textbf{0.76} & \textbf{0.82} & \textbf{0.06} & \textbf{42} & \textbf{17} & \textbf{25}\tabularnewline
8 & men women{*} & 0.75 & 0.86 & 0.10 & 21 & 10 & 11\tabularnewline
9 & workforce & 0.75 & 1.26 & 0.51 & 2 & 1 & 1\tabularnewline
10 & economy & 0.74 & 0.78 & 0.04 & 4 & 2 & 2\tabularnewline
11 & pregnancy & 0.73 & 1.07 & 0.34 & 3 & 2 & 1\tabularnewline
12 & medicaid & 0.73 & 2.43 & 1.71 & 2 & 2 & 0\tabularnewline
13 & facility & 0.71 & 0.92 & 0.21 & 3 & 2 & 1\tabularnewline
14 & policy & 0.70 & 0.83 & 0.14 & 2 & 1 & 1\tabularnewline
15 & gun & 0.69 & 0.85 & 0.16 & 6 & 4 & 2\tabularnewline
\textbf{16} & \textbf{healthcare} & \textbf{0.68} & \textbf{0.74} & \textbf{0.06} & \textbf{116} & \textbf{76} & \textbf{40}\tabularnewline
17 & environmental protection agency & 0.68 & 1.13 & 0.46 & 2 & 1 & 1\tabularnewline
\textbf{18} & \textbf{interest/rate} & \textbf{0.65} & \textbf{0.97} & \textbf{0.32} & \textbf{8} & \textbf{5} & \textbf{3}\tabularnewline
\textbf{19} & \textbf{loan} & \textbf{0.65} & \textbf{0.72} & \textbf{0.07} & \textbf{12} & \textbf{6} & \textbf{6}\tabularnewline
20 & center & 0.64 & 0.79 & 0.15 & 2 & 1 & 1\tabularnewline
21 & democrat{*} & 0.64 & 1.03 & 0.40 & 11 & 4 & 7\tabularnewline
22 & constitution{*} & 0.62 & 0.63 & 0.01 & 8 & 5 & 3\tabularnewline
\textbf{23} & \textbf{worker} & \textbf{0.61} & \textbf{1.23} & \textbf{0.62} & \textbf{16} & \textbf{11} & \textbf{5}\tabularnewline
24 & american{*} & 0.60 & 0.61 & 0.01 & 158 & 86 & 72\tabularnewline
25 & plant & 0.59 & 0.97 & 0.38 & 3 & 2 & 1\tabularnewline
\textbf{26} & \textbf{insurance} & \textbf{0.58} & \textbf{0.67} & \textbf{0.09} & \textbf{16} & \textbf{7} & \textbf{9}\tabularnewline
27 & trade & 0.57 & 0.74 & 0.17 & 2 & 1 & 1\tabularnewline
28 & action & 0.57 & 0.61 & 0.05 & 14 & 6 & 8\tabularnewline
29 & resource & 0.56 & 0.79 & 0.23 & 4 & 3 & 1\tabularnewline
30 & medicine & 0.55 & 0.90 & 0.35 & 3 & 1 & 2\tabularnewline
31 & priority & 0.54 & 0.93 & 0.39 & 3 & 2 & 1\tabularnewline
32 & community{*} & 0.53 & 0.60 & 0.07 & 14 & 10 & 4\tabularnewline
33 & bank & 0.52 & 0.88 & 0.35 & 6 & 4 & 2\tabularnewline
34 & abortion & 0.52 & 0.92 & 0.39 & 10 & 2 & 8\tabularnewline
35 & republican{*} & 0.52 & 1.12 & 0.60 & 40 & 31 & 9\tabularnewline
\textbf{36} & \textbf{medicare} & \textbf{0.52} & \textbf{0.83} & \textbf{0.31} & \textbf{18} & \textbf{12} & \textbf{6}\tabularnewline
\textbf{37} & \textbf{energy} & \textbf{0.52} & \textbf{1.30} & \textbf{0.78} & \textbf{7} & \textbf{3} & \textbf{4}\tabularnewline
\textbf{38} & \textbf{job} & \textbf{0.51} & \textbf{0.65} & \textbf{0.14} & \textbf{117} & \textbf{73} & \textbf{44}\tabularnewline
39 & nation{*} & 0.50 & 0.54 & 0.04 & 28 & 15 & 13\tabularnewline
40 & fund & 0.49 & 0.49 & 0.00 & 19 & 8 & 11\tabularnewline
41 & plan & 0.48 & 0.53 & 0.05 & 20 & 9 & 11\tabularnewline
42 & troop & 0.48 & 0.64 & 0.16 & 3 & 1 & 2\tabularnewline
43 & missile & 0.48 & 0.53 & 0.05 & 3 & 1 & 2\tabularnewline
44 & weapon mass destruction & 0.47 & 0.64 & 0.16 & 3 & 2 & 1\tabularnewline
45 & doctor & 0.47 & 0.79 & 0.31 & 16 & 6 & 10\tabularnewline
46 & company & 0.46 & 0.87 & 0.41 & 17 & 11 & 6\tabularnewline
47 & people{*} & 0.46 & 0.49 & 0.04 & 240 & 135 & 105\tabularnewline
48 & F.D.A. & 0.45 & 0.68 & 0.23 & 8 & 4 & 4\tabularnewline
49 & family{*} & 0.45 & 0.55 & 0.10 & 90 & 63 & 27\tabularnewline
50 & bad (problem) & 0.45 & 0.46 & 0.01 & 2 & 1 & 1\tabularnewline
\hline 
\end{tabular}
}
    \end{center}
\footnotesize
\flushleft
\textbf{Note}: This table shows the full list of most divisive narratives, as measured by the score in column 3, given by average log-odds ratio of the narratives where the entity appears (column 4), minus the log odds ratio of the entity itself (column 5). The other columns give the number of unique narratives (total, Democrat-slanted, and Republican-slanted) where the entity appears. Bold type indicates a policy narrative listed in Table \ref{tab:divisive-entities-main}. Asterisks indicate an identity/symbolic narrative listed in Table \ref{tab:divisive-entities-main}.
\end{table}

\begin{table}[ht!]
    \caption{List of Entities with Least Divisive Narratives}
    \label{tab:least-divisive-entities}
\begin{center}

\begin{tabular}{llcccccc}
\hline 
\multirow{2}{*}{Rank} & \multirow{2}{*}{Entity} & \multirow{2}{*}{Score} & \multicolumn{2}{c}{\underline{Log-Odds Ratio}} & \multicolumn{3}{c}{\underline{\# Narratives}}\tabularnewline
 &  &  & Narratives & Entity & Total & Dem & Repub\tabularnewline
\hline 
416 & population & -0.04 & 0.06 & 0.10 & 1 & 0 & 1\tabularnewline
417 & client & -0.04 & 0.06 & 0.10 & 1 & 0 & 1\tabularnewline
418 & (positive/negative) effect & -0.04 & 0.06 & 0.10 & 1 & 0 & 1\tabularnewline
419 & strong & -0.04 & 0.34 & 0.38 & 1 & 0 & 1\tabularnewline
420 & union & -0.04 & 0.02 & 0.06 & 1 & 0 & 1\tabularnewline
421 & natural resource & -0.04 & 0.02 & 0.06 & 1 & 0 & 1\tabularnewline
422 & shot & -0.04 & 0.06 & 0.10 & 1 & 0 & 1\tabularnewline
423 & patient & -0.04 & 0.23 & 0.27 & 7 & 0 & 7\tabularnewline
424 & reform & -0.04 & 0.81 & 0.85 & 2 & 0 & 2\tabularnewline
425 & opinion & -0.05 & 0.10 & 0.14 & 3 & 0 & 3\tabularnewline
426 & truth & -0.05 & 0.18 & 0.22 & 2 & 0 & 2\tabularnewline
427 & buy & -0.05 & 0.07 & 0.11 & 2 & 0 & 2\tabularnewline
428 & county & -0.05 & 0.17 & 0.22 & 3 & 0 & 3\tabularnewline
429 & consequence & -0.06 & 0.75 & 0.81 & 2 & 0 & 2\tabularnewline
430 & railroad & -0.06 & 0.46 & 0.52 & 2 & 2 & 0\tabularnewline
431 & air force & -0.06 & 0.73 & 0.79 & 2 & 1 & 1\tabularnewline
432 & founder & -0.07 & 0.52 & 0.59 & 3 & 1 & 2\tabularnewline
433 & god & -0.08 & 1.04 & 1.12 & 9 & 1 & 8\tabularnewline
434 & border & -0.08 & 0.66 & 0.73 & 2 & 1 & 1\tabularnewline
435 & profit & -0.10 & 1.10 & 1.20 & 8 & 8 & 0\tabularnewline
436 & adult & -0.12 & 0.37 & 0.49 & 4 & 4 & 0\tabularnewline
437 & force & -0.15 & 0.27 & 0.42 & 2 & 0 & 2\tabularnewline
438 & congressional & -0.15 & 0.75 & 0.90 & 2 & 0 & 2\tabularnewline
439 & (sick/paid/unpaid) leave & -0.17 & 1.32 & 1.49 & 2 & 0 & 2\tabularnewline
440 & U.S.C. & -0.18 & 1.37 & 1.55 & 2 & 0 & 2\tabularnewline
441 & freedom & -0.19 & 0.58 & 0.77 & 4 & 0 & 4\tabularnewline
442 & legislative (election) & -0.19 & 0.58 & 0.77 & 2 & 0 & 2\tabularnewline
443 & word & -0.24 & 0.53 & 0.77 & 3 & 2 & 1\tabularnewline
444 & winners and losers & -0.53 & 0.81 & 1.33 & 2 & 0 & 2\tabularnewline
445 & appointment & -0.57 & 1.44 & 2.01 & 2 & 1 & 1\tabularnewline
\hline 
\end{tabular}
    \end{center}
\footnotesize
\flushleft
\textbf{Note:} This table shows the least divisive narratives, as measured by the score in column 3, given by average log-odds ratio of the narratives where the entity appears (column 4), minus the log odds ratio of the entity itself (column 5). The other columns give the number of unique narratives (total, Democrat-slanted, and Republican-slanted) where the entity appears.
\end{table}

\clearpage

\subsection*{Distinctive Narrative Graphs for Democrats and Republicans}

We present narrative graphs for mostly Republican (Democrat) narratives as measured by log-odds ratios.

\begin{figure}[ht!]
\begin{center}
\caption{Partisan Worldviews -- Democrats}
\label{fig:PartisanWorldviews_Dem}
\includegraphics[width = \textwidth]{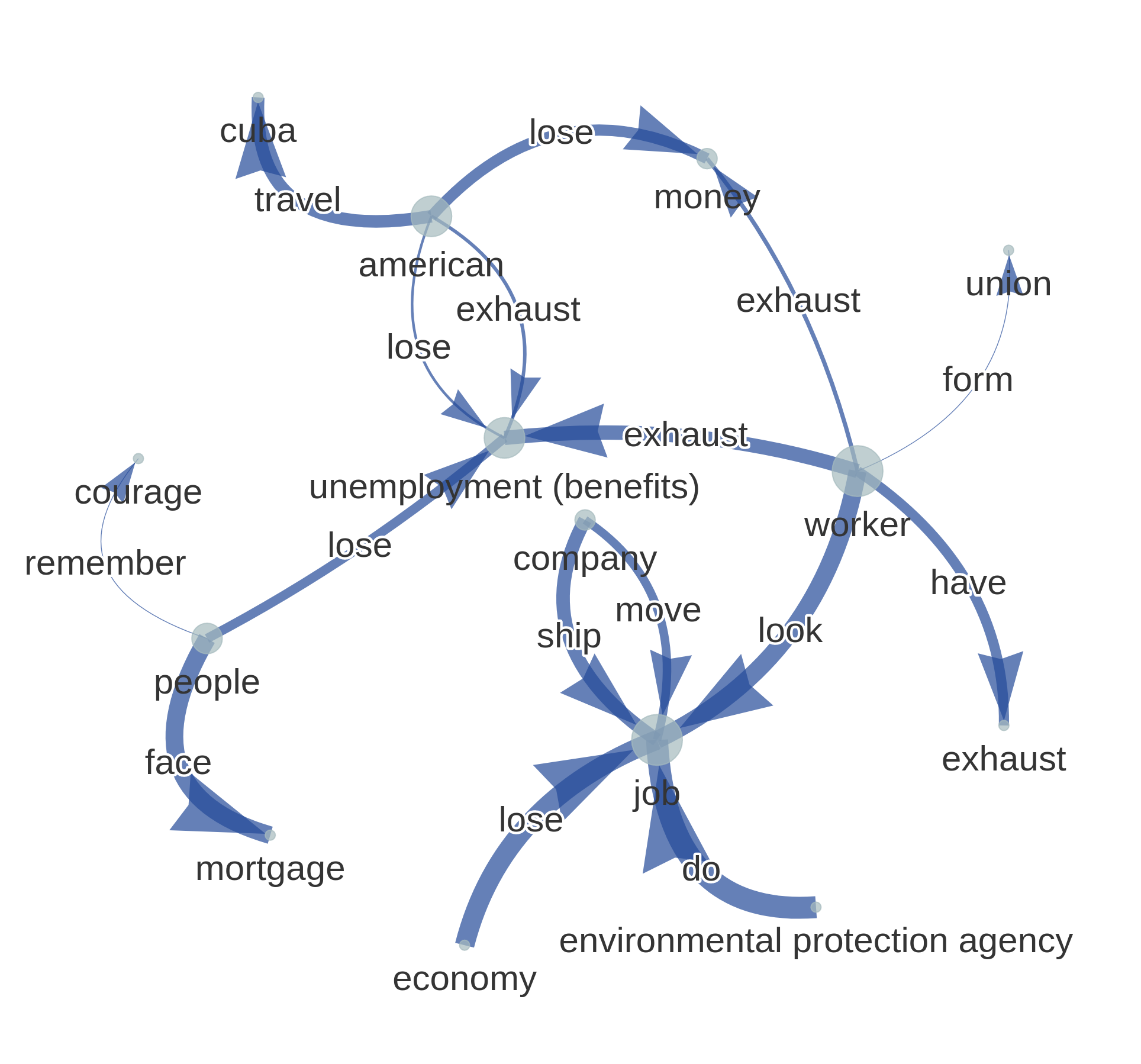}
\end{center}
\footnotesize
\flushleft
\textbf{Note:} This figure presents the 50 most salient Democrat narratives in the U.S. Congressional Record as measured by a log-odds ratio (i.e., whether a narrative is more pronounced by Democrats relative to Republicans). We represent our narrative tuples in a directed multigraph, in which the nodes are entities and the edges are verbs. The network is pruned so as to plot the largest connected subgraph (for this reason, the graph may display less than 50 narratives). The size of the edges is determined by the log odds ratio. The size of nodes is determined by their degree in the network.
\end{figure}

\begin{figure}[ht!]
\begin{center}
\caption{Partisan Worldviews -- Republicans}
\label{fig:PartisanWorldviews_Rep}
\includegraphics[width = \textwidth]{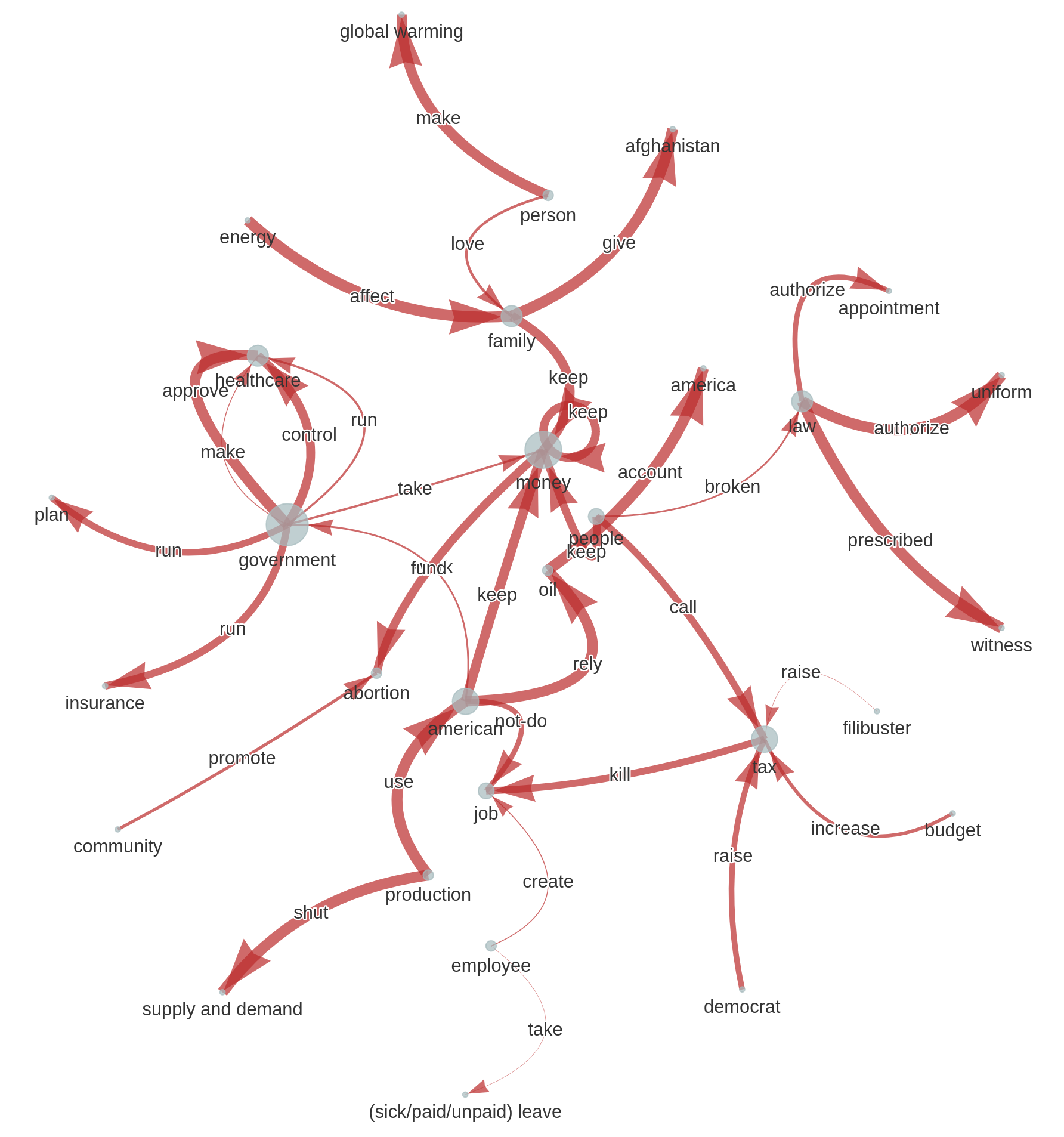}
\end{center}
\footnotesize
\flushleft
\textbf{Note:} This figure presents the 50 most salient Republican narratives in the U.S. Congressional Record as measured by a log-odds ratio (i.e., whether a narrative is more pronounced by Republicans relative to Democrats). We represent our narrative tuples in a directed multigraph, in which the nodes are entities and the edges are verbs. The network is pruned so as to plot the largest connected subgraph (for this reason, the graph may display less than 50 narratives). The size of the edges is determined by the log odds ratio. The size of nodes is determined by their degree in the network.
\end{figure}

\clearpage

\section{Which narratives co-occur together?} \label{app:narrative_cooccurrence}
\setcounter{table}{0}

In many cases, agent-verb-patient tuples are building blocks of a broader political story. In Figure \ref{fig:TerrorNarratives}, for example, we show that two prevalent narratives related to the war in Iraq are: ``Saddam Hussein has weapons of mass destruction'' and ``Saddam Hussein poses a threat''. A more comprehensive story might read: ``Saddam Hussein has weapons of mass destruction and poses a threat to national security. With a military intervention, the United States can make a difference, but at the cost of American lives.'' Similarly, ``people lose job'' is the most common narrative in the corpus, but to which political stories does it belong to? Can we identify narratives that co-occur together to eventually reconstruct these broader political stories? As a first step in this direction, we build on two common approaches in the text-as-data literature below.

\paragraph{Pointwise Mutual Information.}

A simple approach to recover co-occuring narratives is to compute the pointwise mutual information (PMI) between pairs of narratives. For two narratives $n_i$ and $n_j$, the pointwise mutual information is defined as:
\begin{equation}
    PMI(n_i,n_j) = \displaystyle \log\big(\frac{P(n_i \cap n_j)}{P(n_i) P(n_j)}\big)
\end{equation}
If two narratives are independent from one another, than their PMI is equal to zero. The larger the PMI, the more likely two narratives are to co-occur together, adjusting for their overall frequency. To analyze the odds of two narratives co-occurring in the same speech, we compute PMIs for pairs of narratives in the U.S. Congressional Record. 

\paragraph{Narrative Embedding.}

\begin{figure}[ht!]
\caption{Geometry of Embedded Narratives}
\label{fig:narrative_embedding}
\begin{center}
\includegraphics[width = \textwidth]{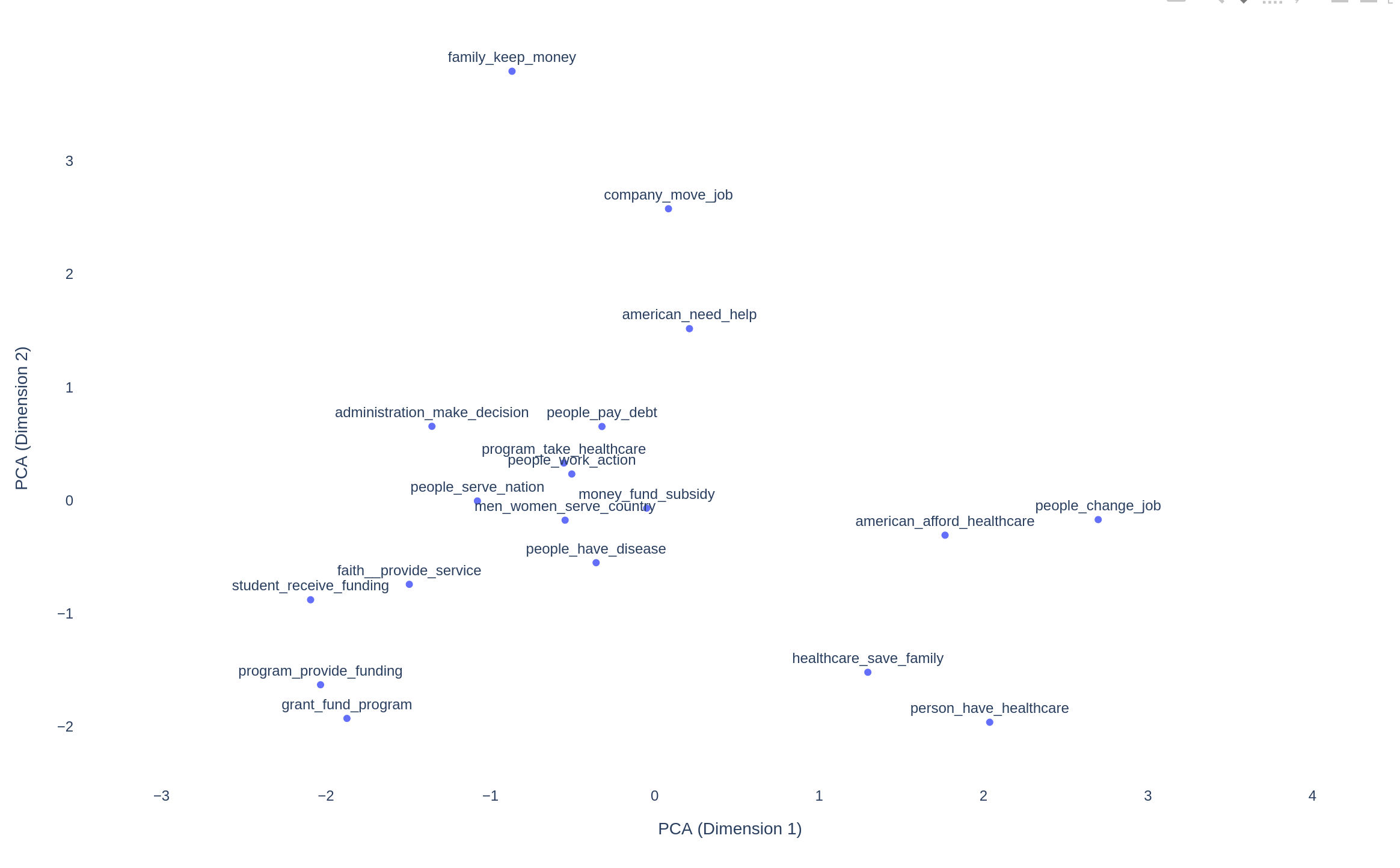}
\end{center}
\footnotesize
\flushleft
\textbf{Note:} This figure shows 20 random narratives in the embedding, projected to two dimensions using PCA.
\end{figure}

A second approach is to train skip-gram embeddings for narratives. As done with word embeddings \citep{MikolovSutskeverChenCorradoDean2013}, this approach locates narratives that appear in similar contexts close to one another in a low-dimensional, dense vectorial space. For the U.S. Congressional Record, we define speeches as documents, narratives as tokens, and then train a Word2Vec model (skip gram with negative sampling) over ten epochs. The window size is set to an arbitrarily large number so that two narratives co-occurring in a speech are always in their respective windows. The dimension of vectors is set to 50. To graphically represent the resulting embedding, Figure \ref{fig:narrative_embedding} plots a random sample of 20 narratives along the two first dimensions of a Principal Component Analysis (PCA), showing how the geometry represents intuitive topical information. Hence, the relationship between two narratives can be computed as the cosine similarity between their respective vectors.

\paragraph{Results.}

Table \ref{tab:coocs} applies these methods to the narrative ``Saddam Hussein poses a threat''. We provide the 20 most associated narratives with our PMI metric as well as the cosine similarity of the narrative embedding vectors. Overall, both methods paint a similar picture. Saddam Hussein posed a threat to national security because of (supposedly possessed) nuclear weapons, the United States won the war but American troops had to make a sacrifice. In fact, Saddam Hussein did not possess nuclear weapons. The administration made a mistake and misled Americans. 

\begin{table}[!ht]
\centering
\caption{Co-occurring Narratives -- War on Terror}
\resizebox{0.9\textwidth}{!}{%
\begin{tabular}{ll}
\toprule
\textbf{Pointwise Mutual Information} & \textbf{Narrative Embedding} \\ 
\midrule
                           iraq pose threat &                            iraq pose threat \\
                         iraq pose jeopardy &        saddam hussein use chemical (weapon) \\
          iraq have weapon mass destruction &                saddam hussein invade kuwait \\
       saddam hussein use chemical (weapon) &                           nation love peace \\
saddam hussein have weapon mass destruction &                       terrorist pose threat \\
               saddam hussein invade kuwait &                           american lose job \\
 saddam hussein use weapon mass destruction &                         people love freedom \\
                          freedom take tree &                             god bless troop \\
            administration mislead american &                            father serve war \\
                 iraqi  take responsibility & saddam hussein have weapon mass destruction \\
                          nation love peace &                        people share concern \\
                       american risk family &                            people need help \\
                          iraqi  take power &                            news face nation \\
                administration make mistake &                          iraq pose jeopardy \\
                        nation love freedom &                          nation take action \\
                      terrorist pose threat &                        men women put family \\
                        people love freedom &                          girl attend school \\
                     men women wear uniform &                    men women make sacrifice \\
                            america win war &                        american hold market \\
                           war torn country &                        person serve country \\
\bottomrule
\end{tabular}
}
\footnotesize
\flushleft
\textbf{Note:} This table presents narratives that tend to co-occur with ``Saddam Hussein poses a threat''. Column (1) lists the 20 narratives with the highest PMI associated to this narrative. Column (2) lists the 20 narratives that are the closest to this narrative in the embedding (as measured by cosine similarity).
\label{tab:coocs}
\end{table}

\clearpage
\section{Narratives on Social Media Data} \label{app:trump_tweet_archive}
\setcounter{table}{0}

Though we illustrate \textsc{relatio}'s workings on the U.S. Congressional Record, the package is not designed to be corpus-specific. In this section, we provide a short analysis of a Twitter corpus.\footnote{Researchers not involved in the development of \textsc{relatio} have applied it to newspaper articles \citep{ottonello2022financial} as well as social media data from different platforms \citep{sipka2021comparing}.} We focus on 32,323 tweets from the \href{https://www.thetrumparchive.com/}{Trump Tweet Archive}. The tweets were written by Donald Trump between August 2011 and October 2020 and amount to 68,616 sentences. 

We clean the text the same way as done for the congressional speeches, with the exception of stopwords, taken directly from the spaCy english model ``en\_web\_core\_md''. We keep the 100 most frequent named entities and specify 100 additional clusters for the embeddings. We then count the resulting narrative tuples.

\begin{figure}
\centering
\caption{Top 50 Trump Narratives}
\includegraphics[width = \textwidth]{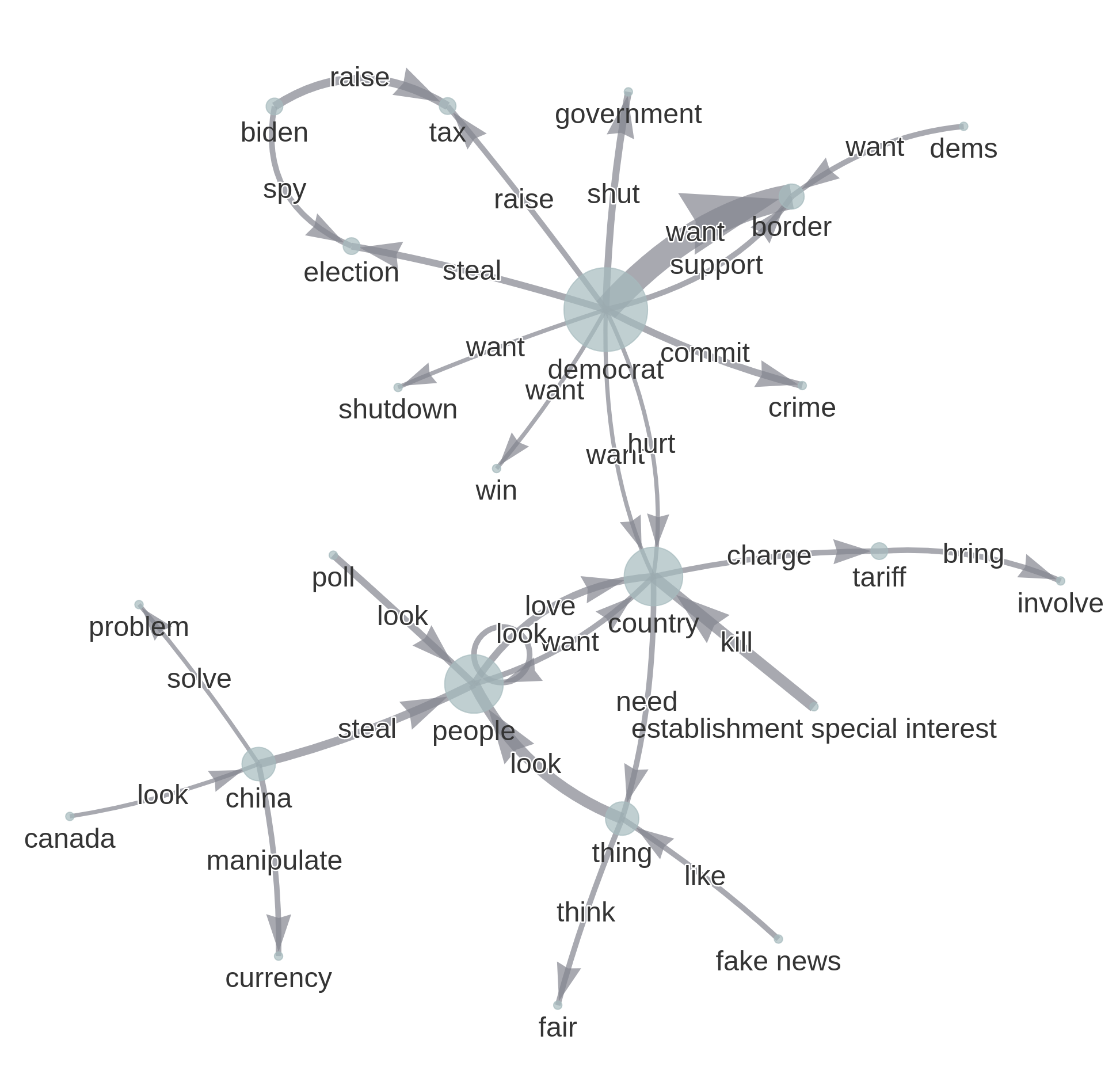}
\label{fig:trump_narratives}
\flushleft
\footnotesize
\textbf{Note:} This figure presents the 50 most frequent narratives from the Trump Tweet Archive. We represent our narrative tuples in a directed multigraph, in which the nodes are entities and the edges are verbs. The network is pruned so as to plot the largest connected subgraph (for this reason, the graph may display less than 50 narratives). The size of the edges is determined by the narrative's frequency. The size of nodes is determined by their degree in the network.
\end{figure}

Figure \ref{fig:trump_narratives} represents Trump's 50 most frequent narratives in the form of a narrative graph. The former president notably claims that ``The Democrats hurt the country'', ``steal the election'' and ``raise taxes''. ``China manipulates its currency'' and ``steals from the people'', while ``the establishment and special interests kill the country''.

The corpus is much smaller than the U.S. Congressional Record. It takes just a few minutes to run the full \textsc{relatio} pipeline on it. Interested readers will find a step-by-step tutorial to extract narratives from this corpus on our \href{https://github.com/relatio-nlp/relatio}{GiHhub repository}. 

\clearpage
\section{List of Analyzed Narratives} \label{sec:FinalNarratives}
\setcounter{table}{0}

Here we list all narratives mined from the Congressional Record corpus. The list includes dimension-reduced narratives appearing at least 50 times in the period 1994-2015. Narratives are sorted by most to least frequent. \\

\noindent people lose job, 
citizen abide law, 
american lose job, 
government run healthcare, 
american have healthcare, 
god bless america, 
people need help, 
god bless troop, 
worker lose job, 
small business create job, 
service connect disability, 
constitutional (thing) balance budget, 
people have healthcare, 
public held debt, 
people pay tax, 
people lose family, 
american pay tax, 
employee provide healthcare, 
people do job, 
men women serve country, 
people take opportunity, 
mortgage back military, 
american lose healthcare, 
american keep money, 
american lose family, 
program provide service, 
people make decision, 
people look job, 
healthcare provide healthcare, 
business create job, 
people have concern, 
american look job, 
family lose love, 
program provide funding, 
healthcare provide service, 
people have job, 
child have healthcare, 
people lose healthcare, 
company do business, 
people have opportunity, 
people want job, 
people keep money, 
american abide law, 
people get healthcare, 
U.S.C. authorize service, 
F.D.A. approve drug, 
people have insurance, 
agency provide service, 
american not-have healthcare, 
child attend school, 
period offer appropriate, 
program provide healthcare, 
government take healthcare, 
person have abortion, 
student attend school, 
republican take power, 
people get job, 
amount bear ratio, 
people not-have healthcare, 
defense submit congressional, 
people love freedom, 
people take healthcare, 
american want job, 
person serve country, 
american work hard, 
someone have opportunity, 
veteran receive healthcare, 
men women put family, 
people create job, 
someone pay attention, 
someone have healthcare, 
men women serve nation, 
people make money, 
government provide service, 
veteran serve country, 
god bless men women, 
someone do nothing, 
doctor make decision, 
republican join democrat, 
problem face country, 
agency administer program, 
agency furnish information , 
people find job, 
oil make profit, 
american give family, 
people commit crime, 
doctor take healthcare, 
law respect faith , 
child get education , 
program provide grant, 
republican do nothing, 
child receive education , 
community provide service, 
budget balance budget, 
american get healthcare, 
democrat take power, 
government make decision, 
american have concern, 
employee sponsor healthcare, 
administration take action, 
american pay attention, 
child have education , 
project create job, 
challenge face nation, 
people pay attention, 
men women risk family, 
service provide service, 
people have idea, 
people serve country, 
constitutional (thing) require balance budget, 
american need help, 
iraq pose threat, 
agency do job, 
american do job, 
program make difference, 
insurance deny coverage, 
american find job, 
challenge face country, 
economy create job, 
coal fire plant, 
program create job, 
student participate program, 
people get money, 
people have problem, 
government take money, 
people need healthcare, 
people give family, 
men women give family, 
republican balance budget, 
republican shut government, 
bank make loan, 
god bless family, 
law enforcement add category, 
budget raise tax, 
government sponsor business, 
american take opportunity, 
problem face nation, 
federally fund program, 
federally fund study, 
federally recognize tribe, 
child receive healthcare, 
saddam hussein pose threat, 
healthcare save family, 
small business provide healthcare, 
people have money, 
family make sacrifice, 
people make choise, 
capital  gain tax, 
people cross border, 
democrat raise tax, 
america have healthcare, 
child have opportunity, 
employee offer healthcare, 
commonwealth perform duty, 
american make sacrifice, 
director mean director, 
god continue america, 
american have opportunity, 
men women wear uniform, 
government do nothing, 
democrat join republican, 
company provide service, 
company hold business, 
person have child, 
family need help, 
budget increase tax, 
missile propel bomb, 
someone have idea, 
agency receive grant, 
people enter country, 
american want healthcare, 
family keep money, 
doctor provide healthcare, 
individual not-attain adult, 
legislative (election) have consequence, 
child need help, 
public create job, 
budget do nothing, 
country face challenge, 
government take property, 
judiciary conduct markup, 
agency take action, 
healthcare offer healthcare, 
family receive funding, 
administration do nothing, 
american create job, 
veteran make sacrifice, 
american do business, 
republican cut medicare, 
child not-have healthcare, 
people risk family, 
person make difference, 
american serve country, 
someone take healthcare, 
men women make sacrifice, 
tax kill job, 
tide lift ship, 
provider provide service, 
plan balance budget, 
america lead world, 
people have choise, 
american receive healthcare, 
family pay tax, 
government run program, 
american have job, 
people buy healthcare, 
program award grant, 
people earn money, 
people buy insurance, 
government take action, 
someone pay tax, 
people not-pay tax, 
company make profit, 
kid get education , 
people put family, 
senior have healthcare, 
corporation  pay tax, 
attorney (judge) make decision, 
corporation  do business, 
center provide service, 
person do job, 
student graduate school, 
nation face challenge, 
senior have prescription drug, 
people violate law, 
government do everything, 
american lack healthcare, 
agreement enters force, 
school provide education , 
government do business, 
government control healthcare, 
doctor provide service, 
person earn dollar, 
raw material  meet road, 
energy create job, 
people not-have insurance, 
people express concern, 
lender make loan, 
job pay life quality, 
employee drop coverage, 
brave men women give family, 
company create job, 
employee provide coverage, 
healthcare offering healthcare, 
people have family, 
parent raise child, 
people buy car, 
parent send child, 
employee sponsor plan, 
person make decision, 
employee take (sick/paid/unpaid) leave, 
service connect veteran, 
program save family, 
child need healthcare, 
agency provide funding, 
people do business, 
american make choise, 
iraqi  take responsibility, 
founder create job, 
people have child, 
american lose unemployment (benefits), 
staff do job, 
people put money, 
american need job, 
nation produce oil, 
director considers appropriate, 
saddam hussein have weapon mass destruction, 
bush sign law, 
crime add category, 
person have healthcare, 
budget cut program, 
family have healthcare, 
company move job, 
business do business, 
physician provide service, 
people work job, 
program receive funding, 
government do job, 
people take risk, 
people lose insurance, 
stimulus package create job, 
country produce oil, 
american know truth, 
defense waive limit, 
people not-have job, 
american make energy, 
someone do job, 
hard earn money, 
people receive money, 
people need job, 
mother have child, 
program save money, 
program help american, 
gun abide law, 
men women serve military, 
people get education , 
senior pay premium, 
(sick/paid/unpaid) leave be guam, 
doctor perform abortion, 
american risk family, 
men women lose family, 
republican put together, 
defense notify congressional, 
individual receive service, 
terrorist kill american, 
child have diabetes, 
people provide service, 
outlay not-exceed payment, 
people have coverage, 
american make decision, 
american suffer diabetes, 
plan do nothing, 
american lose coverage, 
program provide education , 
people break law, 
veteran need healthcare, 
child review study, 
agency submit application, 
senior have choise, 
person provide service, 
american not-do job, 
people make sacrifice, 
people put together, 
american have choise, 
people have skill, 
individual purchase healthcare, 
iran develop nuclear (weapon), 
people need funding, 
employee sponsor coverage, 
god wipe shot, 
person make disaster, 
american need healthcare, 
child participate program, 
law prescribed witness, 
people participate program, 
people graduate school, 
people not-find job, 
government pick winners and losers, 
nation owes appreciation, 
democrat want tax, 
god grant strong, 
american make consumer, 
F.D.A. approve prescription drug, 
senior get prescription drug, 
republican cut program, 
people save money, 
energy affect family, 
person give family, 
government balance budget, 
fund create job, 
program help people, 
american need understand, 
brave men women risk family, 
republican give tax break, 
country support terrorist, 
people make mistake, 
person make dollar, 
agency receive fund, 
republican cut tax, 
country face problem, 
mother raise child, 
child reach future, 
family work hard, 
doctor practice healthcare, 
family send child, 
american sent message, 
program provide food, 
voter cast ballot, 
family take healthcare, 
individual have healthcare, 
people have income, 
employee provide service, 
american travel cuba, 
someone lose job, 
program provide money, 
person lose family, 
church offer prayer, 
people lose money, 
person dedicate family, 
american need (tax) relief, 
school hire student, 
people take money, 
people need service, 
american pay income, 
employee hire worker, 
administration make decision, 
people give money, 
employee drop healthcare, 
fuel drive economy, 
budget reduce deficit, 
grant submit application, 
people make living, 
individual participate program, 
veteran administer law, 
people make contribution, 
employee lose job, 
supreme make decision, 
story detail fight, 
company take opportunity, 
carrier provide service, 
people raise concern, 
constitution require balance budget, 
individual perform service, 
someone make money, 
individual provide service, 
american have diabetes, 
budget increase deficit, 
school participate program, 
vacancy not-affect power, 
employee pay minimum wage, 
people make difference, 
people drive car, 
program meet regulation , 
men women do job, 
person seek abortion, 
child receive service, 
customer sell gun, 
senior need help, 
people work minimum wage, 
people pay money, 
people get insurance, 
american have insurance, 
regulation  kill job, 
family have child, 
people suffer illness, 
people need understand, 
agency participate program, 
business hire worker, 
child get healthcare, 
action have consequence, 
student need help, 
veteran receive service, 
someone get money, 
drug save family, 
veteran receive money, 
person give baby (boomer), 
agency use fund, 
child lose parent, 
citizen pay tax, 
administration request level (spending), 
employee hire immigrant, 
someone seek recognition, 
republican play politics, 
small business offer healthcare, 
government approve healthcare, 
person have corrective action, 
american put food, 
storm cause damage, 
child witness violence, 
people do nothing, 
doctor deliver baby (boomer), 
american hold market, 
people call tax, 
student reduce class size, 
individual receive money, 
person have cancer, 
iran have nuclear (weapon), 
people work living, 
medicare provide healthcare, 
america face challenge, 
student receive grant, 
american pay money, 
american balance budget, 
life quality pay job, 
money pick deductible, 
expenditure not-exceed payment, 
mother take healthcare, 
car get gasoline, 
plan create job, 
american suffer illness, 
republican put plan, 
word have word, 
grant award grant, 
iran acquire nuclear (weapon), 
american want balance budget, 
budget cut deficit, 
policy create job, 
government fund program, 
car carry passenger, 
individual receive funding, 
senior have coverage, 
republican have idea, 
illness threaten family, 
person make contribution, 
F.D.A. regulate tobacco, 
american pay gasoline, 
people have diabetes, 
economy lose job, 
someone know nothing, 
person serve nation, 
country receive funding, 
government raise tax, 
people pay premium, 
people take grant, 
attorney (judge) do job, 
employee provide insurance, 
neighbor help neighbor, 
american make contribution, 
employee participate program, 
republican have plan, 
agency issue regulation , 
mother have diabetes, 
administration make mistake, 
american get tax, 
agency receive funding, 
firefighter lose family, 
medicare operate prescription drug, 
person have pregnancy, 
people have experience, 
people work family, 
pregnancy feel pain, 
plan meet regulation , 
people have education , 
american have government, 
government borrow money, 
judiciary recommend substitute, 
money pay tax, 
republican propose tax, 
physician take healthcare, 
freedom take tree, 
agency have authority, 
republican propose budget, 
attorney (judge) have (legal) statute, 
federally fund center, 
word do justice, 
student take loan, 
american buy healthcare, 
people work government, 
government solve problem, 
program serve child, 
people not-have opportunity, 
family give afghanistan, 
american want government, 
money save money, 
budget cut tax, 
people make dollar, 
student receive education , 
individual attain adult, 
terrorist pose threat, 
consumer have choise, 
grant make grant, 
brave men women put family, 
person serve military, 
american not-find job, 
program help child, 
challenge face america, 
school provide service, 
employee hire illegal immigrant, 
family raise child, 
country need energy, 
family send kid, 
american tighten belt, 
group run ad, 
program require appropriation, 
veteran control small business, 
people start business, 
company make money, 
people make fund, 
someone pay fair, 
child have future, 
tax create job, 
american lose money, 
friend make comment, 
F.D.A. approve consumer, 
government not-create job, 
law make appropriation, 
people work hard, 
people earn minimum wage, 
senior pay pocket, 
people need money, 
people remember courage, 
child suffer diabetes, 
someone take opportunity, 
american get job, 
people feel life quality, 
attorney (judge) follow law, 
appropriation do job, 
democrat do nothing, 
corporation  make profit, 
someone make mistake, 
someone have insurance, 
hard earn dollar, 
agency have (legal) statute, 
bank issue card, 
government provide healthcare, 
money keep money, 
people want government, 
american make facility, 
people have illness, 
veteran get healthcare, 
employee do job, 
veteran have service, 
people make minimum wage, 
homeowner face mortgage, 
brave men women serve country, 
person perform abortion, 
program provide loan, 
program improve healthcare, 
parent make decision, 
nation face threat, 
agency receives grant, 
people lose coverage, 
people have program, 
student have education , 
school need help, 
american afford healthcare, 
defense prescribed regulation , 
republican play game, 
employee make contribution, 
disease threaten family, 
troop do job, 
american have coverage, 
crisis face country, 
constitution give authority, 
fund subsidize deficit, 
american take healthcare, 
american have voice, 
witness give witness, 
budget end medicare, 
people lose unemployment (benefits), 
person commit crime, 
people take job, 
worker join union, 
government create job, 
senior need prescription drug, 
authority provide funding, 
kid have healthcare, 
republican want medicare, 
H.M.O. make decision, 
program have effect, 
farmer need help, 
people love peace, 
mother have cancer, 
grant provide funding, 
people suffer diabetes, 
director award grant, 
tax benefit wealthy, 
people invest money, 
word give education , 
republican do job, 
country face crisis, 
people abide law, 
decision affect family, 
brave men women lose family, 
someone want job, 
appropriation make grant, 
government run deficit, 
business employ people, 
agency provide information , 
budget have deficit, 
someone have problem, 
person commit (legal) charge, 
agency serve school, 
people change job, 
american save money, 
person love family, 
worker form union, 
association give grade, 
community base healthcare, 
someone make decision, 
child enter school, 
men women earn dollar, 
animal lay milk, 
veteran suffer diabetes, 
immigrant enter country, 
people raise family, 
child graduate school, 
people have voice, 
corporation  not-pay tax, 
child have disability, 
social security balance budget, 
american have money, 
someone get job, 
administration impose sanction, 
people filing bankruptcy, 
nothing create job, 
american deserve truth, 
agency provide technical assistance, 
injury fire missile, 
soldier lose family, 
democrat put together, 
government provide funding, 
hard work american, 
constitution give power, 
social security run deficit, 
fund provide service, 
student take opportunity, 
grant fund project, 
law authorize compensation, 
law enforcement do job, 
service administer program, 
father serve war, 
food drug administration regulate tobacco, 
problem face american, 
iraq have weapon mass destruction, 
employee take opportunity, 
action have effect, 
people make future, 
grant use grant, 
people want healthcare, 
attorney (judge) interpret law, 
community promote abortion, 
child have family, 
family lose job, 
crime get gun, 
law authorize uniform, 
worker find job, 
doctor treat medicare, 
insurance raise interest/rate, 
obligation bearing interest/rate, 
agency enter contract, 
person get corrective action, 
people afford healthcare, 
doctor take medicare, 
people get help, 
employee pay premium, 
nation face crisis, 
people serve nation, 
doctor treat patient, 
money justify money, 
people have faith , 
business provide service, 
small business take opportunity, 
american earn money, 
company hold loan, 
project receive funding, 
american take job, 
senior take opportunity, 
military do job, 
someone loses job, 
corrective action save family, 
american get money, 
american put family, 
business hire people, 
people do best, 
service recommend substitute, 
people serve war, 
republican hold prisoner, 
attorney (judge) apply law, 
physician perform abortion, 
small business hire worker, 
america create job, 
american lose faith , 
american purchase healthcare, 
asset back military, 
self-employed executes budget, 
soldier give family, 
individual meet regulation , 
supreme rule unconstitutional, 
iran pose threat, 
people have difficulty, 
fund make funding, 
money pay money, 
program provide fund, 
community receive grant, 
mother give baby (boomer), 
bank provide loan, 
veteran serve nation, 
parent lose child, 
democratic offer alternative, 
physician assist suicide, 
money fund outlay, 
administration propose budget, 
institute conduct study, 
F.D.A. have authority, 
fed raise interest/rate, 
program provide resource, 
outlay exceed payment, 
program benefit american, 
agency share information , 
employee perform job, 
family put food, 
application meet regulation , 
reserve raise interest/rate, 
student receive degree (education), 
threat face nation, 
company employ people, 
people want buy, 
men women defend nation, 
electronic device save family, 
american hold love, 
people lose everything, 
appropriation fund program, 
budget reduces deficit, 
student get education , 
employee make decision, 
agency require information , 
service control small business, 
employee hire people, 
person take healthcare, 
american want together, 
crime commit crime, 
american keep healthcare, 
china join wto, 
person lose job, 
defense submit service, 
people lose love, 
republican want tax, 
tax benefit american, 
kid need help, 
physician deliver baby (boomer), 
news sign text, 
appropriation report appropriation, 
attorney (judge) enforce law, 
people do harm, 
program help poor, 
employee provide institution, 
god bless nation, 
people borrow money, 
american paid tax, 
crime use gun, 
project meet criterion, 
american deserve (tax) relief, 
fact impose tax, 
business make profit, 
small business hire people, 
people keep staff, 
network provide service, 
american face problem, 
american face challenge, 
people want program, 
measure breach budget, 
people want help, 
government have power, 
country have deficit, 
american have enough, 
administration submit budget, 
business provide healthcare, 
child lose healthcare, 
people solve problem, 
tax stimulate economy, 
employee pay overtime, 
company sell consumer, 
public provide service, 
american want action, 
people have exhaust, 
employee hire permanent, 
director determines appropriate, 
person have baby (boomer), 
american have faith , 
worker earn minimum wage, 
student pursue education , 
agency receives fund, 
environmental protection agency regulate emission (pollution), 
american serve nation, 
government estimate contract, 
indian provide service, 
american send money, 
challenge face american, 
people take responsibility, 
attorney (judge) make determination, 
physician practice healthcare, 
justice write opinion, 
democrat have plan, 
country need help, 
project meet regulation , 
country make progress, 
nothing limit authority, 
government not-do nothing, 
student afford school, 
american rely oil, 
production shut supply and demand, 
company pay tax, 
worker support retiree, 
self-employed deduct healthcare, 
child have special (need/rule), 
school do job, 
water resource development authorize river, 
patient receive healthcare, 
service do job, 
people need (tax) relief, 
people do wrong, 
healthcare deny coverage, 
healthcare provide coverage, 
american keep job, 
measure would budget, 
program mean program, 
country sign treaty, 
bush nominate attorney (judge), 
community need help, 
people paid tax, 
business take opportunity, 
people wear clothes, 
people work money, 
government receive funding, 
people have government, 
democrat balance budget, 
people keep healthcare, 
F.D.A. approve facility, 
american pay fair, 
republican put budget, 
people serve military, 
someone have fact, 
iran obtain nuclear (weapon), 
physician make decision, 
mother pay tax, 
person make global warming, 
veteran need help, 
administration do everything, 
country participate program, 
ship carry passenger, 
person risk family, 
nation take action, 
veteran put family, 
appropriate rely payment, 
people live family, 
person earn men women, 
republican create job, 
country face threat, 
kid have opportunity, 
republican use filibuster, 
administration request amount, 
family have income, 
money fund abortion, 
threat face country, 
administration cut program, 
people lose saving, 
administration do job, 
people take action, 
offender commit crime, 
coal fire electricity, 
faith  provide service, 
country have problem, 
law require budget, 
parent make choise, 
policy held debt, 
people make profit, 
defense transfer fund, 
manufacturer collect sale, 
government make promise, 
farmer plant crop, 
small business purchase healthcare, 
law authorize program, 
nation face problem, 
tax have effect, 
agency make decision, 
community make decision, 
grant give priority, 
budget cut education , 
people not-afford healthcare, 
administration threaten veto, 
immigrant pay tax, 
program benefit people, 
people provide healthcare, 
founder start business, 
american have housing, 
american have service, 
national preserve social security, 
money fund program, 
american provide family, 
federally fund project, 
patient save family, 
education  award grant, 
parent take child, 
someone have job, 
interest/rate authorize agency, 
innocent lose family, 
ship launch missile, 
individual pay tax, 
oil account america, 
parent have child, 
law authorize appointment, 
parent protect child, 
nothing do nothing, 
insurance drop coverage, 
country take opportunity, 
mother lose job, 
hero give family, 
people try help, 
people have power, 
legislative (election) call word, 
community provide funding, 
people pay debt, 
budget add trillion, 
veteran find job, 
american want reform, 
government run insurance, 
child have child, 
law prohibit practice, 
people have life quality, 
person make workforce, 
force base defense, 
someone raise staff, 
law authorize project, 
people do everything, 
staff put together, 
war torn country, 
plan provide (tax) relief, 
country lose job, 
budget reach together, 
budget report budget, 
american make ultimate sacrifice, 
agency award contract, 
people have opinion, 
kid graduate school, 
girl attend school, 
flood cause damage, 
senior have medicare, 
someone need help, 
men women fight war, 
american make difference, 
republican give tax, 
agency carry program, 
country have healthcare, 
physician provide healthcare, 
american deserve nothing, 
people receive funding, 
people make family, 
employee use (sick/paid/unpaid) leave, 
terrorist enter country, 
funding orphan child, 
juvenile commit violent crime, 
family adopt child, 
administration take position, 
job require skill, 
news face nation, 
people have cosponsor, 
republican end medicare, 
applicant submit application, 
program provide help, 
money fund subsidy, 
student leave school, 
story wrench heart, 
message transmit concurrent, 
news face country, 
plan have effect, 
person terminate pregnancy, 
patient need healthcare, 
medicare have prescription drug, 
people work small business, 
decision have effect, 
job pay minimum wage, 
people put food, 
child lose family, 
doctor make healthcare, 
person choose abortion, 
worker pay tax, 
terrorist attack country, 
american hold accountable, 
soldier pierce tank, 
small business pay tax, 
doctor practice offense, 
people support family, 
community participate program, 
someone get tax, 
institution provide service, 
people have confusion, 
american access healthcare, 
american have ballot, 
family make decision, 
raw material  hit road, 
fund pay profit, 
people leave welfare, 
iraq invade kuwait, 
government have authority, 
student enter school, 
U.S.C. authorize uniform, 
someone commit crime, 
family balance budget, 
someone have experience, 
law provide authority, 
child eat lunch, 
medicaid provide healthcare, 
national conduct study, 
government issue identification, 
veteran receive compensation, 
law provide remedy, 
city  mean city , 
grant provide service, 
administration propose tax, 
american do nothing, 
military make sacrifice, 
supreme declare unconstitutional, 
government take opportunity, 
america need help, 
small business create america, 
senior join H.M.O., 
family save money, 
people keep job, 
firefighter open fire, 
government issue picture, 
people move job, 
air force request application, 
internal revenue code strike period, 
mother have baby (boomer), 
employee sponsor insurance, 
corporation  pay fair, 
bank lend money, 
government pay money, 
kid start smoking, 
someone do business, 
program outlive effectiveness, 
program provide profit, 
someone have heart, 
nation sign treaty, 
god rest love, 
company have employee, 
individual lose job, 
worker do job, 
student make difference, 
government save money, 
agency conduct analysis, 
budget cut medicaid, 
people go job, 
parent care child, 
administration provide information , 
small business get loan, 
american do best, 
money outweigh money, 
physician treat patient, 
unemployment (benefits) look job, 
attorney (judge) make law, 
people get coverage, 
people balance budget, 
individual make decision, 
employee create job, 
program serve american, 
medicare cover service, 
american work money, 
railroad provide transportation, 
money get money, 
program need reform, 
filibuster raise tax, 
america need energy, 
agency make determination, 
wealthy pay fair, 
people help people, 
student receive funding, 
party reach agreement, 
program help family, 
veteran take opportunity, 
natural resource recommend substitute, 
people kill people, 
nuclear (weapon) pose threat, 
people have common sense, 
child lack healthcare, 
program take healthcare, 
people wear uniform, 
O.P.E.C. increase production, 
person have diabetes, 
market take healthcare, 
people have disease, 
senior pay deductible, 
republican cut education , 
program have success, 
police give family, 
consumer cause harm, 
child start smoking, 
american use site, 
person make choise, 
oil hit gasoline, 
republican take action, 
parent take healthcare, 
individual lose family, 
small business pool resource, 
child get good start, 
someone paid attention, 
police do job, 
child reach adult, 
government provide information , 
school use fund, 
healthcare improve healthcare, 
people preexist condition, 
employee not-provide healthcare, 
person make payment, 
program provide technical assistance, 
person commit offense, 
american deserve healthcare, 
american receive money, 
employee offering healthcare, 
republican want repeal (something), 
people want tax, 
people share concern, 
veteran receive recognition, 
senior make choise, 
people want america, 
senior afford prescription drug, 
job require education , 
nation love freedom, 
people not-get job, 
family leave welfare, 
parent send kid, 
appropriation provide funding, 
(sick/paid/unpaid) leave be columbia, 
father have diabetes, 
family face mortgage, 
government run plan, 
law authorize activity, 
capital  gain interest/rate, 
american put money, 
veteran have healthcare, 
insurance provide coverage, 
america win war, 
money make difference, 
program provide essential , 
veteran participate program, 
university conduct study, 
authority make decision, 
measure not-increase deficit, 
doctor acquire disease, 
student use technology, 
bank provide service, 
american work government, 
problem face america, 
veteran serve war, 
veteran seek healthcare, 
person need help, 
american obtain healthcare, 
senior depend medicare, 
people have plan, 
county mean county, 
winners and losers appropriate fund, 
family take opportunity, 
employee provide money, 
family make choise, 
facility provide service, 
founder endow men women, 
person have heart, 
men women serve arm force, 
people have appreciation, 
tobacco cause cancer, 
county lose population, 
people make america, 
juvenile commit crime, 
amtrak provide service, 
saddam hussein use chemical (weapon), 
person lay family, 
individual take opportunity, 
fund carry funding, 
saddam hussein use weapon mass destruction, 
government have responsibility, 
community use fund, 
party raise soft money, 
nation lead world, 
small business provide service, 
agency use information , 
veteran earn money, 
bureaucrat make decision, 
american not-afford healthcare, 
employee have healthcare, 
application decrease appropriation, 
party seek remedy, 
american feel pain, 
agency promulgate regulation , 
oil pay fee, 
student leave discipline, 
individual buy healthcare, 
people have feeling, 
major issue face country, 
commitment have witness, 
budget provide (tax) relief, 
soldier serve country, 
grant fund program, 
saddam hussein invade kuwait, 
person control small business, 
country lead world, 
medicare run money, 
worker exhaust money, 
people pay income, 
country make fund, 
family receive money, 
agency receive amount, 
american make fund, 
job support family, 
agency collect information , 
plan amends plan, 
disability receive education , 
african make contribution, 
judiciary conduct promulgate regulation, 
someone have diabetes, 
child need service, 
provider offer service, 
employee perform function, 
father write constitution, 
money hold bag (trash), 
person enter contract, 
medicare manage plan, 
people promote welfare, 
person graduate school, 
university do study, 
school receive funding, 
bible fulfil self-employed, 
student afford education , 
individual commit crime, 
medicare provide prescription drug, 
corporation  take opportunity, 
program have (positive/negative) effect, 
student pay interest/rate, 
individual make legislative (election), 
people save family, 
family lose healthcare, 
people get loan, 
grant provide grant, 
government receive grant, 
idea have merit, 
program do nothing, 
store collect sale, 
weapon pierce tank, 
person receive amount, 
american lose insurance, 
company ship job, 
worker have exhaust, 
people buy consumer, 
fund provide funding, 
someone get healthcare, 
republican do everything, 
mother lose mother, 
mother graduate school, 
american sell consumer, 
american work minimum wage, 
reconciliation carry recommendation, 
supreme have (legal) statute, 
prescription drug save family, 
republican raise tax, 
school educate child, 
country head direction, 
community perform abortion, 
worker exhaust unemployment (benefits), 
people kill american, 
treasury strikingfor purpose, 
offender serve (legal) sentence, 
railroad provide service, 
healthcare take healthcare, 
patient receive patient, 
people want spend, 
american work family, 
doctor see medicare, 
agency desire grant, 
american lose concern, 
republican take money, 
individual bearing interest/rate, 
government give money, 
healthcare work american, 
republican gain power, 
government incur money, 
employee do business, 
transportation issue certification, 
business have employee, 
people have ability, 
plan cut tax, 
senior get healthcare, 
american owe appreciation, 
people not-pay attention, 
company participate program, 
faith  base community, 
service connect condition, 
program have bipartisan (support), 
people drive mile , 
american pay dollar, 
O.P.E.C. cut production, 
american have life quality, 
air force indicate appointment, 
government have money, 
people work american, 
people fight war, 
people leave country, 
american support action, 
law authorize study, 
attorney (judge) represent client, 
program reduce disease, 
administration enforce law, 
consumer make instruct, 
american fight war, 
project provide water, 
applicant receive grant, 
grant receive grant, 
government make payment, 
program provide community, 
baby (boomer) reach adult, 
people try job, 
american have idea, 
someone know someone, 
child have food, 
democrat have idea, 
small business need help, 
people face mortgage, 
government protect citizen, 
people sign petition, 
republican provide (tax) relief, 
school make decision, 
program help nation, 
people have trouble, 
money paid amount, 
federally assist housing, 
servicemembers lose family, 
budget receive recommendation, 
republican propose medicare, 
senior rely medicare, 
person obtain abortion, 
disaster affected area, 
republican have opportunity, 
transaction lack abuse, 
environmental protection agency do job, 
regulation  have effect, 
budget pay debt, 
F.D.A. approve plant, 
people use money, 
veteran suffer illness, 
senior not-afford prescription drug, 
troop make sacrifice, 
trade create job, 
people not-have money, 
constitution vested power, 
administration make bad (problem), 
police lose family, 
patient get child, 
american rely program, 
oil reach gasoline, 
nation make progress, 
good start serve child, 
people use drug, 
american have confusion, 
men women defend country, 
people work action, 
american deserve government, 
people have courage, 
people bear staff, 
program help community, 
people establish justice, 
republican make promise, 
healthcare make healthcare, 
budget do job, 
american make life quality, 
american buy car, 
program benefit child, 
injustice motivate crime, 
people make comment, 
veteran need service, 
citizen do extraordinary, 
corporation  make charity, 
illegal immigrant commit crime, 
small business afford healthcare, 
judiciary do job, 
plan have merit, 
social security pay money, 
person receive money, 
F.D.A. approve medicine, 
people provide family, 
student pay loan, 
people have freedom, 
national submit intelligence, 
democrat have power, 
program provide housing, 
business pay tax, 
violent crime serve (legal) sentence, 
healthcare save money, 
people pay mortgage, 
law confer power, 
people not-want job, 
medicare pay doctor, 
people have resource, 
resource do job, 
national make recommendation, 
employee provide notice, 
program help farmer, 
people commit suicide, 
student bring gun, 
business employ american, 
measure provide authority, 
people make judgment, 
bush threaten veto, 
child receive food, 
american want soluton, 
medicine save family, 
doctor save family, 
oil create job, 
unemployment (benefits) find job, 
america pay tax, 
child brought gift, 
child have succeed, 
american exhaust unemployment (benefits), 
someone have concern, 
girl have baby (boomer), 
government pay debt, 
worker change job, 
tax reduce deficit, 
water provide water, 
school reduce class size, 
family attend school, 
person perform service, 
program serve people, 
company provide healthcare, 
constitution prohibit america, 
county provide service, 
people take loan, 
people establish constitution, 
people watch network, 
employee perform service, 
good start provide service, 
government work american, 
airline lose job, 
employee provide education , 
pesticide cause disease, 
child watch network, 
someone drive car, 
victim receive compensation, 
people have nothing, 
people have commitment, 
disability limit major issue, 
iraqi  take power, 
iraq pose jeopardy, 
someone want tax, 
defendant award punitive damage, 
employee maintain plan, 
america have opportunity, 
parent have choise, 
appropriation provide fund, 
judiciary take action, 
budget take healthcare, 
eligible participate program, 
school receive fund, 
bank do business, 
worker lose healthcare, 
refinery make profit, 
american take grant, 
money amount taxable, 
program achieve goal, 
nation face fight, 
budget make bad (problem), 
director issue information , 
(sick/paid/unpaid) leave be virgin , 
ethanol blend gasoline, 
senior receive healthcare, 
kid not-have healthcare, 
farmer lose crop, 
law provide protection, 
business provide job, 
republican offer budget, 
attorney (judge) promulgate regulation , 
nation love peace, 
men women write constitution, 
social security strike period, 
worker look job, 
someone serve country, 
diabetes affect american, 
program help the needy, 
plan reduce deficit, 
american send child, 
congressional budget define money, 
export create job, 
people attend school, 
china manipulate dollar, 
program do job, 
farmer have crop, 
irs target group, 
production use american, 
american provide service, 
budget set priority, 
folk do job, 
citizen have healthcare, 
nation need energy, 
administration mislead american, 
homeland attack unit, 
business estimate tax, 
people get tax, 
lender have lender, 
child have illness, 
someone get everything, 
opinion show american, 
government not-do job, 
small business employ workforce, 
child have cancer, 
student pursue degree (education), 
witness limit medicine, 
small business provide job, 
crime use weapon, 
group provide service, 
disaster cause damage, 
program carry program, 
american buy insurance, 
employee sponsor saving, 
employee offer coverage, 
toxic (thing) cause cancer, 
employee provide employee, 
project provide service, 
people face problem, 
application contain information , 
men women do nothing, 
nation make commitment, 
someone want buy, 
people stop smoking, 
american do everything, 
government make commitment, 
guard save family, 
company bring job, 
administration have plan, 
people broken law, 
agency conduct program, 
self-employed employ individual, 
director take action, 
plan save money, 
plant lose job, 
program receive grant, 
employee put money, 
people loan money, 
american drive car, 
car cross border, 
american produce energy, 
small business receive loan, 
katrina hit ocean, 
burmese contain export, 
someone have child, 
federally fund healthcare, 
republican refuse action, 
lender participate program, 
people have disability, 
budget increase debt, 
people lose concern, 
national provide service, 
government make healthcare, 
employee have employee, 
police put family, 
person make healthcare, 
student get loan, 
american earn minimum wage, 
administration negotiate trade, 
soldier make sacrifice, 
american scratch staff, 
agency give priority, 
brave men women serve nation, 
nothing affect authority, 
program receives funding, 
people get service, 
brave men women make sacrifice, 
men women lose job, 
child receive funding, 
family afford healthcare, 
people send kid, 
community base community, 
someone do everything, 
bush nominate circuit (courts), 
student earn degree (education), 
someone buy gun, 
unemployment (benefits) exhaust money, 
uniform make sacrifice, 
student teach child, 
person get healthcare, 
uniform serve country,

\clearpage
\section{List of entities labeled as procedural or noise terms} \label{app:procedural_and_noise}
\setcounter{table}{0}

We present here the entities that we consider to be procedural or not interpretable after dimension reduction. We name them by the most frequent term in their cluster. These entities are excluded in our main analysis. The list below shows all such excluded entities.

absolutely right, accept, accomplish, accurate, adam, address, adjustment, administrative office, advisory, afford, agreeable, akin, alan, allegiance, amaze, anderson, andrew, angel, approach, argument, aristide, arlen, available, ayes, baca, ball, barbara, bartlett, basic, be, beautiful, behind, ben, benjamin, berkley, berman, biggert, billy, bipartisanship, blm, bob, bobby, body, bolton, bonner, bonnie, boswell, bottom, brian, briefing, brook, brownfields, buck, buckley, building, bumper, bureau, burn, burton, byron, cabinet, cafta, candidate, cardoza, carl, carolyn, certain, chamber, chandler, chapter, charles, charlie, cheap, cheney, chenoweth, chris, christopher, chuck, clark, classify annex, clear, close, cloture, coat, code, colin powell, colloquy, columbus, commenters, complicate, compromise, comptroller, conferee, conference, confirm, congressional caucus, congressional gold medal, congressman, consent, conservation, consideration, continue, corp, corrine, costello, counsel, couple, cranston, damato, dan, daniel, danny, date, dave, david, day, de la garza, debate, debbie, demint, dennis, department, designate, desjarlais, determine, dialogue, dianne, diaz balart, dice, dick, dick cheney, different, dirksen office building, disabled, disagree, disappoint, discus, distinguish, division, do, doable, document, don, double, downstream, draconian, draft, dreier, due, ed, eddie bernice, edward, elizabeth, emerge, enforce, engage, enter, entity, eric, erisa, essence, establish, estimate, estrada, exactly, exceed, excite, executive, exist, extension, extent, extraneous material, extremely important, eye, factor, familiar, fannie, far, far proceeding, federal, filner, final, fine, finish, first, first responder, fit, follow, foot, formal, forward, frank, frank frank, frank gallegly, fred, freddie, fy, ga, gallegly, gary, gate, generalparagraph, george, gibbon, ginny, glosson, go, government accountability office, government printing office, governor, graf, gram, grisham, half, ham, happen, happy, harry, hearing, helm, helpful, henry, herger, herseth, high, hinchey, hj re, holden, hollen, homeownership, hour equally divide, hour workweek, howard, hr, hussein, ii, iii, implement, implication, important, inc, inspire, interior, introduce, ironic, irresponsible, item, jack, james, james madison, jeff, jerry, jesse, jim, jimmy, jo ann, joe, johnny, join, joint, joint chief staff, jon, joseph, journal last day, jr, judicial conference, kansa, keith, kent, kucinich, lahood, larry, late, latourette, ledbetter, letter, linda, list, lng, local coordinate entity, loretta, lot, lungren, ma, madam, madison, mail, main street, majority, manner, manzullo, mark, martha, martin, mary, max, mccotter, mcdonald, mcmorris, md, medically necessary, mel, michael, miguel estrada, mike, milosevic, minute, minute remain, mitch, mn, modify, moment, multiple, myrick, na, naive, nancy, nc, nd, net neutrality, newark, newt, nez, nick, nj, nomination, non, norton, np, nt do, numbered, numerous, objection, observation, offer, office, ok, olver, open, oppose, orrin, outside, overall, oversight, overwhelm, owen, page, park, parliamentarian, pas, pat, patrick, patten, patty, paulson, percent, perez, pete, peter, petitioner, petraeus, phil, pitt, plate, point, point order, politically correct, polluter, popular, post, powell, preside officer, president, president tempore, procedure, process, projection, proponent, provision, qualify, quayle, question, quick, randy, rank, rapidly, re, read, ready, recommit, reconsider, redesignate, redesignating, reduction, regular order, rehberg, rehnquist, rein, remain available expend, remark, reno, repeat, report, representative, requisite number word, revolve fund, reyes, richard, rick, riggs, robert, rodney, roger, ron, rothman, rule, rule administration, rulemaking, rumsfeld, russell, russell office building, sam, sander, save, schedule, schmidt, sean, secure, select, select intelligence, select reserve, sen, sense, separate, session, sex offender, shay, shuler, side, sj re, skelton, solomon, somalia, sorry, sotomayor, southern, southwick, specie, speech, spouse, st, stack, standard, starr, statement, station, status quo, stearns, stennis, steve, straight, string, structure, stump, subparagraph, subtitle, sufficient second, sullivan, sunset, supplemental, support, susan, sutton, table, ted, territory, th, thanks, thing, thornton, tim, tim myrick, time, tn, todd, tom, tomorrow, tonight, tradition, tsa, tx, unacceptable, underfunded, underlie, unique, unprecedented, upon, va, valid, ve get, version, view, vitally important, wait, walter, wasserman, way mean, wayne, whatever, whichever, whole, whose, william, willing do, withdraw, woolsey, worthwhile, xiv, xx, xxi, xxii, year, year old, zero, zoe, 

\clearpage
\section{List of Stopwords} \label{sec:stopwords}
\setcounter{table}{0}

We remove common words in English, procedural terms specific to the U.S. Congressional Record, numbers, as well as all the names of Congress members and U.S. States. The complete list is below: \\

i
me
my
myself    
we 
us
our            
ours           
ourselves      
you            
your          
yours         
yourself       
yourselves    
he             
him           
his           
himself      
she            
her            
hers           
herself       
it             
its            
itself         
they           
them           
their          
theirs         
themselves     
what
which
who
whom
this
that
these
those
a
an
the
and
but
if
or
because
as
until
while
of
at
by
for
with
about
against
between
into
through
during
before
after
above
below
to
from
up
down
in
out
on
off
over
under
again
further
then
once
here
there
when
where
why
how
all
any
both
each
few
more
most
other
some
such
no
nor
not
only
own
same
so
than
too
very
much
actual
abraham
aderholt
adler
akaka
alexander
allard
allen
amash
amodei
andrews
applegate
archer
arcuri
armey
ashcroft
ayotte
bacchus
bachmann
bachus
baesler
baird
baker
baldacci
baldwin
ballance
ballenger
barber
barca
barcia
barkley
barletta
barlow
barr
barrasso
barrett
barrow
barton
bass
bateman
baucus
bayh
bean
beatty
beauprez
becerra
begich
beilenson
bell
benishek
bennet
bennett
bentivolio
bentley
bentsen
bera
bereuter
berry
bevill
biden
bilbray
bilirakis
bingaman
bishop
black
blackburn
blackwell
blagojevich
bliley
blumenauer
blumenthal
blunt
blute
boccieri
boehlert
boehner
bonamici
bond
bonilla
bonior
bono
boozman
boren
borski
boucher
boustany
boxer
boyd
boyda
bradley
brady
braley
breaux
brewster
bridenstine
bright
brooks
broun
browder
brown
brownback
brownley
bryan
bryant
buchanan
bucshon
bumpers
bunn
bunning
burgess
burns
burr
burris
bustos
butterfield
buyer
byrd
byrne
callahan
calvert
camp
campbell
canady
cannon
cantor
cantwell
cao
capito
capps
capuano
cardenas
cardin
carnahan
carney
carper
carson
carter
cartwright
case
casey
cassidy
castle
castor
castro
cazayoux
chabot
chafee
chaffetz
chambliss
chapman
chiesa
childers
chocola
christensen
chrysler
chu
cicilline
clarke
clay
clayton
cleaver
cleland
clement
clinger
clinton
clyburn
coats
coble
coburn
cochran
coffman
cohen
cole
coleman
collins
combest
conaway
condit
connolly
conrad
conyers
cook
cooksey
cooley
coons
cooper
coppersmith
corker
cornyn
corzine
costa
cotton
courtney
coverdell
cox
coyne
craig
cramer
crane
crapo
crawford
cremeans
crenshaw
crowley
cruz
cubin
cuellar
culberson
cummings
cunningham
dahlkemper
daines
danforth
danner
darden
daschle
davis
dayton
deal
deconcini
defazio
degette
delahunt
delaney
delauro
delay
dellums
denham
dent
derrick
deutch
deutsch
dewine
dickey
dingell
dixon
djou
dodd
doggett
dole
domenici
donnelly
dooley
doolittle
dorgan
doyle
drake
driehaus
duckworth
duffy
duncan
dunn
durbin
durenberger
edwards
ehlers
ehrlich
ellison
ellmers
ellsworth
emanuel
emerson
engel
english
ensign
enyart
enzi
eshoo
esty
etheridge
evans
everett
ewing
exon
faircloth
fallin
farenthold
farr
fattah
fawell
fazio
feeney
feingold
feinstein
ferguson
fields
fincher
fingerhut
fischer
fish
fitzgerald
fitzpatrick
flake
flanagan
fleischmann
fleming
fletcher
flores
foley
forbes
ford
fortenberry
fossella
foster
fowler
fox
foxx
frahm
frankel
franken
franks
frelinghuysen
frisa
frist
frost
fudge
funderburk
furse
gabbard
gallego
gallo
ganske
garamendi
garcia
gardner
garrett
gejdenson
gekas
gephardt
geren
gerlach
gibbons
gibbs
gibson
giffords
gilchrest
gillibrand
gillmor
gilman
gingrey
gingrich
glenn
glickman
gohmert
gonzalez
goode
goodlatte
goodling
goodwin
gordon
gorton
gosar
goss
gowdy
graham
gramm
grams
grandy
granger
grassley
graves
grayson
green
greene
greenwood
gregg
griffin
griffith
grijalva
grimm
grucci
gunderson
guthrie
gutierrez
gutknecht
hagan
hagel
hahn
hall
halvorson
hamburg
hamilton
hanabusa
hancock
hanna
hansen
hare
harkin
harman
harper
harris
hart
hartzler
hastert
hastings
hatch
hatfield
hayes
hayworth
heck
hefley
heflin
hefner
heineman
heinrich
heitkamp
heller
helms
hensarling
herrera
higgins
hill
hilleary
hilliard
himes
hinojosa
hirono
hoagland
hobson
hochbrueckner
hodes
hoeffel
hoekstra
hoeven
hoke
holding
hollings
holt
honda
hooley
horn
horsford
hostettler
houghton
hoyer
hudson
huelskamp
huffington
huffman
hughes
huizenga
hulshof
hultgren
hunter
hurt
hutchinson
hutchison
hutto
hyde
inglis
inhofe
inouye
inslee
isakson
israel
issa
istook
jackson
janklow
jefferson
jeffords
jeffries
jenkins
jindal
johanns
john
johnson
johnston
jones
jordan
joyce
kagen
kaine
kanjorski
kaptur
kasich
kassebaum
kaufman
keating
keller
kelly
kempthorne
kennedy
kennelly
kerns
kerrey
kerry
kildee
kilmer
kilpatrick
kilroy
kim
kind
king
kingston
kinzinger
kirk
kirkpatrick
kleczka
klein
kline
klink
klobuchar
klug
knollenberg
kohl
kolbe
kopetski
kosmas
kratovil
kreidler
kuhl
kuster
kuykendall
kyl
labrador
lafalce
lamborn
lampson
lancaster
lance
landrieu
langevin
lankford
lantos
largent
larsen
larson
latham
latta
laughlin
lautenberg
lazio
leach
leahy
lee
lehman
levin
levy
lewis
lieberman
lightfoot
lincoln
linder
lipinski
livingston
lloyd
lobiondo
loebsack
lofgren
long
longley
lott
lowenthal
lowey
lucas
luetkemeyer
lugar
lujan
lummis
luther
lynch
machtley
mack
maffei
mahoney
majette
maloney
manchin
mann
manton
marchant
marino
markey
marshall
martinez
martini
mascara
massa
massie
matheson
mathews
matsui
mazzoli
mccain
mccandless
mccarthy
mccaskill
mccaul
mcclintock
mccloskey
mccollum
mcconnell
mccrery
mccurdy
mcdermott
mcgovern
mchale
mchenry
mchugh
mcinnis
mcintosh
mcintyre
mckeon
mckinley
mckinney
mcleod
mcmahon
mcmillan
mcnerney
mcnulty
meadows
meehan
meek
meeks
melancon
menendez
meng
merkley
messer
metcalf
meyers
mfume
mica
michaud
michel
mikulski
miller
mineta
minge
minnick
mitchell
moakley
molinari
mollohan
montgomery
moore
moorhead
moran
morella
moynihan
mullin
mulvaney
murkowski
murphy
murray
murtha
musgrave
myers
nadler
napolitano
natcher
neal
nelson
nethercutt
neugebauer
neumann
ney
nickles
noem
nolan
northup
norwood
nugent
nunes
nunn
nunnelee
nussle
nye
obama
oberstar
obey
olson
ortiz
orton
osborne
ose
otter
owens
oxley
packard
packwood
palazzo
pallone
pappas
parker
pascrell
pastor
paul
paulsen
paxon
payne
pearce
pease
pell
pelosi
penny
perlmutter
perriello
perry
peters
peterson
petri
phelps
pickering
pickett
pickle
pingree
pittenger
pitts
pocan
poe
polis
pombo
pomeroy
pompeo
porter
portman
posey
poshard
pressler
price
pryce
pryor
putnam
quigley
quillen
quinn
radanovich
radel
rahall
ramstad
rangel
ravenel
redmond
reed
regula
reichert
reid
renacci
renzi
reynolds
ribble
rice
richardson
richmond
ridge
riegle
rigell
riley
risch
rivers
robb
roberts
roby
rockefeller
rodgers
rodriguez
roe
roemer
rogan
rogers
rohrabacher
rokita
rooney
ros
rose
roskam
ross
rostenkowski
roth
rothfus
roukema
rowland
royce
rubio
ruiz
runyan
ruppersberger
rush
ryan
ryun
sabo
salazar
sali
salmon
sanchez
sanders
sandlin
sanford
sangmeister
santorum
sarbanes
sarpalius
sasser
sawyer
saxton
scalise
scarborough
schaefer
schaffer
schakowsky
schatz
schauer
schenk
schiff
schneider
schock
schrader
schrock
schroeder
schultz
schumer
schwartz
schwarz
schweikert
scott
seastrand
sekula
sensenbrenner
serrano
sessions
sestak
sewell
shadegg
shaheen
sharp
shaw
shays
shelby
shepherd
sherman
sherwood
shimkus
shows
shuster
simmons
simon
simpson
sinema
sires
sisisky
skaggs
skeen
slattery
slaughter
smith
snowbarger
snowe
snyder
sodrel
solis
souder
southerland
space
specter
speier
spence
spratt
stabenow
stenholm
stevens
stewart
stivers
stockman
stokes
strickland
studds
stupak
stutzman
sundquist
sununu
swalwell
sweeney
swett
swift
synar
takano
talent
tancredo
tanner
tate
tauscher
tauzin
taylor
tejeda
terry
tester
thomas
thompson
thornberry
thune
thurman
thurmond
tiahrt
tiberi
tierney
tipton
titus
tonko
toomey
torkildsen
torres
torricelli
traficant
tsongas
tucker
turner
udall
unsoeld
upton
valadao
valentine
van
vargas
veasey
vela
velazquez
vento
visclosky
vitter
voinovich
volkmer
vucanovich
wagner
walberg
walden
walker
wallop
walorski
walsh
walz
wamp
ward
warner
warren
waters
watson
watt
watts
waxman
webb
weber
webster
weiner
welch
weldon
weller
wellstone
wenstrup
westmoreland
wexler
weygand
wheat
white
whitehouse
whitfield
whitten
wicker
williams
wilson
wise
wittman
wofford
wolf
womack
woodall
wu
wyden
wynn
yarmuth
yoder
yoho
young
zeliff
zimmer
aye
section
subsection
sec
act 
roll
yea
yes
nai
no
none
nay
mr
mrs
senator
gentleman
gentlemen
gentlewoman
gentlewomen
gentlelady
gentleladies
sir
committee
speaker
chairman
ask
yield
will
shall
can
say
let
adjourn
speak
etc
thank
absent
tell
per
part
now
month
today
con
pro
thereabout
thereafter
thereagainst
thereat
therebefore
therebeforn
thereby
therefor
therefore
therefrom
therein
thereinafter
thereof
thereon
thereto
theretofore
thereunder
thereunto
thereupon
therewith
therewithal
hereabout
hereafter
hereat
hereby
herein
hereinafter
hereinbefore
hereinto
hereof
hereon
hereto
heretofore
hereunto
hereunder
hereupon
herewith
whereabouts
whereas
whereafter
whereat
whereby
wherefore
wherefrom
wherein
whereinto
whereof
whereon
whereto
whereunder
whereupon
wherever
wherewith
wherewithal
bill
house
member
senate
amendment
section
subsection
committee
subcommittee
line
article
vote
insert
clerk
speaker
district
state
adjourn
monday 
tuesday
wednesday
thursday
friday
saturday
sunday
january
february
march
april
may
june
july
august
september
october
november
december
adjourn
board
general
legislature
legislation
parliament
amend
reject
appeal
appear
answer
appoint
ask
author
calendar
chair
chairman
colleague
commission
congress
court
file
motion
floor
leader
object
paragraph
unanimous
pending
revise
quorum
permission
secretary
resolution
pass
record
print
current
one
two
three
four
five
six
seven
eight
nine
ten
eleven
twelve
thirteen
fourteen
fifteen
sixteen
seventeen
eighteen
nineteen
twenty
thirty
fourty
fifty
sixty
seventy
eighty
ninety
hundred
hundreds
thousand
thousands
alabama
alaska
arizona
arkansas
california
colorado
connecticut
delaware
florida
georgia
hawaii
idaho
illinois
indiana
iowa
kansas
kentucky
louisiana
maine
maryland
massachussetts
michigan
minnesota
mississippi
missouri
montana
nebraska
nevada
hampshire
new
jersey
mexico
york
carolina
north
south
east
west
dakota
ohio
oklahoma
oregon
pennsylvania
island
rhode
tennessee
texas
utah
vermont
virginia
washington
wisconsin
wyoming
united

\end{document}